\pdfoutput=1
\documentclass[11pt,twoside,a4paper,cmspaper,final,collab]{cms-tdr}
\def\svnVersion{6b9a0c5}\def\svnDate{2023/09/14}\def\cmsCernNoTag{CERN-EP-2023-004}\def\cmsCernDate{\today}\def\cmsMessage{Published in the European Physical Journal C as \href{http://dx.doi.org/10.1140/epjc/s10052-023-11952-7}{\doi{10.1140/epjc/s10052-023-11952-7}.}}
\begin{document}\cmsNoteHeader{HIG-21-007}

\providecommand{\cmsTable}[1]{\resizebox{\textwidth}{!}{#1}}
\newlength\cmsFigWidi\ifthenelse{\boolean{cms@external}}{\setlength{\cmsFigWidi}{0.23\textwidth}}{\setlength{\cmsFigWidi}{0.4\textwidth}}
\newlength\cmsFigWidviii\ifthenelse{\boolean{cms@external}}{\setlength{\cmsFigWidviii}{0.57\textwidth}}{\setlength{\cmsFigWidviii}{0.62\textwidth}}
\newlength\cmsFigWidix\ifthenelse{\boolean{cms@external}}{\setlength{\cmsFigWidix}{0.48\textwidth}}{\setlength{\cmsFigWidix}{0.65\textwidth}}
\newcommand{\singleMuCr}{\ensuremath{\PGm + \text{jets}}\xspace}
\newcommand{\doubleMuCr}{\ensuremath{\PGm\PGm + \text{jets}}\xspace}
\newcommand{\singleEleCr}{\ensuremath{\Pe + \text{jets}}\xspace}
\newcommand{\doubleEleCr}{\ensuremath{\Pe\Pe + \text{jets}}\xspace}
\newcommand{\singlePhotonCr}{\ensuremath{\PGg + \text{jets}}\xspace}
\newcommand{\vjets}{\ensuremath{\PV + \text{jets}}\xspace}
\newcommand{\ttbarjets}{\ensuremath{\ttbar + \text{jets}}\xspace}
\newcommand{\wjets}{\ensuremath{\PW + \text{jets}}\xspace}
\newcommand{\zjets}{\ensuremath{\PZ + \text{jets}}\xspace}
\newcommand{\ljets}{\ensuremath{\Pell + \text{jets}}\xspace}
\newcommand{\lljets}{\ensuremath{\Pell\Pell + \text{jets}}\xspace}
\newcommand{\doubleMuMass}{\ensuremath{m_{\Pgm\Pgm}}\xspace}
\newcommand{\doubleEleMass}{\ensuremath{m_{\Pe\Pe}}\xspace}
\newcommand{\mTsup}[1]{\ensuremath{\mT^{#1}}\xspace}
\newcommand{\sieie}{\ensuremath{\sigma_{\eta\eta}\xspace}}
\newcommand{\mtMuon}{\ensuremath{\mTsup{\Pgm}}\xspace}
\newcommand{\mtElectron}{\ensuremath{\mTsup{\Pe}}\xspace}
\newcommand{\mjj}{\ensuremath{m_{\text{jj}}}\xspace}
\newcommand{\ttH}{\ensuremath{\ttbar\PH}\xspace}
\newcommand{\ggH}{\ensuremath{\cPg\cPg\PH}\xspace}
\newcommand{\VH}{\ensuremath{\PV\PH}\xspace}
\newcommand{\ZH}{\ensuremath{\PZ\PH}\xspace}
\newcommand{\ZZpair}{\ensuremath{\PZ\PZ^*}\xspace}
\newcommand{\omegaTilde}{\ensuremath{\tilde{\omega}_{\text{min}}}\xspace}
\newcommand{\binvRunTwoObs}{0.16\xspace}
\newcommand{\binvRunTwoExp}{0.09\xspace}
\newcommand{\binvAlltimeObs}{0.15\xspace}
\newcommand{\binvAlltimeExp}{0.08\xspace}
\newcommand{\combinedGofPValue}{12\%\xspace}
\newcommand{\bhinvlim}{\ensuremath{{\mathcal{B}(\PH \to \text{inv})}}\xspace}
\newcommand{\hinv}{\ensuremath{\PH \to \text{inv}}\xspace}
\newcommand{\zinv}{\ensuremath{\PZ \to \text{inv}}\xspace}
\newcommand{\smhinv}{\ensuremath{\PH \to \ZZpair \to 4\PGn}\xspace}
\newcommand{\htmiss}{\ensuremath{H_\mathrm{T}^\text{miss}}\xspace}
\newcommand{\llost}{\ensuremath{\Pell_{\text{lost}}}\xspace}
\newcommand{\ptmisstrack}{\ensuremath{p_{\mathrm{T},\text{track}}^{\text{miss}}}\xspace}
\newcommand{\ptvecmisstrack}{\ensuremath{\vec{p}_{\mathrm{T},\text{track}}^{\text{miss}}}\xspace}
\newcommand{\mindphiAj}{\ensuremath{\abs{\Delta\phi_\text{min}({\ptvecmiss, \vec{p}_\mathrm{T,1234})}}}\xspace}
\newcommand{\vecrecoil}{\ensuremath{\overrightarrow{\text{recoil}}}\xspace}
\newcommand{\ptjone}{\ensuremath{\vec{p}^{~\text{j}}_\mathrm{T,1}}\xspace}
\newcommand{\ptjtwo}{\ensuremath{\vec{p}^{~\text{j}}_\mathrm{T,2}}\xspace}
\newcommand{\ptbone}{\ensuremath{\vec{p}^{~\text{b}}_\mathrm{T,1}}\xspace}
\newcommand{\ptbtwo}{\ensuremath{\vec{p}^{~\text{b}}_\mathrm{T,2}}\xspace}
\newcommand{\sigmadmnuc}{\ensuremath{\sigma^{\text{SI}}_{\text{{DM\mbox{-}nucleon}}}}\xspace}

\newlength\cmsTabSkip\setlength{\cmsTabSkip}{1ex}
\newcommand{\MG}{\MADGRAPH{}5}
\cmsNoteHeader{HIG-21-007}
\title{A search for decays of the Higgs boson to invisible particles in events with a top-antitop quark pair or a vector boson in proton-proton collisions at \texorpdfstring{$\sqrt{s} = 13\TeV$}{sqrt(s) = 13 TeV}}

\author*[cern]{Olivier Davignon, David Anthony, Eshwen Bhal, Jim Brooke, Aaron Bundock, Henning Flaecher, Maciej Glowacki, Ben Krikler, Sudarshan Paramesvaran, Rob White}

\date{\today}

\abstract{A search for decays to invisible particles of Higgs bosons produced in association with a top-antitop quark pair or a vector boson, which both decay to a fully hadronic final state, has been performed using proton-proton collision data collected at ${\sqrt{s}=13\TeV}$ by the CMS experiment at the LHC, corresponding to an integrated luminosity of 138\fbinv. The 95\% confidence level upper limit set on the branching fraction of the 125\GeV Higgs boson to invisible particles, \bhinvlim, is 0.54 (0.39 expected), assuming standard model production cross sections. The results of this analysis are combined with previous \bhinvlim searches carried out at ${\sqrt{s}=7}$,~8, and 13\TeV in complementary production modes. The combined upper limit at 95\% confidence level on \bhinvlim is 0.15 (0.08 expected).}

\hypersetup{
  pdfauthor={CMS Collaboration},
  pdftitle={Search for Higgs boson decays to invisible particles produced in association with a top-quark pair or a vector boson in proton-proton collisions at \texorpdfstring{$\sqrt{s}=13~\mathrm{TeV}$} and combination across Higgs production modes},
  pdfsubject={CMS},
  pdfkeywords={CMS, Higgs, invisible, ttH, VH}
}

\titlerunning{A search for decays of the Higgs boson to invisible particles in events with a top-antitop quark pair...}

\maketitle 

\section{Introduction}
\label{sec:Introduction}

The Higgs boson (\PH)~\cite{PhysRevLett.13.321, Higgs:1964ia, PhysRevLett.13.508, PhysRevLett.13.585, Higgs:1966ev, Kibble:1967sv} of mass 125\GeV was discovered by the \mbox{ATLAS} and CMS Collaborations in 2012 \cite{ATLAS:2012yve,CMS:2012qbp, CMS:2013btf}. Since then its properties, including its coupling to other standard model (SM) particles, have been extensively studied using proton-proton (${\Pp\Pp}$) collision data from the CERN LHC collected at ${\sqrt{s}=7}$, 8, and 13\TeV with the \mbox{ATLAS} \cite{ATLAS:2022vkf} and CMS \cite{CMS:2022dwd} detectors. Properties of the Higgs boson can be exploited to probe for signs of behaviour beyond the SM (BSM). In the SM, the decay of the Higgs boson to an invisible final state (\hinv) is only possible via \smhinv, with a branching fraction of 0.1\%~\cite{ParticleDataGroup:2022pth}. Several BSM theories predict a larger branching fraction to invisible final states, \bhinvlim~\cite{SHROCK1982250, Belanger:2001am, Datta:2004jg, Dominici:2009pq}, namely in Ref.~\cite{Argyropoulos:2021sav} and references therein. For example, in a scenario where the Higgs boson connects the SM and dark matter (DM) sectors~\cite{Kanemura:2010sh, Djouadi:2011aa, Baek:2012se, Djouadi:2012zc, Beniwal:2015sdl, DiFranzo:2015nli}, \bhinvlim is enhanced as the Higgs boson can decay to a pair of DM particles of mass $m_{\text{DM}}<m_{\PH}/2$.

Direct searches for \hinv have been performed by the \mbox{ATLAS}~\cite{ATLAS:2017nyv, ATLAS:2019cid, ATLAS:2021kxv, ATLAS:2021gcn, ATLAS-VBF-Run2, ATLAS:2022ygn} and CMS~\cite{CMS:2016dhk, SUS-19-009_PAPER,SUS-19-011_PAPER,SUS-20-002_PAPER,CMS:2020ulv,CMS_MonojetV,VBF-Run2-paper} Collaborations using data collected during Run~1 (2011--2012) and Run~2 (2015--2018). These target channels in which the Higgs boson is produced via vector boson fusion (VBF), gluon-gluon fusion (\ggH), and in association with either a vector boson (\VH, where \PV stands for either a \PW or \PZ boson) or with a \ttbar quark pair (\ttH). The current most stringent constraint on \bhinvlim set by the CMS experiment is via the VBF channel using Run~1 and Run~2 data, which reports a 95\% confidence level~(\CL) upper limit of 0.18 (0.10 expected)~\cite{VBF-Run2-paper}.

\begin{figure}[hbt]
\centering
    \includegraphics[width=\cmsFigWidi]{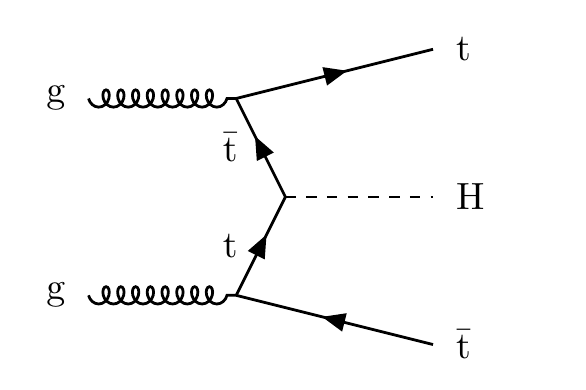}
    \includegraphics[width=\cmsFigWidi]{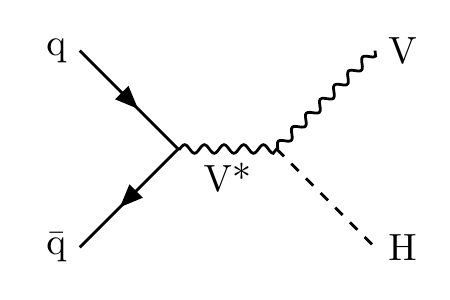}
    \caption{Representative LO Feynman diagrams for the SM Higgs boson production channels \ttH and \VH.}
    \label{fig:SM_feynman}
\end{figure}

In this paper, a search for an invisibly decaying Higgs boson, produced in association with a \ttbar quark pair or a \PV boson, where the associated particles decay to a fully hadronic final state, is reported. Representative leading order (LO) Feynman diagrams for \ttH and \VH are presented in Fig.~\ref{fig:SM_feynman}. The search in the \VH channel looks only at topologies in which the presence of the \PV boson is inferred from well separated decay products, complementing the previous \VH search with merged decay products arising from boosted \PV bosons~\cite{CMS_MonojetV}. The search uses LHC ${\Pp\Pp}$ collision data collected during the years 2016--2018, corresponding to a total integrated luminosity of 138\fbinv at $\sqrt{s}=13\TeV$. This is the first time that these final states have been used by the CMS experiment to search for the \hinv process using data from 2016--2018.

The missing transverse momentum, \ptvecmiss, is the transverse component of the negative vector sum of all reconstructed particle momenta in an event, and has a magnitude \ptmiss. There are two main sources of events resulting in \ptvecmiss. The first is events with invisible \PZ boson decays and visible jets (\zinv). The second is referred to as the lost lepton background, \llost, where \Pell stands for either an \Pe or \PGm. This includes events from \ttbarjets and \wjets processes where one or more leptons are misreconstructed, excluded by the phase space selection, or fall outside the detector acceptance. Control regions (CRs) enriched in these background sources, requiring either one lepton, one photon, or two same-flavour opposite-sign leptons, are used to constrain these backgrounds from data. The hadronic recoil is defined as the vectorial sum of the \ptvecmiss and the \pt of any selected charged lepton(s) or photon in an event, and its magnitude is used as the discriminating variable to separate the \hinv signal from backgrounds. The 95\% \CL upper limit on \bhinvlim is extracted from a fit to the hadronic recoil distribution of selected events, performed across the signal regions (SRs) and CRs. In the SRs, the hadronic recoil is equivalent to the \ptmiss, while in the CRs it effectively measures the \pt of the \PV boson or photon. The exclusion of leptons and photons ensures good correspondence between SRs and CRs.

This paper is organised as follows: Section~\ref{sec:CMS} is a brief description of the CMS detector. The simulated samples used in this analysis are summarised in Section~\ref{sec:Simulated_samples}. Section~\ref{sec:Event_reconstruction} describes the event reconstruction and object definitions used in this analysis, while the event selection and event categorisation are detailed in Section~\ref{sec:Event_selection_and_categorisation}. The data CRs used for estimating the SM backgrounds are introduced in Section~\ref{sec:Data_control_regions}. Section~\ref{sec:Statistical_interpretation} describes the statistical procedure used to constrain the backgrounds and extract the signal. The results of the search are presented in Section~\ref{sec:Results}. The results of combining this search with other CMS searches for invisibly decaying Higgs bosons are described in Section~\ref{sec:Combination}, and the results are summarised in Section~\ref{sec:Conclusion}.

\section{The CMS detector}
\label{sec:CMS}

The CMS apparatus is a multipurpose, nearly hermetic detector, designed to trigger on~\cite{CMS:2020cmk,CMS:2016ngn} and identify electrons, muons, photons, and (charged and neutral) hadrons~\cite{CMS:2020uim,CMS:2018rym,CMS:2014pgm}.
The central feature of the CMS apparatus is a superconducting solenoid, of $6~\mathrm{m}$ internal diameter. Within the field volume are the silicon pixel and strip tracker, the crystal electromagnetic calorimeter (ECAL), and the brass-scintillator hadron calorimeter (HCAL). Muons are measured in gas-ionisation chambers embedded in the steel flux-return yoke of the magnet. Besides the barrel and endcap detectors, CMS has extensive forward calorimetry, performed on high $\eta$ objects in the HCAL forward calorimeter, which is located $11.2~\mathrm{m}$ from the interaction region along the beam axis. A global ``particle-flow" (PF) algorithm~\cite{CMS:2017yfk} aims to reconstruct all individual particles in an event, combining information provided by all subdetectors. The reconstructed particles are used to build \PGt leptons, jets, and \ptmiss~\cite{CMS:2018jrd,CMS:2016lmd,CMS:2019ctu}.

The first level of the CMS trigger system, composed of custom hardware processors, uses information from the calorimeters and muon detectors to select the most interesting events, at a rate of roughly $100~\mathrm{kHz}$. The high-level trigger (HLT) processor farm performs event reconstruction similar to that of the full CMS reconstruction, but optimised for speed. This decreases the event rate from around $100\unit{kHz}$ to around $1\unit{kHz}$, before data storage~\cite{CMS:2016ngn}.

The procedures for calculating the integrated luminosity recorded by the CMS detector for each data-taking year are documented in Refs.~\cite{CMS:2021xjt,CMS-PAS-LUM-17-004,CMS-PAS-LUM-18-002} for 2016--2018, respectively.

A more detailed description of the CMS experiment can be found in Ref.~\cite{CMS:2008xjf}.

\section{Simulated samples}
\label{sec:Simulated_samples}

Monte Carlo (MC) simulated events are used to model signal and background contributions in all analysis regions, except for quantum chromodynamics (QCD) multijet production processes, which are estimated from data using a dedicated control sample and simulation-based transfer factors. The method for estimating QCD multijet production is detailed in Section~\ref{sec:QCD_estimation}. In all cases, MC samples are produced using either \POWHEG version 1.0 or higher~\cite{Oleari:2010nx} or \MGvATNLO version~2.4.2 or higher~\cite{Alwall:2014hca} matrix element (ME) generators. The ME is encoded with the maximum amount of information available for a hard scattering event. The parton-level simulation provided by the ME generators is interfaced with \PYTHIA version~8~\cite{Sjostrand:2014zea} to model the shower and hadronisation of partons in the initial and final states, along with the underlying event description, using the tune CUETP8M1 (CP5) when simulating events for the 2016 (2017 and 2018) data-taking periods~\cite{PythiaTune}. The propagation of all final state particles through the CMS detector is simulated using the \GEANTfour~\cite{Geant4} toolkit. Samples for 2016 make use of the NNPDF3.0 LO or next-to-LO (NLO) parton distribution functions (PDFs)~\cite{PDFs}, whereas samples for the years 2017 and 2018 use the NNPDF3.1 next-to-NLO (NNLO) PDFs.

Processes featuring \hinv occurring in \ttH, \VH, VBF, and \ggH channels are modelled by \POWHEG version~2.0~\cite{POWHEGttH, POWHEGVBF, POWHEGVH, POWHEGggF} at NLO in QCD. These samples require the SM Higgs boson to decay to four neutrinos (\smhinv) resulting in \bhinvlim$=1$. The cross sections are appropriately normalised to the corresponding SM predictions computed at NLO (\ttH), NNLO (\VH, VBF), and next-to-NNLO (\ggH) accuracy in QCD, and to NLO accuracy in electroweak~(EW) corrections~\cite{deFlorian:2016spz}. Background \ZH processes with the Higgs boson decaying to $\PQb\overline{\PQb}$ and the associated \PZ boson decaying to $\PQb\overline{\PQb}$, ${\Pell\Pell}$, and \qqbar (where q represents a light or charm quark) are generated at NLO using \MGvATNLO with the FxFx~\cite{Frederix:2012ps} matching scheme for 2016 samples, and with \POWHEG version~2.0~\cite{Zanoli:2021iyp} for 2017 and 2018 samples.

The \vjets processes are generated at LO in QCD using \MGvATNLO with up to four partons in the final state using the MLM~\cite{Mangano:2006rw} matching scheme between hard scatters and parton showers. These processes are generated in bins of hadronic transverse energy, \HT, which is the magnitude of the \ptvec sum of all jets reconstructed at generator level. The LO simulation of \vjets processes is corrected to account for missing higher-order diagrams with K-factors derived from \MGvATNLO-generated NLO QCD \vjets processes with up to two partons. These K-factors are extracted as a function of boson \pt and the \pt of the leading jet in the event. These K-factors are extracted as a function of boson \pt and the \pt of the leading jet in the event, and typically vary between 0.5 and 1.5 depending on the boson \pt.

The \singlePhotonCr processes are generated at NLO in QCD with \MGvATNLO, using a binning based on the \pt of the photon. The binning scheme for this sample is defined at the ME level to increase the statistical precision in the phase space regions probed by this analysis. 

Processes including $t$-channel single \PQt quarks, and \ttbar pairs with up to two additional partons in ME computation are generated at NLO with \POWHEG version~2.0~\cite{POWHEG:doubletop,Alioli:2009je}. Single \PQt quarks produced in the $s$ channel are modelled using \MGvATNLO, and also in the $tW$ channel using \POWHEG version~1.0~\cite{POWHEG:singletop}. The \PQt quark \pt spectrum in \ttbar processes is corrected to match the spectrum obtained from the NNLO~QCD~$+$~NLO~EW simulation, following Ref.~\cite{Czakon:2017wor}. Rare ${\ttbar\PX}$~$+$~jets backgrounds cover processes where \ttbar is produced in association with a boson \PX (\PGg,~\PV, or a visibly decaying \PH), generated at NLO. The ${\ttbar\PGg}$~$+$~jets, ${\ttbar\PW}$~$+$~jets, and ${\ttbar\PZ}$~$+$~jets samples are generated using \MGvATNLO, with subsequent decays generated using \textsc{MadSpin}~\cite{Artoisenet:2012st} to account for spin correlations in the former two cases. The \ttH~$+$~jets sample, where the \PH decays to visible states, is generated using \POWHEG.

Diboson ${\PZ\PZ}$ and ${\PW\PZ}$ production processes are generated at LO using \PYTHIA, while the ${\PW\PW}$ process is simulated at NLO in QCD using the \POWHEG version~2.0~\cite{POWHEG:WW}. The QCD multijet samples are generated at LO using \MGvATNLO in exclusive ranges of \HT in order to increase the statistical precision in the phase-space probed by this analysis.

\section{Event reconstruction}
\label{sec:Event_reconstruction}

During LHC runs, each beam crossing results in several ${\Pp\Pp}$ collisions in the detector. Additional ${\Pp\Pp}$ interactions within the same or nearby bunch crossing, known as pileup, make PF object reconstruction more challenging. The reduction of the effect of pileup relies on mitigation techniques~\cite{CACCIARI2008119} that filter energy deposits associated with pileup vertices and remove objects not associated with the primary interaction vertex (PV). The PV is the vertex associated with the hardest scattering in the event, according to tracking information, as described in Ref.~\cite{CMS-TDR-15-02}. All simulated samples from Section~\ref{sec:Simulated_samples} are reweighted to match the pileup distribution observed in data. In the SR, the final state is required to contain jets, a sizeable hadronic recoil, and no isolated leptons or photons. Candidate leptons and photons are selected with ${\pt>10\GeV}$ and pseudorapidity ${\abs{\eta} < 2.4}$ for muons~\cite{CMS:2018rym}, ${\pt>10\GeV}$ and either ${\abs{\eta} < 1.44}$ or ${1.57 < \abs{\eta} < 2.5}$ for electrons~\cite{CMS:2020uim}, ${\pt>20\GeV}$ and ${\abs{\eta} < 2.3}$ for hadronically decaying tau leptons~\cite{CMS:2018jrd}, and ${\pt>15\GeV}$ and either ${\abs{\eta} < 1.44}$ or ${1.57 < \abs{\eta} < 2.5}$ for photons~\cite{CMS:2020uim}. These selection criteria are optimised to reject background contributions, mainly from QCD processes. Other selection criteria depend on the isolation of the lepton or photon from hadronic interactions in the detector within a cone of small (tight isolation) or large (loose isolation) radius. Loose identification and isolation criteria are used to veto candidate events in the SR that contain leptons or photons. The veto efficiencies are $>$99, $\simeq$95, and $\simeq$90\% for loose muons, electrons, and photons, respectively. The SR background contributions are estimated using \singleMuCr, \singleEleCr, \doubleMuCr, \doubleEleCr, and \singlePhotonCr CRs. Tight and loose identification and isolation criteria are used to select and count muons, electrons, and photons in the CRs, enhancing the purity at little expense to the efficiency. These achieve typical selection efficiencies of $\simeq$95, 70, and 70 ($\simeq$98, 95 and 90)\%, for tight (loose) muons, electrons, and photons, respectively.

Jets are reconstructed by clustering all PF candidates originating from the PV with the anti-\kt jet clustering algorithm~\cite{AntiKt,Cacciari:2011ma}, using a distance parameter ${R=0.4}$ (AK4). Jet momentum is determined as the vectorial sum of all particle momenta in the jet, and is found from simulation to be, on average, within 5 to 10\% of the true momentum over the whole \pt spectrum and detector acceptance. Charged-hadron subtraction~\cite{CMS:2016lmd} is then applied to remove charged particles from pileup vertices~\cite{CMS_PAS_JME-14-001}. To ensure the measured jet energy matches that of the particle level jets, jet energy corrections (JEC) derived from simulation as functions of \pt and $\eta$ are applied. Further corrections are applied due to residual discrepancies in the jet energy scale (JES) between data and simulated samples~\cite{CMS:2016lmd}. Additionally, each jet must pass selection criteria to remove jets adversely affected by instrumentation or reconstruction failure. The jet energy resolution (JER) in simulated samples is smeared to match the measured resolution, which is typically 15--20\% at 30\GeV, 10\% at 100\GeV, and 5\% at 1\TeV~\cite{CMS:2016lmd}. The AK4 jets are required to have ${\pt>30\GeV}$ and ${\abs{\eta}<5.0}$, and those with loose leptons and photons located within a cone of ${\Delta R<0.4}$ of the jet direction are removed.

The AK4 jets that originate from the hadronisation of a bottom quark (\PQb-tagged jets) are identified using the \textsc{DeepCSV} algorithm, which correctly identifies \PQb jets with ${\pt>20\GeV}$ with a probability of 80\% and has a charm or light jet mistag probability of 10\%~\cite{Sirunyan:2017ezt}. Simulated events containing \PQb jets are corrected to be in agreement with the data by deriving corrections from data control samples that contain \PQb jets.

Pileup effects are mitigated at the reconstructed particle level using the pileup per particle identification algorithm (PUPPI) \cite{Sirunyan:2020foa,Bertolini:2014bba} by defining a local shape variable that can discriminate between particles originating from the PV and from pileup. Charged particles originating from pileup are discarded. For neutral particles, a local shape variable is computed based on the information from charged particles in their vicinity that originate from the PV within the tracker acceptance, and information from both charged and neutral particles outside this acceptance. The momenta of neutral particles are then rescaled based on the probability that they originated from the PV as deduced from the local shape variable~\cite{Sirunyan:2020foa}.

When a high \pt \PQt quark or \PV boson decays hadronically, a large set of collimated particles cross the detector. These can be clustered within a single jet of radius $R=0.8$ (AK8) using the anti-\kt algorithm. In order to reduce pileup effects, PUPPI PF candidates are used to seed the AK8 jet finder. The main feature that distinguishes hadronically decaying \PQt quarks or \PV bosons from the quark or gluon fragmentation is the jet mass. To improve the resolution, the modified mass-drop tagger algorithm~\cite{Larkoski:2014wba,Dasgupta:2013ihk,Butterworth:2008iy} (also known as the soft-drop algorithm, SD) with the angular exponent ${\beta=0}$, soft cutoff threshold ${z_{\text{cut}}<0.1}$, and characteristic radius ${R_{0}=0.8}$~\cite{Thaler:2010tr} is applied to each AK8 jet to remove soft and wide-angle radiation. In addition, a deep neural network (DNN) classifier called the \textsc{DeepAK8}~\cite{CMS-PAS-JME-18-002} algorithm is employed by assigning a set of numerical scores to each reconstructed AK8 jet corresponding to the probabilities that it originates from particular final states of V boson decays, for example ${\PZ\to\PQb\PAQb}$, ${\PZ\to\qqbar}$, ${\PW\to\PQc\PQs}$, rather than from QCD multijet processes. For this analysis, reconstructed AK8 jets originating from \PQt quarks (\PW bosons) are selected by requiring ${\pt>400~(200)\GeV}$, SD mass $m_{\text{SD}}$ between 120 and 210 (65 and 120)\GeV, and a DeepAK8 probability score for \PQt quarks (\PW bosons) larger than between 72.5 and 83.4 (91.8 and 92.5)\% depending on the year of data-taking. The resulting \PQt quark (\PW boson) tagging efficiency at the ${\pt>400~(200)\GeV}$ threshold limit is estimated from simulation as 28 (25)\% with a 1\% mistag rate from QCD jets. Simulated events containing AK8 jets are corrected to agree with the data using data-derived correction factors, and dedicated JEC are also applied~\cite{CMS-PAS-JME-18-002}.

The calculation of energy sums such as the hadronic recoil, \ptvecmiss, and \htvecmiss, which is the negative \ptvec sum of jets reconstructed at the HLT level with a \ptmiss threshold of 20\GeV applied, are based on AK4 jets, therefore JEC are propagated through the use of the \ptvec-corrected jets.

\section{Event selection and categorisation}
\label{sec:Event_selection_and_categorisation}

In this analysis the signal is extracted from a combined fit to the hadronic recoil distribution of events in SRs and CRs as defined for the \ttH and VH categories. The CRs are used to estimate the contributions of different SM processes in each SR. Where possible, the CRs have kinematic requirements identical to the SR, and leptons or a photon are used in the CR definition, but otherwise ignored in the calculation of event observables. The \singleEleCr and \singleMuCr CRs, enriched in \wjets and \PQt quark background processes, are used to derive corrections to \llost contributions predicted by simulation. The \doubleEleCr, \doubleMuCr, and, in the case of the \VH category, \singlePhotonCr CR samples are used to derive corrections to the expected contribution from \zjets production, where the \PZ boson decays to a pair of neutrinos. A QCD multijet enriched CR (hadronic sideband) is also used to estimate hadronic backgrounds in the SR.

\subsection{Trigger requirements}
\label{subsec:triggers}

Events of interest are collected via a suite of triggers that are applied to variables calculated using PF candidates reconstructed at the level of the HLT. The trigger requirements vary amongst analysis regions and data-taking periods. Events in the SR, hadronic sideband, and muon CRs are collected using HLT selection criteria on \ptmiss and the missing \HT, \htmiss, which is the magnitude of \htvecmiss. Muons are not considered in the calculation of PF~\ptmiss and PF~\htmiss to allow the same trigger to be used in the SR and the muon CRs, with a typical efficiency of $>$90\% for \ptmiss $>250\GeV$. The use of the combined \ptmiss and \htmiss triggers in the muon CRs instead of single-muon triggers corresponds more closely to the selection in the SR and minimises selection biases. Trigger thresholds increase with time due to the increase in instantaneous luminosity during Run~2. In 2016, the \ptmiss and \htmiss thresholds vary between 90 and 120\GeV. In 2017 and 2018, these thresholds are 120\GeV. During data-taking in 2017, additional corrections were applied to account for the effect of ECAL endcap noise at high $\abs{\eta}$ on PF \ptmiss measurements. Additionally, for 2016 and 2017 data-taking periods, there was an inefficiency arising from a gradual shift in the timing of the ECAL trigger inputs in the region ${\abs{\eta}>2.0}$~\cite{CMS:2020cmk}. This resulted in events containing an electron or photon (jet) with ${\pt>50~(100)\GeV}$ having an efficiency loss of up to 20\%, depending on \pt and $\eta$. Correction factors for this trigger inefficiency are obtained from 2016 and 2017 data and applied to simulation samples as a function of $\eta$.

Events in the \singleEleCr and \doubleEleCr CRs from the 2016, 2017, and 2018 data sets are required to pass a tight (loose) single-electron trigger with \pt thresholds of 27, 35, and 32 (105, 115, and 115)\GeV, respectively. The low-threshold single-electron triggers require the electron candidate to pass a tight isolation condition, while the high-threshold trigger imposes a looser selection on the isolation to improve the efficiency at high \pt. Photon events are required to pass a single-photon trigger with a \pt threshold of 175 (200)\GeV without any isolation condition for the 2016 (2017 and 2018) data sets. Simulated electron or photon events are accepted if they pass exactly one of the above trigger requirements, and the efficiency of this selection is corrected with data-derived efficiency correction factors.

\subsection{Offline selection}
\label{subsec:Baseline_selection}

In order to select events with a large amount of jet activity and sizeable hadronic recoil, a further offline selection is applied to all regions. To improve the purity of the signal, large missing energy is desirable, therefore events require the hadronic recoil, \htmiss, and \HT to be greater than 200\GeV. Furthermore, the largest \pt of an AK4 jet in an event, \ptjone, is required to be greater than 80\GeV. To ensure consistency amongst different estimators of the hadronic recoil, the recoil as calculated from PF candidates in \ptmiss and from PF jets in \htmiss must satisfy ${\htmiss/\text{recoil}<1.2}$ and azimuthal separation ${\abs{\Delta\phi(\vecrecoil, \htvecmiss)}<0.5}$. To further improve the quality of events, a selection is made on \ptmisstrack, which is equivalent to \ptmiss but calculated using only charged PF particles, and therefore is expected to be well-aligned with the hadronic recoil direction. Requirements of ${\ptmisstrack>60\GeV}$ and azimuthal separation ${\abs{\Delta\phi(\vecrecoil, \ptvecmisstrack)}<1}$ are applied in the SR and hadronic sideband. The kinematic selection for all regions is optimised according to the Asimov significance between signal (S) and background (B) yields assuming a background systematic uncertainty $\Delta\text{B}$ of 5\% or 10\%~\cite{Cowan:2010js}. The peaks of the distribution for a given variable corresponds to its selection threshold.

In order to facilitate the combination of this analysis with the results from other \hinv searches, additional selections are introduced to reduce the potential event overlap. A veto is implemented to ensure orthogonality with the VBF phase space, through a veto on events with leading (subleading) AK4 jets with ${\abs{\eta_{1}}~(\abs{\eta_{2}})>2.4}$, and an inversion of the kinematic selection employed by the VBF \hinv analysis~\cite{VBF-Run2-paper}. This removes events containing two AK4 jets with $\ptjone>80\GeV$ and the subleading jet \pt, \ptjtwo, to be greater than 40\GeV, where the jets are from opposite detector hemispheres (${\eta_{1}\eta_{2} < 0}$), have a large \mjj ($>$200\GeV), small azimuthal separation (${\Delta\phi_{\text{jj}} < 1.5}$), and a large $\eta$ gap (${\abs{\eta_{\mjj}} > 1.0}$). Moreover, orthogonality to leptonic \ttH decays is ensured in the single-lepton CRs by requiring the transverse mass of the combined single-lepton and hadronic recoil system, defined as
\begin{equation}
m_\mathrm{T}^{\Pell} = \sqrt{2 p^{\Pell}_{\mathrm{T}} (\text{recoil}) [1 - \cos{(\phi(\vec{p}^{\Pell}_{\mathrm{T}}) - \phi(\vecrecoil))}]},
\end{equation}
to be lower than 110\GeV. Orthogonality between leptonic \ttH decays in the dilepton CRs is ensured by requiring the invariant mass of the charged lepton pair, $m_{\Pell\Pell}$, to be lower than 120\GeV in these CRs. Selecting on the invariant masses of lepton pairs also suppresses the \ttH signal contamination in the CRs. Overlap between the ggH/boosted~\VH \hinv analysis and the resolved \VH category of this analysis is rendered negligible by explicitly removing events from the low-purity boosted \VH category defined in Ref.~\cite{CMS_MonojetV} if they contain exactly two AK4 jets with an invariant mass, \mjj, forming a dijet candidate with $65 < \mjj < 120\GeV$. No corresponding selection is necessary for the resolved \VH category as a result, while there is negligible change to the sensitivity of the boosted \VH category.

During significant periods of data-taking in 2018, the HCAL portion corresponding to the region $-1.57<\phi<-0.87$, $-3.0<\eta<-1.39$ was not functional. Events from 2018 with $-1.8<\phi(\vecrecoil)<-0.6$ are vetoed if they contain jets within the affected region, which removes $\approx$65\% of the total data from the affected region. To ensure good correspondence between data and simulation, the simulation is reweighted to account for the efficiency loss. A summary of the offline requirements are provided in Table~\ref{tab:common_sel}.

\begin{table*}[htb]
    \centering
    \topcaption{Offline selection applied to all categories and regions in this analysis to improve signal purity and reduce overlap with the phase space of other \hinv searches.}
    \begin{tabular}{ccc}
        Variable & Selection & Purpose \\\hline
        recoil & $>200\GeV$ & \multirow{3}{*}{Signal purity} \\
        \htmiss & $>200\GeV$ & \\
        \ptjone & $>80\GeV$ & \\ [\cmsTabSkip]
        
        $\htmiss/\text{recoil}$ & $<1.2$ & \multirow{2}{*}{Event quality} \\
        $\abs{\Delta\phi(\vecrecoil, \htvecmiss)}$ & $<0.5 $ & \\ [\cmsTabSkip]
        
        $\abs{\eta_1}$, $\abs{\eta_2}$ & $<2.4$ & \multirow{4}{*}{Analysis orthogonalisation} \\
        VBF signal & Veto (inversion on signal selection) & \\
        $m_{T}^{\Pell}$ & $<110\GeV$ & \\
        $m_{\Pell\Pell}$ & $<120\GeV$ & \\        
    \end{tabular}
    \label{tab:common_sel}
\end{table*}

\subsection{Signal regions}
\label{subsec:Categorisation}

The search focuses on three types of hadronic final states: those with boosted \PQt quarks and/or boosted \PW bosons reconstructed with dedicated merged jet algorithms; those with one or more \PQb jets and no boosted \PQt quark or \PW boson, targetting the bulk of hadronic \ttH events; and those with two resolved jets with the \mjj compatible with that of a \PW or \PZ boson. The latter complements the boosted \VH channel analysed in Ref.~\cite{CMS_MonojetV}.

Events are categorised into boosted and resolved \ttH, and resolved \VH topologies. The \ttH category requires that at least five AK4 jets and one \PQb jet are present. The boosted \ttH topology requires that at least one AK8 jet is reconstructed and either \PQt- or \PW-tagged, and is subcategorised by the AK8 jet and \PQb jet multiplicities. Events without such \PQt- or \PW-tagged jets are categorised as belonging to a resolved \ttH topology, with further selections on the leading AK4 jet (leading or subleading \PQb jet) \ptvec and the hadronic recoil, $\abs{\Delta\phi(\vecrecoil, \ptjone)}$ ($\abs{\Delta\phi(\vecrecoil, \ptbone)}$ or $\abs{\Delta\phi(\vecrecoil, \ptbtwo)}$) applied to discriminate between \ttH and \ttbarjets processes. Finally, the remaining events are allocated to the resolved \VH topology category if they have exactly two AK4 jets with \mjj between 65 and 120\GeV, compatible with a \PW or \PZ boson decay. The resolved \VH subcategories are separated according to the \PQb jet multiplicity. Subcategories are also defined based on \ptjtwo to suppress QCD multijet background. The subcategory definitions are summarised in Table~\ref{tab:non_VBF_categories}. The intended outcome of this categorisation is a set of event samples with high purity for a given production mode, and minimal background contamination or signal cross-contamination.

\begin{table*}[hbtp]
    \centering
    \topcaption{Categorisation of the \ttH and \VH production modes in the analysis. No additional selections are applied to the boosted \ttH subcategories.}
    \begin{tabular}{cccccccc}
        Category & Subcategory & $n_\text{j}$ & $n_{\PQb}$ & $n_{\PQt}$ & $n_{\PW}$ & \ptjtwo (\GeVns) & Other \\
        \hline
        \multirow{6}{*}{Boosted \ttH} & 2Boosted1b & $\geq 5$ & 1 & \multicolumn{2}{c}{2} & \multirow{6}{*}{$> 80$} & \multirow{6}{*}{\NA}\\
        & 2Boosted2b & $\geq 5$ & $\geq 2$ & \multicolumn{2}{c}{2} & & \\
        & 1t1b & $\geq 5$ & 1 & 1 & 0 & &\\
        & 1t2b & $\geq 5$ & $\geq 2$ & 1 & 0 & &\\
        & 1W1b & $\geq 5$ & 1 & 0 & 1 & &\\
        & 1W2b & $\geq 5$ & $\geq 2$ & 0 & 1 & &\\ [\cmsTabSkip]
        
        \multirow{4}{*}{Resolved \ttH} & 5j1b & 5 & 1 & 0 & 0 & \multirow{4}{*}{$> 80$} & $\abs{\Delta\phi(\vecrecoil, \ptbone)} > 1.0$, \\
        & 6j1b & $\geq 6$ & 1 & 0 & 0 & & $\abs{\Delta\phi(\vecrecoil, \ptjone)} > \pi/2$\\ [\cmsTabSkip]
        
        & 5j2b & 5 & $\geq 2$ & 0 & 0 & & $\abs{\Delta\phi(\vecrecoil, \ptbone)} > 1.0$, \\
        & 6j2b & $\geq 6$ & $\geq 2$ & 0 & 0 & & $\abs{\Delta\phi(\vecrecoil, \ptbtwo)} > \pi/2$ \\ [\cmsTabSkip]
        
        \multirow{3}{*}{\VH} & 2j0b & 2 & 0 & 0 & 0 & \multirow{3}{*}{$> 30$} & \multirow{3}{*}{$65 < m_\text{jj} < 120\GeV$} \\
        & 2j1b & 2 & 1 & 0 & 0 & \\
        & 2j2b & 2 & 2 & 0 & 0 & \\
    \end{tabular}
    \label{tab:non_VBF_categories}
\end{table*}

A requirement on \mindphiAj, defined as the minimum azimuthal separation between the hadronic recoil and the momentum direction of any of the four highest \pt jets, of $>$0.5 is applied to suppress QCD multijet events where the hadronic recoil is aligned with a jet. A parameter \omegaTilde is designed to suppress events where missing energy is the result of a jet \pt mismeasurement, and is especially effective in categories with no \PQb jets. For the $i$th jet in the event, $\omega_{i}$ is defined as $\arctan{(H_{\mathrm{T}, \text{min}}^{\text{miss}} / p_{\mathrm{T},i})}$, where $p_{\mathrm{T},i}$ is the \pt of jet $i$, and $H_{\mathrm{T}, \text{min}}^{\text{miss}}$ is the minimum value of \htmiss that can be obtained by changing the value of $p_{\mathrm{T},i}$. The value of $\omega_{i}$ minimised over $i$ is \omegaTilde. A detailed derivation of this variable is given in Ref.~\cite{sakuma2019alternative}. QCD multijet events in the SR are further suppressed by requiring $\omegaTilde > 0.3$. Requirements to suppress QCD events are applied in the SR only for \ttH categories, and to both SR and CRs in the \VH categories in order to ensure good correspondence amongst the regions. The selections applied to \omegaTilde and \mindphiAj are not applied in the CRs used for background estimation of the \ttH categories, where the hadronic recoil does not stem from jet mismeasurement. This is to increase event counts in the CRs, particularly in the boosted \ttH categories.

The hadronic recoil in \ttH production is closely aligned with the direction of the Higgs boson typically. In \ttbar events, the \ptvecmiss is usually parallel or antiparallel to the direction of the leading \PQb jet, as the \PQt quarks are produced back-to-back. Therefore, the angles between the direction of the hadronic recoil and the leading or subleading jet or \PQb jet \ptvec directions provide additional features for \ttbar background suppression in the resolved \ttH categories. The angular variables $\abs{\Delta\phi(\vecrecoil, \ptjone)}$, $\abs{\Delta\phi(\vecrecoil, \ptbone)}$, and $\abs{\Delta\phi(\vecrecoil,\ptbtwo)}$ are the most sensitive discriminators between \ttH and \ttbar. The selection based on these angular variables has been optimised by maximising the combined expected sensitivity of the \ttH analysis and is summarised in Table~\ref{tab:non_VBF_categories}.

\section{Control regions and background estimation}
\label{sec:Data_control_regions}

The analysis makes use of the \singleMuCr and \singleEleCr CRs to estimate \llost background contributions, which are mainly from \ttbarjets, single \PQt quark, and \wjets events. The background contributions from \zinv, which include ${\PZ\PZ}$, ${\ttbar\PZ}$, and Drell-Yan (DY) contributions, are estimated from the \doubleMuCr, \doubleEleCr, and \singlePhotonCr CRs. Hadronic backgrounds in the SR such as QCD multijet contributions are estimated using a transfer factor method applied to a QCD enriched sideband CR.

\subsection{\texorpdfstring{Estimation of \llost and \zinv backgrounds}%
                            {Background estimation of lost lepton and Z to invisible}}
\label{sec:llostandzinv_estimation}

The \singleMuCr (\singleEleCr) CR is defined by requiring exactly one tightly-isolated muon (electron) with $\pt>20 (40)\GeV$. Both CRs require ${50 < m_\mathrm{T}^{\Pell} < 110\GeV}$. The single-lepton CRs are used to constrain the \llost background, which is the main source of background in the \ttH and \VH2j2\PQb categories. In the \ttH category, the \llost contribution arises mainly from \ttbar, single \PQt quark, and ${\ttbar\PV}$ processes, while in the \VH category it is from \PW~$+$~jet events.

In the \doubleMuCr (\doubleEleCr) CR, one tightly-isolated muon (electron) with ${\pt >20~(40)\GeV}$, and one loose muon (electron) with the opposite charge and $\pt>10 (10)\GeV$ are required with invariant mass, \doubleMuMass (\doubleEleMass), compatible with a \PZ boson. For the \ttH (\VH) category, the invariant mass is required to be between 75 and 105 (60 and 120)\GeV. The processes $\PZ\to\PGn\PAGn$ and ${\PZ\to\Pell\Pell}$ are kinematically nearly identical, largely due to lepton universality, hence the dilepton regions can be used to constrain the \zinv background and minimise theoretical uncertainties. This is important for the \zinv background, which dominates the \VH category and contributes to the \ttH category especially at high hadronic recoil. In the \ttH category, events are selected for which ${\Delta\phi(\vecrecoil, \ptvecmisstrack)>\pi/2}$, which reduces the \ttbarjets background and favours DY production in the dilepton CRs. 

The \singlePhotonCr CR is used for background estimation in the \VH category only, and requires exactly one loose photon with ${\pt>230\GeV}$. This region is used to constrain the \zinv background as the event kinematics and topologies are similar for \zjets and \singlePhotonCr events, improving the sensitivity to the \VH signal primarily at high hadronic recoil compared to the dilepton CRs because of the larger number of events.

Photons can usually be discriminated from other sources of ECAL deposits using the properties of the deposits themselves, such as isolation in ECAL and HCAL, or the shape of the electromagnetic showers. However, occasionally other particles will be incorrectly identified as photons, for example where a jet is misidentified as a photon in QCD multijet events. In order to estimate the contribution from misidentified photons in the \singlePhotonCr CR, a purity measurement is performed. The purity is defined as the fraction of reconstructed photon candidates that correspond to genuine isolated photons originating from the PV in the event. The photon purity is measured in data based on the lateral width \sieie~\cite{CMS:2015xaf}, which parametrises the shape of the energy deposit associated with the photon in the ECAL. The characteristic \sieie~distribution from genuine photons peaks at $\sieie<1$, while the distribution due to misidentified photons possesses a less pronounced peak with a much broader decline for $\sieie>1$. A template fit to the \sieie~distribution is performed, where for genuine photons simulated \singlePhotonCr events are used to build the signal templates, while for misidentified photons a data sample enriched in misidentified photon events is obtained by inverting the isolation requirements in the \singlePhotonCr CR. The purity is defined as the fraction of genuine photons extracted from the fit that pass the \sieie~selection. The photon purity is measured separately in bins of $p_{\mathrm{T}}^{\PGg}$ and for each data-taking period and varies between 1.5 and 4.5\%. The contamination is the fraction of misidentified photons in the \singlePhotonCr CR, and is estimated at around 4\% for ${p_{\mathrm{T}}^{\PGg}>200\GeV}$. The QCD multijet contribution in the \singlePhotonCr CR is then estimated by weighting events in data for each $p_{\mathrm{T}}^{\PGg}$ bin by the corresponding contamination. A 25\% systematic uncertainty is attributed to the QCD multijet background normalisation, and is estimated by performing the procedure for different \sieie~binning in the template fit, which accounts for any mismodelling of the simulated \sieie~distribution. The statistical uncertainty in the photon purity estimate in each $p_{\mathrm{T}}^{\PGg}$ bin is found to be much smaller than the systematic one. The full requirements for the analysis CRs are shown in Table~\ref{tab:CR_defs}.

\begin{table*}[htbp]
    \centering
    \topcaption{Summary of all CR requirements, excluding selections suppressing the QCD multijet background, and excluding the requirement of ${\Delta\phi(\protect\vecrecoil,\ptvecmisstrack)>\pi/2}$ applied to the \ttH category in the dilepton CRs. No mass requirements are imposed in the \singlePhotonCr.}
    \begin{tabular}{{c}{c}{c}{c}{c}}
    Control region & Category & $n_{\text{object}}$ reqs. & Mass reqs. (\GeVns) & \pt reqs. (\GeVns) \\\hline
    \multirow{2}{*}{\singleMuCr} & \ttH & \multirow{2}{*}{$n_{\Pgm} = 1$} & \multirow{2}{*}{$50 < \mtMuon < 110$ } & \multirow{2}{*}{$p^{~\PGm}_{\mathrm{T,1}} > 20$}\\
    & \VH & & & \\ [\cmsTabSkip]
    
    \multirow{2}{*}{\singleEleCr} & \ttH & \multirow{2}{*}{$n_{\Pe} = 1$} & \multirow{2}{*}{$50 < \mtElectron < 110$} & \multirow{2}{*}{$p^{~\Pe}_{\mathrm{T,1}} > 40$}\\
    & \VH & & & \\ [\cmsTabSkip]
    
    \multirow{2}{*}{\doubleMuCr} & \ttH & \multirow{2}{*}{$n_{\PGm} = 2$} & $75 < \doubleMuMass < 105$ & \multirow{2}{*}{$p^{~\PGm}_{\mathrm{T,1}} > 20$, $p^{~\PGm}_{\mathrm{T,2}} > 10$}\\
    & \VH & & $60 < \doubleMuMass < 120$  & \\ [\cmsTabSkip]
    
    \multirow{2}{*}{\doubleEleCr} & \ttH & \multirow{2}{*}{$n_{\Pe} = 2$} & $75 < \doubleEleMass < 105$ & \multirow{2}{*}{$p^{\mathrm{~\Pe}}_{\mathrm{T,1}} > 40$, $p^{\mathrm{~\Pe}}_{\mathrm{T,2}} > 10$}\\
    & \VH & & $60 < \doubleEleMass < 120$ & \\ [\cmsTabSkip]
    
    \singlePhotonCr & VH & $n_{\PGg} = 1$ & \NA & $p_{\mathrm{T}}^{\PGg} > 230$\\
    \end{tabular}
    \label{tab:CR_defs}
\end{table*}

\subsection{Residual backgrounds from QCD multijet production}
\label{sec:QCD_estimation}

The event selection aims to reduce background contributions from QCD multijet production as much as possible by requiring ${\mindphiAj>0.5}$ and ${\omegaTilde>0.3}$, although a QCD multijet background enriched sideband is used to estimate any remaining background contribution with the help of a transfer factor between sideband and SR, which is derived from simulation. The sideband is defined with an identical selection to that of the SR, but with an inversion on the requirements on \mindphiAj and \omegaTilde, such that $\mindphiAj<0.5$ and more stringently $\omegaTilde < 0.2$. The criteria for \omegaTilde is determined by optimising the sideband to be as QCD multijet-enriched as possible while ensuring the SR has negligible QCD multijet background. For the \VH category, the \mjj requirement is also inverted in order to have the sideband sufficiently populated.

The SRs in both the \ttH and \VH categories suffer from limited simulated QCD multijet event counts, so it is not possible to reliably define a transfer factor for each SR bin in individual subcategories. Within the statistical precision of the simulated QCD multijet samples, the shape of the hadronic recoil and relative population of the \ttH subcategories are observed not to depend on \omegaTilde and \mindphiAj. Therefore, the expected QCD sideband yields are integrated over all \ttH subcategories and hadronic recoil intervals, and over hadronic recoil intervals for each \VH category, in the sideband and SR, to construct the transfer factors. The resulting hadronic recoil distributions are used to predict the relative QCD multijet background in each subcategory and hadronic recoil interval.

The estimated QCD multijet background yield in the \ttH SR for subcategory $i$ and hadronic recoil interval $j$, $N^{\text{QCD, SR}_{\ttH}}_{i,j}$, is given by
\begin{linenomath}
\ifthenelse{\boolean{cms@external}}
{
\begin{multline}
N^{\text{QCD, SR}_{\ttH}}_{i,j} = \sum_p \sum_q (N^{\text{data, CR}_{\ttH}}_{p,q}\\
 - N^{\text{EW, CR}_{\ttH}}_{p,q}) \text{TF}^{\ttH}_{\text{QCD}} f^{\ttH}_{c_{i}} f^{\ttH}_{m_{j}},
    \label{eqn:qcd_pred}
\end{multline}
}
{
\begin{equation}
N^{\text{QCD, SR}_{\ttH}}_{i,j} = \sum_p \sum_q (N^{\text{data, CR}_{\ttH}}_{p,q} - N^{\text{EW, CR}_{\ttH}}_{p,q}) \text{TF}^{\ttH}_{\text{QCD}} f^{\ttH}_{c_{i}} f^{\ttH}_{m_{j}},
\label{eqn:qcd_pred}
\end{equation}
}
\end{linenomath}
where EW refers to processes that are not QCD multijet, summation indices $p$ and $q$ are the subcategory and hadronic recoil bins, respectively, $\text{TF}^{\ttH}_{\text{QCD}}$ is the QCD multijet simulation transfer factor defined as the ratio between the expected QCD multijet background contribution in the SR and the sideband, and $f^{\ttH}_{c_{i}}$ and $f^{\ttH}_{m_{j}}$ are the fractions of simulated QCD multijet events in each subcategory and hadronic recoil bin, respectively.

In the \VH category, the sideband regions are defined for each subcategory, as the number of simulated QCD multijet events is sufficient to derive the hadronic recoil fractions $f_{m_{j}}$ separately for each subcategory. The method is otherwise analogous to that of \ttH, given by Eq.~\ref{eqn:qcd_pred}.

The results of the QCD prediction aggregated over data sets from the 2016--2018 period are found to be small in comparison to background contributions from \llost and \zinv processes. In addition to the statistical uncertainties, a 100\% systematic uncertainty is assigned to the predicted background yields from QCD multijet production. The actual uncertainty in the QCD prediction is measured at around 50\%, derived by calculating the QCD contribution in the entire \ttH category for a signal-depleted validation region analogous to the SR but requiring $0.2<\omegaTilde<0.3$ and ${\mindphiAj}>0.5$, and comparing the estimate to data. It is inflated to 100\% to be more conservative when handling the individual \ttH subcategories that are limited by event counts at larger hadronic recoil, which was found to have negligible impact on the final fit.

\section{Statistical interpretation}
\label{sec:Statistical_interpretation}

A maximum likelihood fit method is used to obtain an upper limit on \bhinvlim. The fit is performed simultaneously across each year, region, category, and hadronic recoil interval, with systematic uncertainties acting as nuisance parameters in the fit correlated to varying degrees across year and category.

\subsection{Likelihood model}
\label{subsec:Likelihood_model}

The limits on \bhinvlim are extracted via a simultaneous binned maximum likelihood fit to the hadronic recoil distributions obtained in the SR and CRs. The likelihood can be written as
\begin{gather}
\mathcal{L}=\mathcal{L}_{\text{SR}}~\mathcal{L}_{\PGm}~\mathcal{L}_{\Pe}~\mathcal{L}_{\PGm\PGm}~\mathcal{L}_{\Pe\Pe}~\mathcal{L}_{\PGg},
\label{eqn:full_likelihood}
\end{gather}
where $\mathcal{L}_{\text{SR}}$ is the likelihood function for the SR (boosted \ttH, resolved \ttH, \VH), and $\mathcal{L}_{\PGm}$, $\mathcal{L}_{\Pe}$, $\mathcal{L}_{\PGm\PGm}$, $\mathcal{L}_{\Pe\Pe}$, and $\mathcal{L}_{\PGg}$ designate the likelihood functions for the \singleMuCr, \singleEleCr, \doubleMuCr, \doubleEleCr, and \singlePhotonCr CRs, respectively. The likelihood function for the SR is defined as
\begin{equation}
\label{eqn:lh_sr}
\mathcal{L}_\text{SR}=\prod_{\text{cat}=i}^{n_{\text{cat}}}\prod_{\text{recoil}=j(i)}^{n_{\xi(i)}}\text{Poisson}(n_\text{obs}^{i,j}\mid n^{i,j}_\text{pred}),
\end{equation}
with
\begin{equation}
\label{eqn:lh_sr_breakdown}
\begin{aligned}
n^{i,j}_\text{pred} & = \hat{\mu} s^{i,j} \rho_{s}^{i,j}\\
 & + b^{i,j}_{\llost} I^{i,j} \rho_{\llost}^{i,j}\\
 & + b^{i,j}_{\zinv} L^{i,j} \rho_{\zinv}^{i,j}\\
 & + b^{i,j}_\text{QCD} \rho^{i,j}_\text{QCD},
\end{aligned}
\end{equation}
where the symbols are defined in Table~\ref{tab:FitParamsSR}. The signal strength, $\hat{\mu}$, is interpreted as the maximum likelihood estimator for \bhinvlim, where the signal prediction assumes that $\bhinvlim=1$. The fit also includes additional free parameters $I^{i,j}$ and $L^{i,j}$, which depend on category $i$, hadronic recoil bin $j$, and the number of recoil bins in each category $n_{\xi(i)}$. The first of these parameters, $I^{i,j}$, simultaneously scales the normalisation of the \llost background in the SR and the sum of the \wjets, \ttbar~$+$~jets, and single \PQt quark backgrounds, $\text{X}^{i,j}_{\PQt,\PW}$, in the \singleMuCr and \singleEleCr CRs. The second of these parameters, $L^{i,j}$, simultaneously scales the normalisation of the \zinv background in the SR (Z(${\PGn\PAGn}$)~$+$~jets and ${\ttbar\PZ(\PGn\PAGn)}$) and the sum of the \PGg~$+$~jets, DY~$+$~jets, ${\ttbar\PZ}$~$+$~jets, and multiboson backgrounds, $\text{X}^{i,j}_{\PZ/\PGg}$, in the \doubleMuCr, \doubleEleCr, and \singlePhotonCr CRs.

\begin{table*}[htbp]
\centering
\topcaption{Meaning of the symbols used in Eqs.~\ref{eqn:lh_sr} and \ref{eqn:lh_sr_breakdown} that define the likelihood function.}
\begin{tabular}{ll}
 Symbol & Meaning \\ 
 \hline
 $\hat{\mu}$ & Signal strength estimator of \bhinvlim \\ 
 $s^{i,j}$ & Simulation predicted number of signal events in bin $i,j$ of the SR\\
 $\rho_{s}^{i,j}$ & Systematic uncertainties affecting signal prediction in bin $i,j$  of the SR\\
 $b^{i,j}_{\llost}$ & Simulation predicted number of \llost events in bin $i,j$ of the SR\\
 $I^{i,j}$ & Normalisation parameter for the \llost estimation in bin $i,j$\\
 $\rho_{\llost}^{i,j}$ & Systematic uncertainties affecting the \llost background in bin $i,j$ of the SR\\
 $b^{i,j}_{\zinv}$ & Simulation predicted number of \zinv events in bin $i,j$ of the SR\\
 $L^{i,j}$ & Normalisation parameter for the \zinv estimation in bin $i,j$\\
 $\rho_{\zinv}^{i,j}$ & Systematic uncertainties affecting the \zinv background in bin $i,j$ of the SR\\
 $b^{i,j}_\text{QCD}$ & Predicted number of QCD events in bin $i,j$  of the SR\\
 $\rho^{i,j}_\text{QCD}$ & Systematic uncertainties of the QCD component in bin $i,j$  of the SR\\
\end{tabular}
\label{tab:FitParamsSR}
\end{table*}

The likelihood for the \singleMuCr and \singleEleCr CRs is given by
\begin{linenomath}
\ifthenelse{\boolean{cms@external}}
{
\begin{multline}
\mathcal{L}_{\PGm,\Pe}=\prod_{\text{cat}=i}^{n_\text{cat}}\prod_{\text{recoil}=j(i)}^{n_{\xi(i)}}\text{Poisson}(n_\text{obs}^{i,j}\mid \\
\text{X}^{i,j}_\text{\PQt,\PW}I^{i,j}\rho_{\PQt,\PW}^{i,j} + \text{X}^{i,j}_\text{other} \rho_\text{other}^{i,j}),
\end{multline}
}
{
\begin{equation}
\mathcal{L}_{\PGm,\Pe}=\prod_{\text{cat}=i}^{n_\text{cat}}\prod_{\text{recoil}=j(i)}^{n_{\xi(i)}}\text{Poisson}(n_\text{obs}^{i,j}\mid\text{X}^{i,j}_{\PQt,\PW}I^{i,j}\rho_{\PQt,\PW}^{i,j} + \text{X}^{i,j}_\text{other} \rho_\text{other}^{i,j}),
\end{equation}
}
\end{linenomath}
and for the \doubleMuCr, \doubleEleCr, and \singlePhotonCr CRs is given by
\begin{linenomath}
\ifthenelse{\boolean{cms@external}}
{
\begin{multline}
\mathcal{L}_{\PGm\PGm,\Pe\Pe,\PGg}= \prod_{\text{cat}=i}^{n_\text{cat}}\prod_{\text{recoil}=j(i)}^{n_{\xi(i)}} \\
\text{Poisson}(n_\text{obs}^{i,j}\mid\text{X}^{i,j}_{\PZ/\PGg}L^{i,j}\rho_{\PZ/\PGg}^{i,j} + \text{X}^{i,j}_\text{other} \rho_\text{other}^{i,j}),
\end{multline}
}
{
\begin{equation}
\mathcal{L}_{\PGm\PGm,\Pe\Pe,\PGg}=\prod_{\text{cat}=i}^{n_\text{cat}}\prod_{\text{recoil}=j(i)}^{n_{\xi(i)}}\text{Poisson}(n_\text{obs}^{i,j}\mid\text{X}^{i,j}_{\PZ/\PGg}L^{i,j}\rho_{\PZ/\PGg}^{i,j} + \text{X}^{i,j}_\text{other} \rho_\text{other}^{i,j}),
\end{equation}
}
\end{linenomath}
where X$^{i,j}$ is the sum of background yields, and $\rho^{i,j}$ refers to the associated systematic uncertainty.

Because of the low event counts in the dilepton CRs, the subcategory yields are summed into the boosted and resolved \ttH categories. For the boosted \ttH category, the \doubleMuCr and \doubleEleCr CR yields are summed together to form a single \lljets CR. Furthermore, in the boosted and resolved \ttH subcategories, $I^{i,j}$ are shared across subcategories, therefore $i$ takes only two values corresponding to the boosted and resolved \ttH classes.

\subsection{Systematic uncertainties}
\label{subsec:Systematic_uncertainties}

The model on which the maximum likelihood fit is based is inclusive of experimental and theoretical uncertainties. These are modelled as nuisance parameters, which are typically constrained by a template fit where there is a dependence on the hadronic recoil distribution, but are otherwise constrained by a log-normal distribution for those that affect the overall normalisation of a given process.

Theoretical uncertainties related to the PDF parameters and missing higher order corrections in the QCD and EW perturbative expansions are estimated by following the procedure outlined in Ref.~\cite{deFlorian:2016spz} for \ttH and \VH processes, and in Ref.~\cite{Lindert:2017olm} for \vjets and \singlePhotonCr processes. Systematic uncertainties related to the PDF, and the renormalisation and factorisation scales, are treated as independent nuisance parameters but are correlated across years in the fit.

A photon normalisation uncertainty of 40\% is included in the \singlePhotonCr CR, to cover uncertainties in the translation between the \singlePhotonCr and ${\PZ\to\Pell\Pell}$ yields, and is only correlated between 2017 and 2018 samples given the \singlePhotonCr sample for 2016 is generated with a different tune.

Data-derived correction factors are applied to simulated events containing \PQb, \PQt, and \PW jets, and therefore the systematic uncertainties due to the limited precision in these corrections are propagated to the simulated samples. These are referred to as tagging uncertainties, and also account for the uncertainties in the tagging efficiencies and misidentification probabilities. The tagging methods and uncertainty propagation are consistent between years, and therefore are correlated across years in the fit.

The uncertainty in the combined PF \ptmiss and \htmiss trigger efficiency is computed using the \singleMuCr and \doubleMuCr CRs. These are independent of the \ptmiss and \htmiss data sets in the SR, ensuring an unbiased measurement of the uncertainty. This uncertainty is measured at 2\%, and is applied independently in each year of data-taking due to variations in the trigger performance. The same uncertainty is measured in the electron and photon trigger efficiency, and is similarly uncorrelated between years. An additional trigger inefficiency uncertainty due to the mistiming of ECAL trigger inputs detailed in Section~\ref{subsec:triggers} is applied to the data-taking years 2016 and 2017.

The uncertainty in the integrated luminosity varies between 1.2--2.5\% depending on the data-taking year, with an overall uncertainty of 1.6\% for the 2016--2018 period \cite{CMS:2021xjt,CMS-PAS-LUM-17-004,CMS-PAS-LUM-18-002}. The uncertainty is applied with correlated and uncorrelated components across years.

The uncertainties considered in the analysis are presented in Table~\ref{tab:systematicsRun2} with the pre-fit ranges corresponding to the maximum and minimum deviations of the event yields from their nominal values across each region, year of data-taking, category, recoil bin, and all SM background processes, when the respective systematic uncertainty is changed within $\pm$1~standard deviation. Systematic uncertainties not specified above are typically assumed to be uncorrelated from year to year when performing the fit. Those for which the source of the systematic uncertainty is identical for each year are treated as correlated. All systematic uncertainties are correlated across regions.

The overall experimental uncertainty is found to be dominated by \PW tagging for the \ttH and \PQb tagging for the \VH categories in the SR. The lepton and photon candidate efficiencies for identification, isolation, and reconstruction, and uncertainties in the JER, JES, and trigger efficiencies also make significant contributions. The theoretical uncertainty is dominated by variations in the renormalisation scale, factorisation scale, and PDF for \vjets processes, although these are particularly sensitive to the high exclusive jet multiplicity characterising the \ttH and \VH categories.

\begin{table*}[hbtp]
\centering
\topcaption{The ranges corresponding to the maximum and minimum deviations of the event yields from their nominal values, provided where applicable across each region, year of data-taking, category, recoil bin, and all SM background processes, when the respective systematic uncertainty is changed within $\pm$1~standard deviation.}
\cmsTable{\begin{tabular}{{l}{c}{c}{c}{c}{c}{c}{c}}
Systematic uncertainties on background yields (pre-fit) & \multicolumn{2}{c}{Signal region} & \multicolumn{2}{c}{\ljets} & \multicolumn{2}{c}{\lljets} & \singlePhotonCr \\
  & \ttH cat. & \VH cat. & \ttH cat. & \VH cat. & \ttH cat. & \VH cat. & \VH cat. \\
\noalign{\hrule height 0.5pt}
\multicolumn{1}{c}{Theoretical uncertainties} & & & & & & & \\
Fact. scale \vjets (QCD) & $<$1.0-7.7 \% & $<$1.0-19 \% & $<$1.0-2.6 \% & $<$1.0-11 \% & $<$1.0-20 \% & $<$1.0-22 \% & 6.0 \% \\
Ren. scale \vjets (QCD) & $<$1.0-7.2 \% & $<$1.0-8.6 \% & $<$1.0-3.6 \% & $<$1.0-10 \% & $<$1.0-14 \% & 2.0-11 \% & 12 \% \\
PDF \vjets & $<$1.0-9.1 \% & 2.0-23 \% & $<$1.0-3.1 \% & $<$1.0-15 \% & $<$1.0-23 \% & $<$1.0-26 \% & 8.0 \% \\
Ren. \& Fact. scale \ttH (QCD) & $<$1.0-1.7 \% & $<$1.0 \% & $<$1.0-1.4 \% & $<$1.0-1.4 \% & $<$1.0 \% & $<$1.0 \% & \NA \\
Ren. \& Fact. scale \ttbar (QCD) & 7.8-15 \% & 2.5-9.3 \% & 6.4-17 \% & $<$1.0-6.3 \% & $<$1.0-5.8 \% & $<$1.0-5.8 \% & \NA \\
NNLO QCD \& NLO EW \PQt quark \pt reweighting (inc. PDF) & $<$1.0-3.1 \% & $<$1.0-1.2 \% & $<$1.0-4.0 \% & $<$1.0-3.9 \% & $<$1.0 \% & $<$1.0 \% & \NA \\
Ren. \& Fact. scale VV (QCD) & $<$1.0 \% & $<$1.0 \% & $<$1.0 \% & $<$1.0 \% & $<$1.0 \% & $<$1.0 \% & $<$1.0 \% \\
\ttH \& \VH cat. cross section (QCD) & 5.8-9.2 \% & $<$1.0-3.8 \% & \NA & \NA & \NA & \NA & \NA \\
\ttH \& \VH cat. cross section (PDF \& $\alpha_{s}$) & 3.6 \% & 1.6-1.8 \% & \NA & \NA & \NA & \NA & \NA \\
Initial-state radiation  & 2.0 \% & 3.0-6.0 \% & 2.0 \% & $<$1.0-4.2 \% & 2.0 \% & 6.0 \% & $<$1.0-4.0 \% \\
Final-state radiation & 5.0 \% & 3.0-5.0 \% & 2.0-2.2 \% & $<$1.0-3.1 \% & 4.6-5.0 \% & 5.0 \% & 2.0-3.0 \% \\
Photon normalisation & \NA & \NA & \NA & \NA & \NA & \NA & 40 \% \\
[\cmsTabSkip]
\multicolumn{1}{c}{Experimental uncertainties} & & & & & & & \\
Integrated luminosity & 1.2-2.5 \% & 1.2-2.5 \% & 1.2-2.5 \% & 1.2-2.5 \% & 1.2-2.5 \% & 1.2-2.5 \% & 1.2-2.5 \% \\
\PQt-tagging & 3.2-6.5 \% & \NA & 2.1-5.7 \% & \NA & \NA & \NA & \NA \\
\PW-tagging & 7.8-18 \% & \NA & 7.1-18 \% & \NA & \NA & \NA & \NA \\
\PQb-tagging & 8.2-12 \% & 8.2-22 \% & 6.5-11 \% & 2.4-11 \% & 5.6-8.7 \% & 1.6-9.6 \% & 6.6-9.0 \% \\
Electron identification \& isolation & \NA & \NA & 3.7-11 \% & 4.7-9.6 \% & $<$1.0-15 \% & $<$1.0-20 \% & \NA \\
Electron reconstruction & \NA & \NA & $<$1.0-1.8 \% & $<$1.0 \% & 1.0-1.5 \% & $<$1.0-1.4 \% & \NA \\
Muon identification & \NA & \NA & $<$1.0-1.0 \% & $<$1.0-1.0 \% & $<$1.0-1.8 \% & $<$1.0-1.9 \% & \NA \\
Muon isolation & \NA & \NA & $<$1.0 \% & $<$1.0 \% & $<$1.0 \% & $<$1.0 \% & \NA \\
Lepton veto & $<$1.0 \% & $<$1.0 \% & \NA & \NA & \NA & \NA & \NA \\
Photon identification \& isolation & \NA & \NA & \NA & \NA & \NA & \NA & 2.4-12 \% \\
Photon reconstruction & \NA & \NA & \NA & \NA & \NA & \NA & $<$1.0 \% \\
Pileup & 1.4-8.8 \% & $<$1.0-4.5 \% & $<$1.0-4.8 \% & $<$1.0-4.7 \% & $<$1.0-2.1 \% & $<$1.0-7.9 \% & $<$1.0-3.3 \% \\
Trigger inefficiency & $<$1.0-12 \% & $<$1.0-1.4 \% & $<$1.0-3.4 \% & $<$1.0-2.4 \% & $<$1.0-1.6 \% & $<$1.0-1.5 \% & $<$1.0-0.3 \% \\
Trigger & 2.0 \% & 2.0 \% & 2.0 \% & 2.0 \% & 2.0 \% & 2.0 \% & 2.0 \% \\
Tau lepton veto & $<$1.0 \% & $<$1.0 \% & $<$1.0-1.0 \% & $<$1.0-2.4 \% & $<$1.0 \% & $<$1.0 \% & $<$1.0 \% \\
JER & 2.4-3.6 \% & $<$1.0-1.1 \% & 1.7-3.0 \% & $<$1.0-1.5 \% & $<$1.0-3.5 \% & $<$1.0-1.4 \% & $<$1.0-2.9 \% \\
JES & $<$1.0-6.3 \% & $<$1.0-2.9 \% & $<$1.0-5.0 \% & $<$1.0-2.2 \% & $<$1.0-6.7 \% & $<$1.0-2.8 \% & $<$1.0-3.8 \% \\
QCD prediction & 100 \% & 100 \% & \NA & \NA & \NA & \NA & \NA \\
\end{tabular}}
\label{tab:systematicsRun2}
\end{table*}

\section{Results}
\label{sec:Results}

\begin{figure*}[htbp!]
    \centering
    \includegraphics[width=0.98\textwidth]{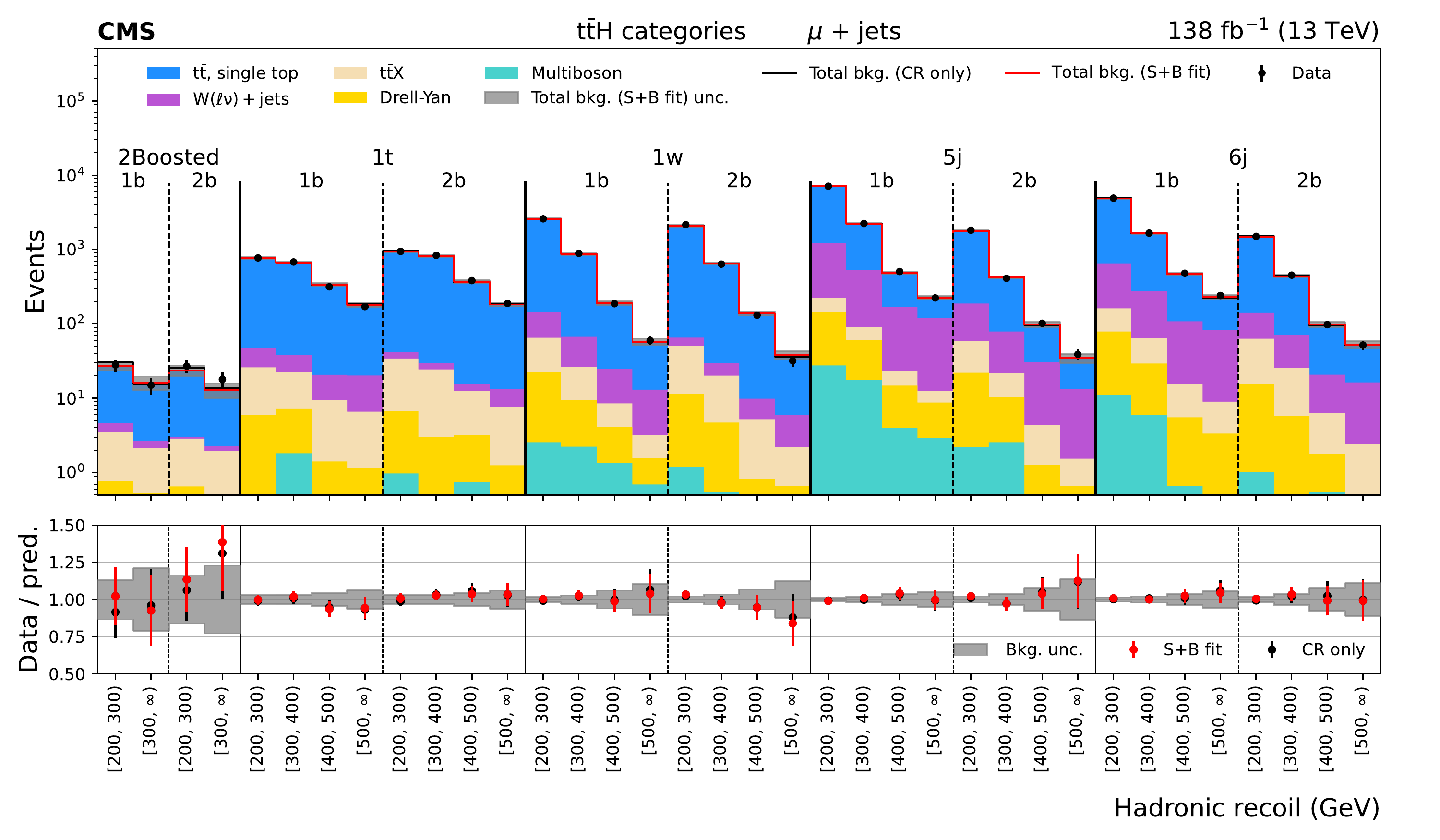}
    \includegraphics[width=0.98\textwidth]{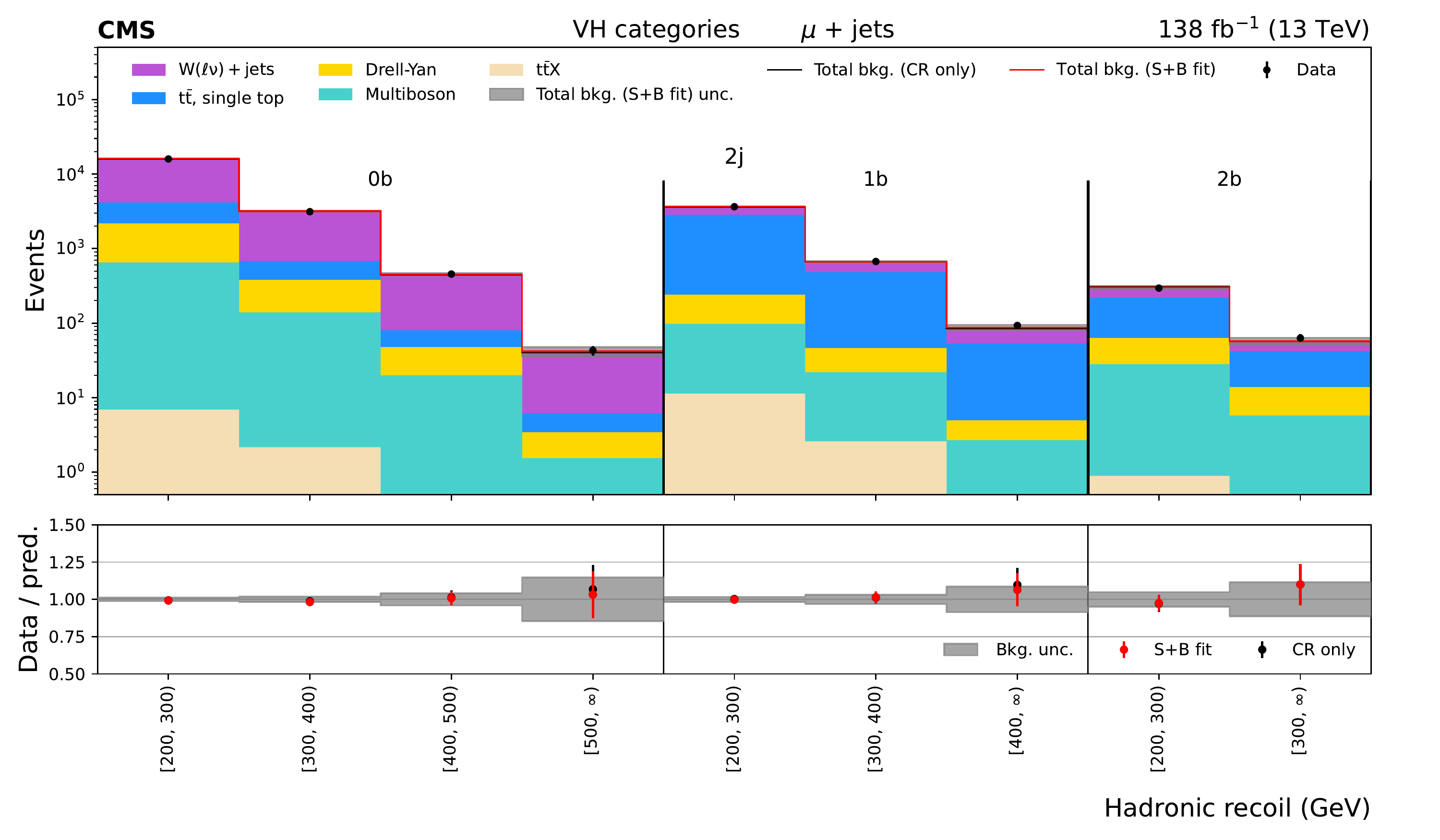}
    \caption{Distributions of hadronic recoil in the \ttH (upper plot) and \VH (lower plot) categories for the \singleMuCr CR. The black histogram shows the total background (bkg.) prediction from a CR only, B-only fit, while the red histogram shows the yields from a CR$+$SR S$+$B fit. The uncertainty in the predicted background (Bkg. unc.) accounts for both systematic and simulated statistical contributions.}
    \label{fig:SingleMu}
\end{figure*}

\begin{figure*}[htbp!]
    \centering
    \includegraphics[width=0.98\textwidth]{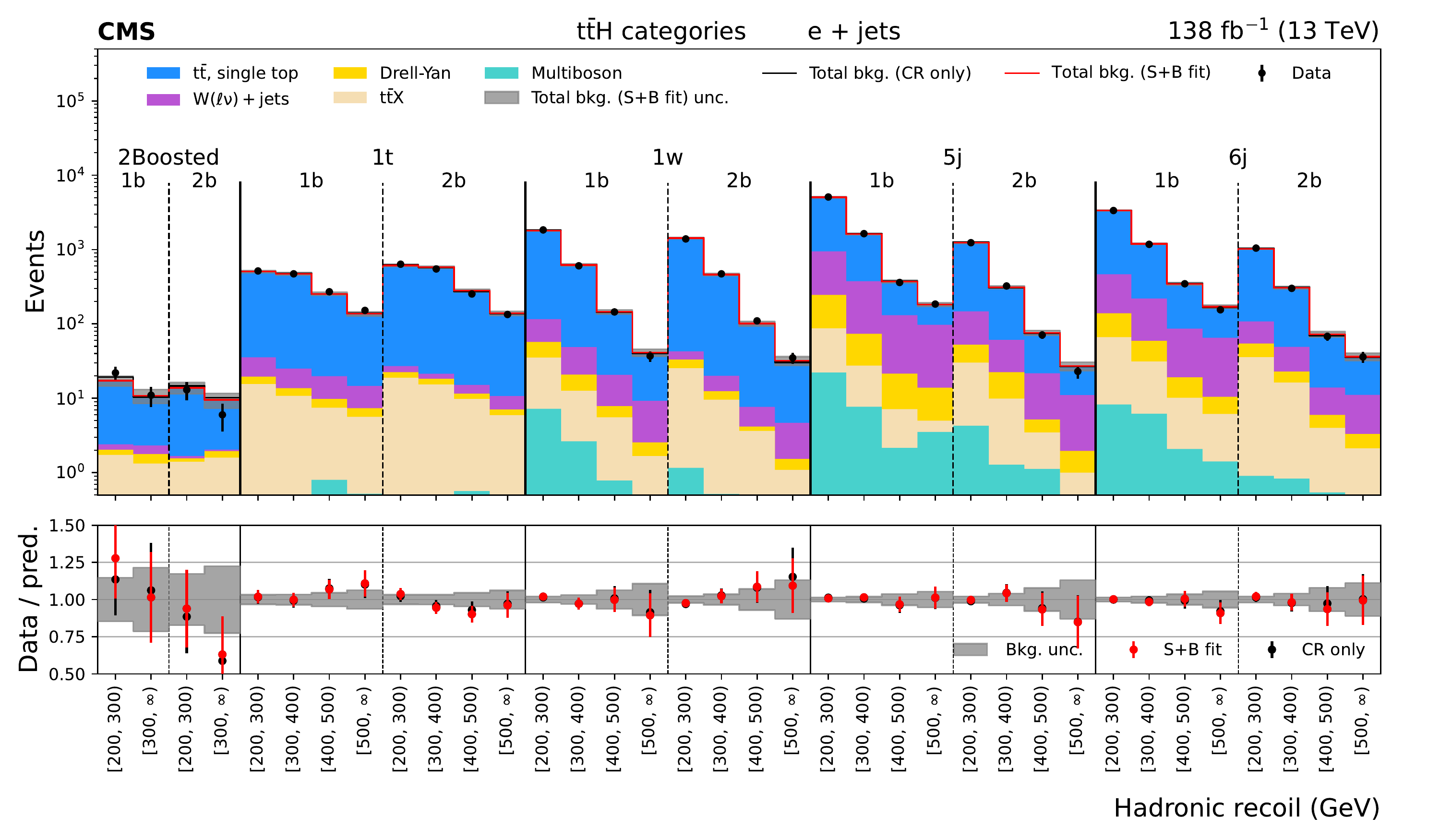}
    \includegraphics[width=0.98\textwidth]{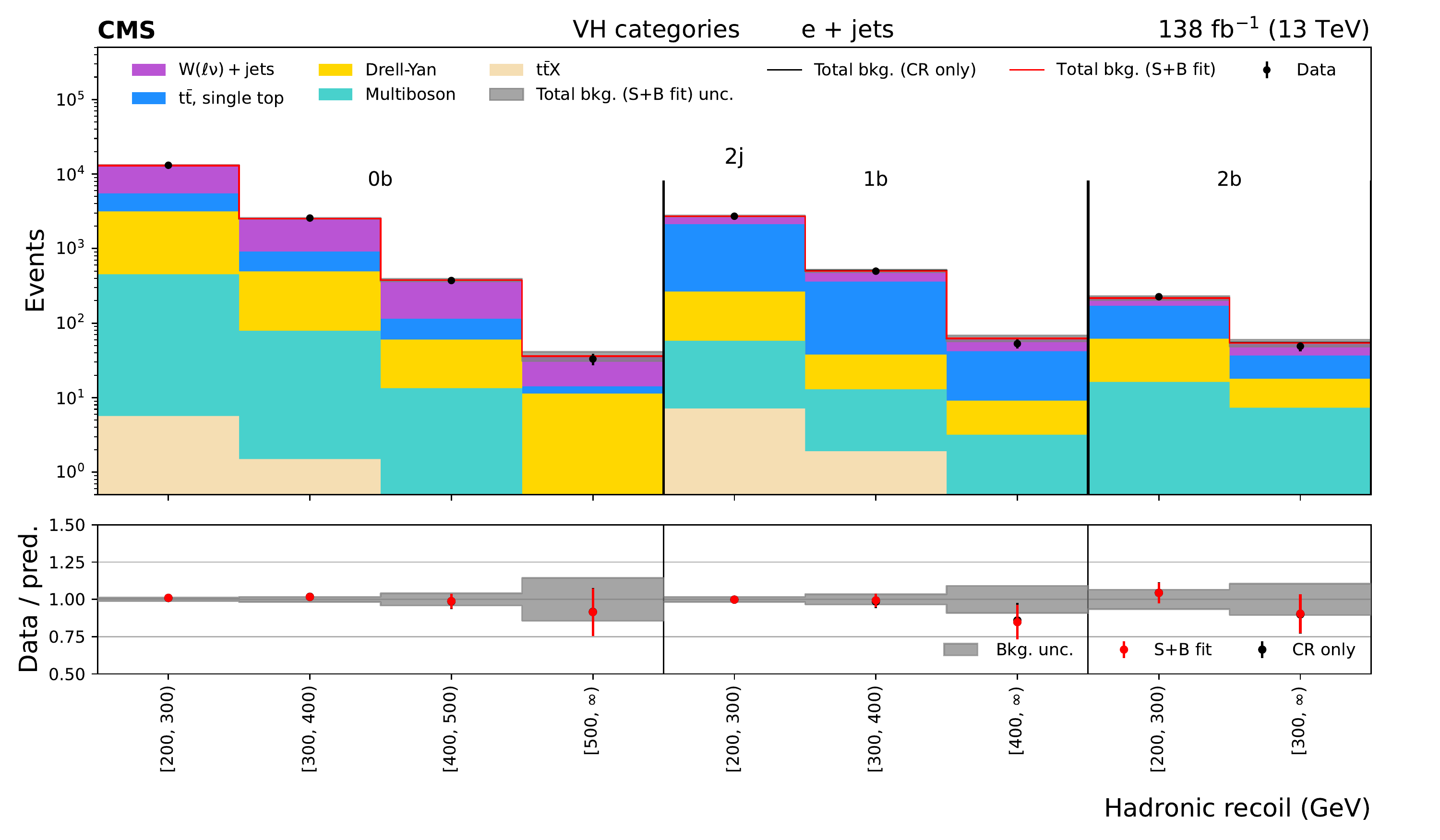}
    \caption{Distributions of hadronic recoil in the \ttH (upper plot) and \VH (lower plot) categories for the \singleEleCr CR. The black histogram shows the total background (bkg.) prediction from a CR only, B-only fit, while the red histogram shows the yields from a CR$+$SR S$+$B fit. The uncertainty in the predicted background (Bkg. unc.) accounts for both systematic and simulated statistical contributions.}
    \label{fig:SingleEle}
\end{figure*}

\begin{figure*}[htbp!]
    \centering
    \includegraphics[width=0.98\textwidth]{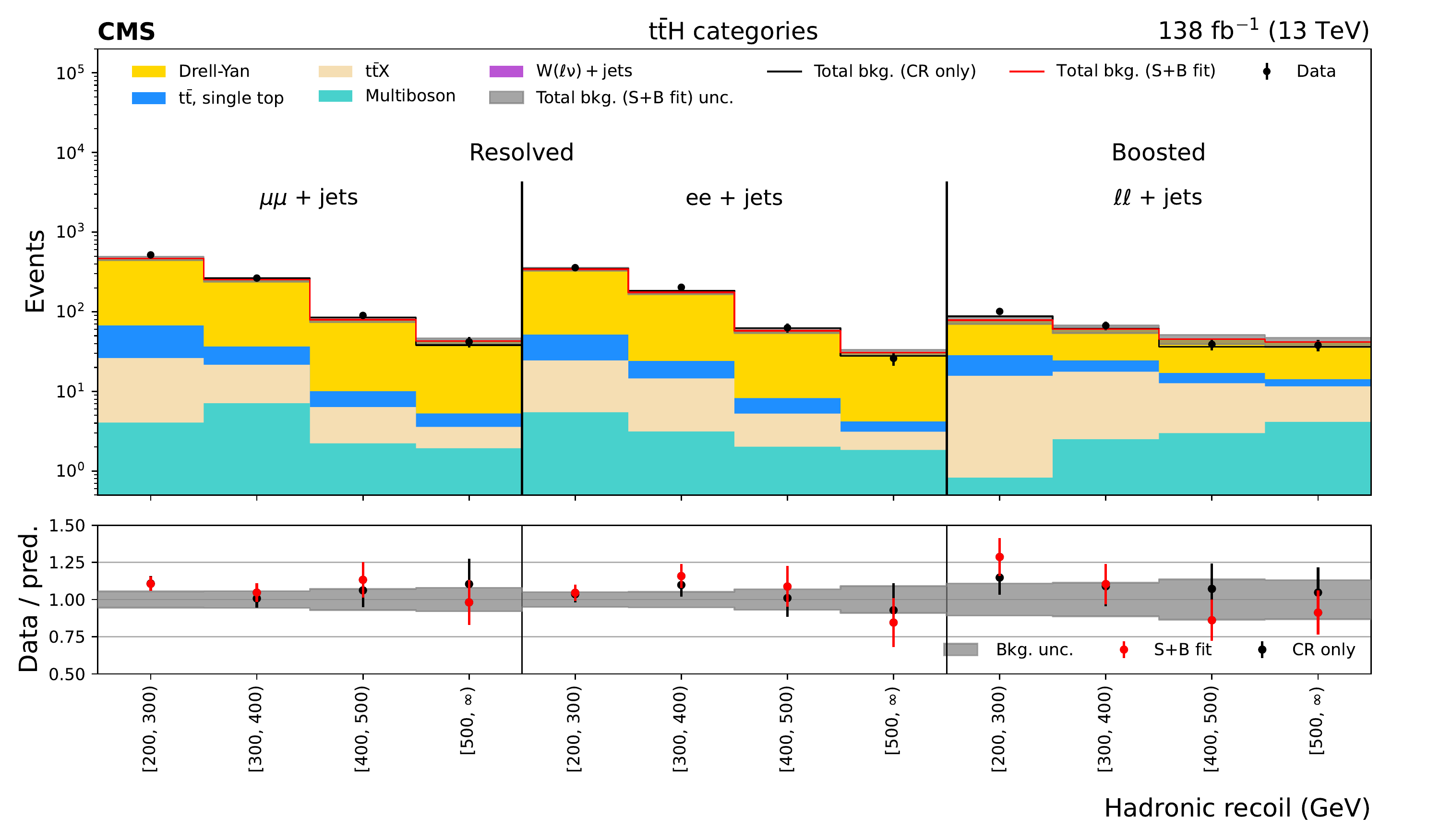}
    \includegraphics[width=0.98\textwidth]{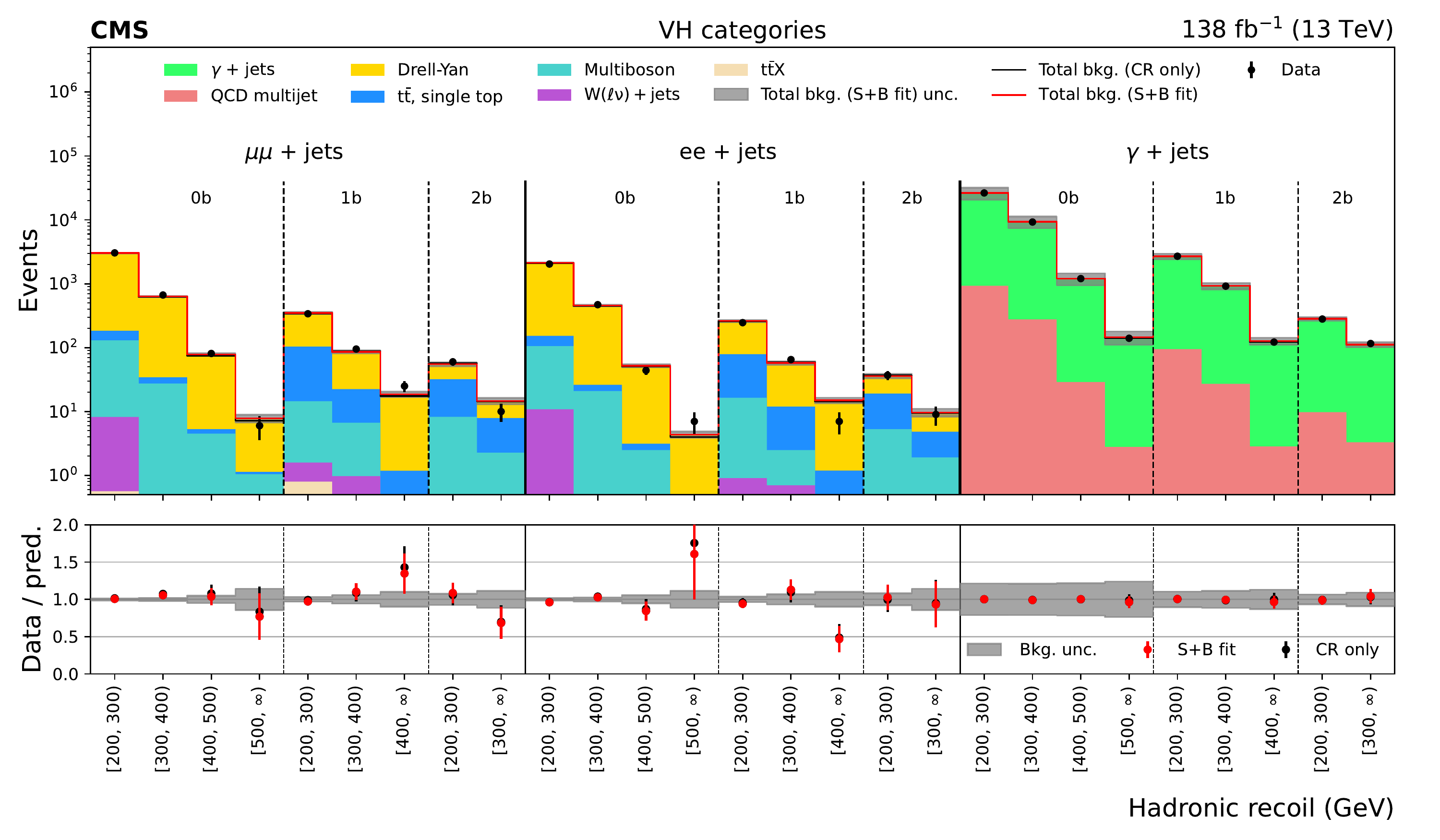}
    \caption{Distributions of hadronic recoil in the \ttH category for the \doubleMuCr, \doubleEleCr, and \lljets CRs (upper plot), and the \VH category for the \doubleMuCr, \doubleEleCr, and \singlePhotonCr CRs (lower plot). The black histogram shows the total background (bkg.) prediction from a CR only, B-only fit, while the red histogram shows the yields from a CR$+$SR S$+$B fit. The uncertainty in the predicted background (Bkg. unc.) accounts for both systematic and simulated statistical contributions.}
    \label{fig:otherreg}
\end{figure*}

\begin{figure*}[htbp!]
    \centering
    \includegraphics[width=0.98\textwidth]{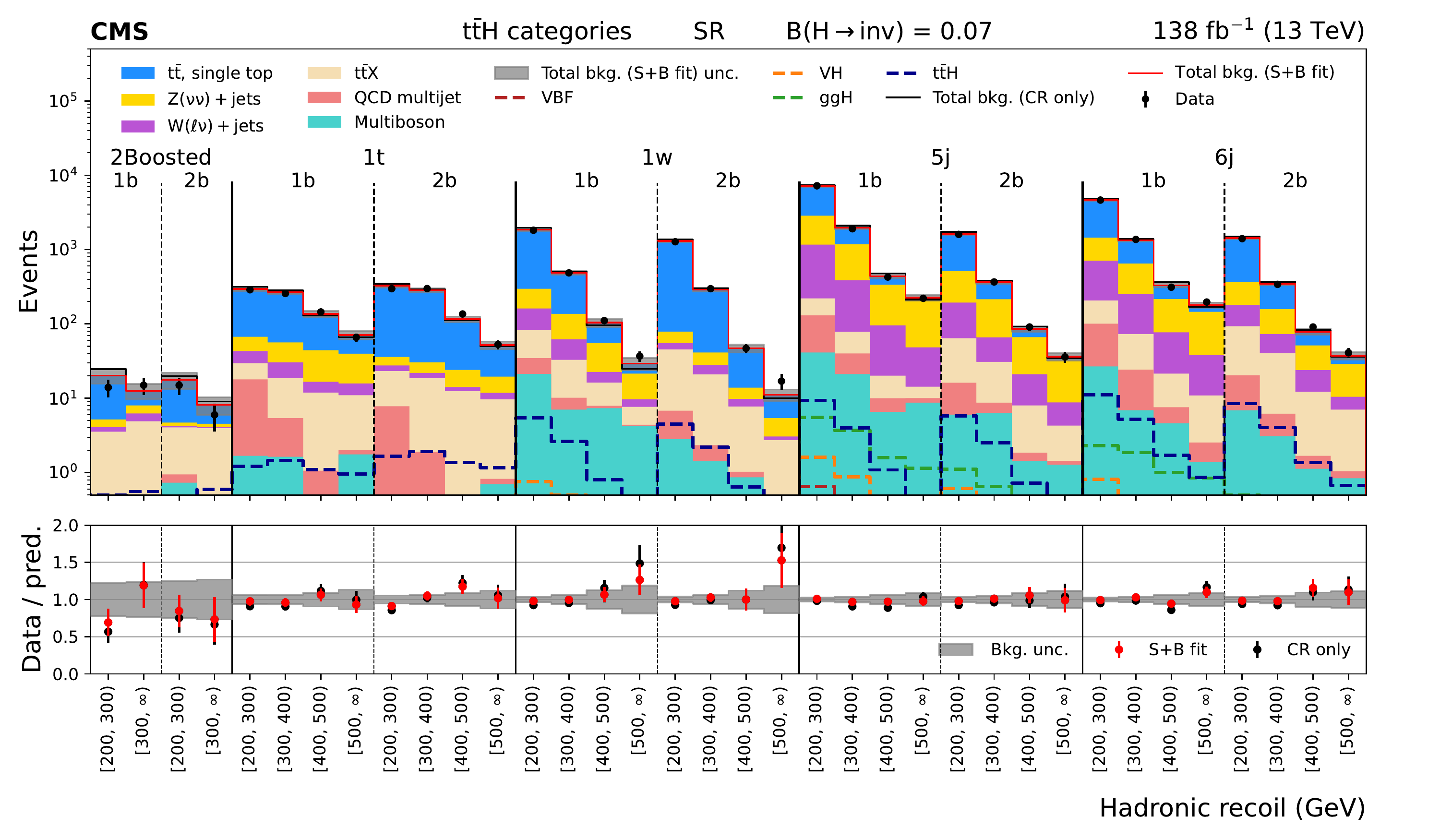}
    \includegraphics[width=0.98\textwidth]{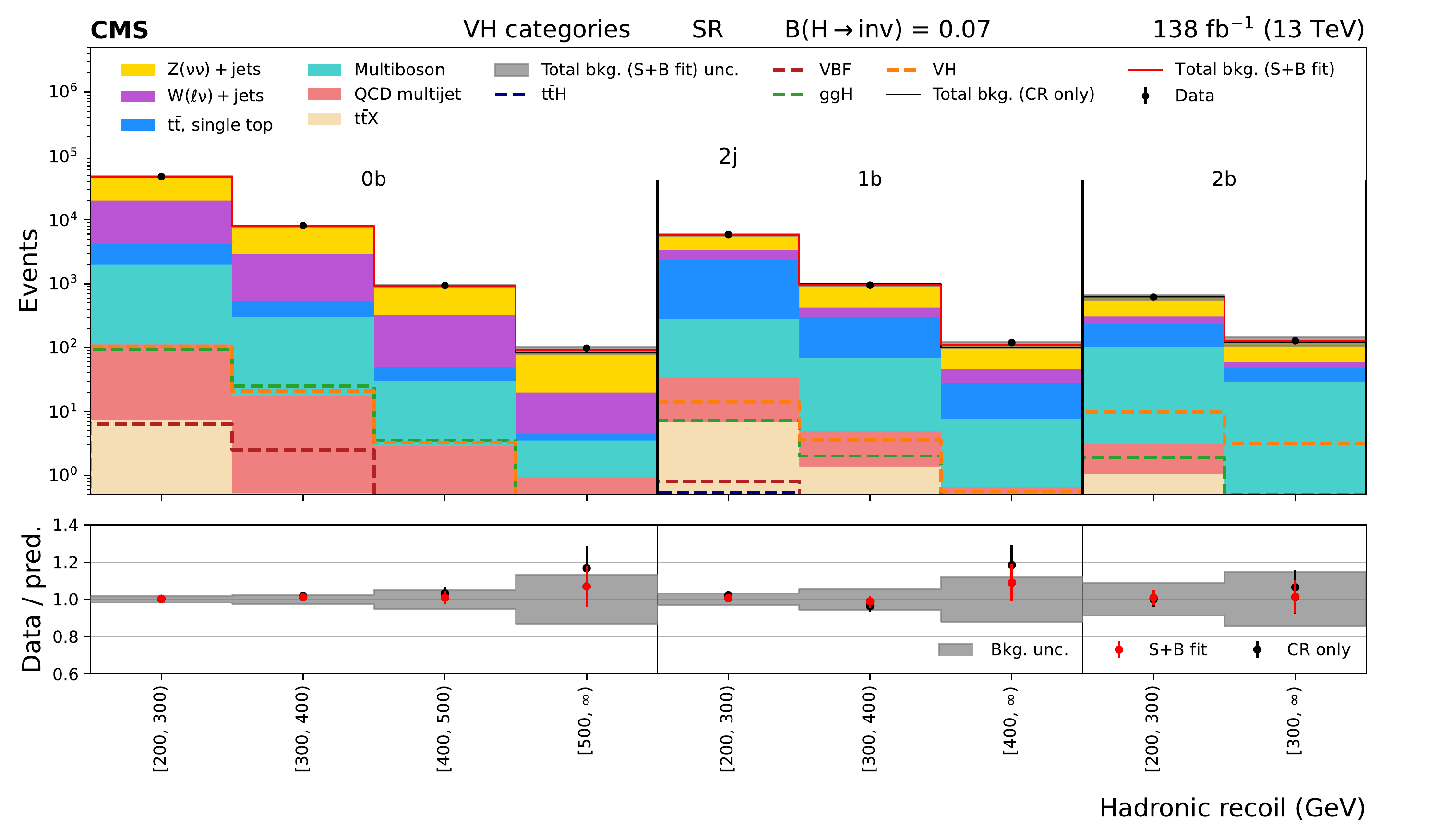}
    \caption{Distributions of hadronic recoil in the \ttH (upper plot) and \VH (lower plot) categories for the SR, showing the signal contributions from \ttH, \VH, \ggH, and VBF weighted by $\bhinvlim=0.07$. The black histogram shows the total background (bkg.) prediction from a CR only, B-only fit, while the red histogram shows the yields from a CR$+$SR S$+$B fit. The uncertainty in the predicted background (Bkg. unc.) accounts for both systematic and simulated statistical contributions.}
    \label{fig:SRMRP}
\end{figure*}

The hadronic recoil distributions across all \ttH and \VH subcategories are shown in Figs.~\ref{fig:SingleMu}~to~\ref{fig:SRMRP}. The predicted background yield from the fit to the CRs only is shown with the result of a fit including the data in the SR. The agreement between the data and simulation is presented below each distribution, with the uncertainty in the predicted background uncertainty (Bkg. unc.) accounting for both systematic and simulated statistical contributions. Figure~\ref{fig:SingleMu}~(\ref{fig:SingleEle}) shows the \singleMuCr (\singleEleCr) CR yields for the \ttH and \VH categories, respectively, aggregated over 2016--2018. In these CRs, \llost background from \ttbar, $\PW\to\Pell\PGn$, and single \PQt quark production dominates, with smaller contributions from multiboson and ${\ttbar\PX}$ processes. The \doubleMuCr, \doubleEleCr, \lljets (only for \ttH), and \singlePhotonCr (only for \VH) CR distributions used for the prediction of backgrounds stemming from \zinv decays are shown in Fig.~\ref{fig:otherreg} for 2016--2018. In addition, the total SM background prediction in the SR, consisting of \llost, \zinv, and QCD backgrounds, is shown for the \ttH and \VH category in Fig.~\ref{fig:SRMRP}. The SR distributions contain all the Higgs boson production modes in the fitted \bhinvlim signal, including the \ggH and VBF contamination in the \ttH and \VH categories, with the prevalence of the \ggH process due to its high production cross section. The post-fit event yields for each subcategory and recoil bin in the SR are tabulated in Table~\ref{tab:SRYieldsTable}. For these results, a fit assuming \bhinvlim$=0$ such that only SM background contributions are considered (B-only) is performed simultaneously using only the CRs (CR only), which are independent of the SR, or across both SR and CRs (CR$+$SR). A fit across all regions, including signal and background contributions (S$+$B fit), is also performed, in which the signal contribution is weighted by the best-fit signal strength, \bhinvlim. In all cases, uncertainties are inclusive of statistical and systematic contributions.

\begin{table*}[htb!]
\centering
\topcaption{Total post-fit yields in the SRs in each recoil bin and analysis category obtained by summing the contributions from the individual data-taking periods. B-only fits are performed for either CR$+$SR or CR only cases. The extracted signal yields from an S$+$B fit are also reported, where the signal strength is weighted by ${\bhinvlim=0.07}$.}
\cmsTable{\begin{tabular}{{l}{l}{c}{c}{c}{c}{c}{c}{c}}
   & \multicolumn{1}{c}{} & {\llost} &  {\zinv} & {QCD} & \multicolumn{2}{c}{Total background} & & Signal \\
& {} &  CR only &  CR only &  CR only & CR only & CR + SR & Data & $\bhinvlim=0.07$ \\
Subcategory & Hadronic recoil & B-only fit & B-only fit & B-only fit & B-only fit & B-only fit & & S$+$B fit \\
\noalign{\hrule height 0.5pt}
\ttH 1t1b & [200, 300) & 251.1 $\pm$ 9.5 & 35.2 $\pm$ 4.1 & 23.1 $\pm$ 16.8 & 309.4 $\pm$ 19.8 & 295.5 $\pm$ 11.6 & 288.0 $\pm$ 17.0 & 1.0 $\pm$ 0.8 \\
 & [300, 400) & 235.2 $\pm$ 9.5 & 35.7 $\pm$ 5.0 & 5.2 $\pm$ 4.2 & 276.1 $\pm$ 11.5 & 268.1 $\pm$ 9.1 & 257.0 $\pm$ 16.0 & 1.3 $\pm$ 1.0 \\
 & [400, 500) & 97.5 $\pm$ 5.3 & 27.6 $\pm$ 4.9 & 0.9 $\pm$ 0.6 & 126.1 $\pm$ 7.2 & 135.5 $\pm$ 6.7 & 145.0 $\pm$ 12.0 & 1.0 $\pm$ 0.8 \\
 & [500, $\infty$) & 37.5 $\pm$ 2.9 & 26.1 $\pm$ 4.9 & 0.3 $\pm$ 0.3 & 63.9 $\pm$ 5.7 & 70.1 $\pm$ 5.1 & 66.0 $\pm$ 8.1 & 0.9 $\pm$ 0.7 \\
\ttH 1t2b & [200, 300) & 312.5 $\pm$ 12.0 & 19.0 $\pm$ 2.2 & 10.9 $\pm$ 8.6 & 342.4 $\pm$ 14.9 & 328.1 $\pm$ 10.5 & 298.0 $\pm$ 17.3 & 1.4 $\pm$ 1.2 \\
 & [300, 400) & 265.9 $\pm$ 10.7 & 20.2 $\pm$ 2.7 & 2.5 $\pm$ 1.7 & 288.6 $\pm$ 11.2 & 287.1 $\pm$ 9.3 & 299.0 $\pm$ 17.3 & 1.6 $\pm$ 1.3 \\
 & [400, 500) & 93.6 $\pm$ 5.1 & 15.4 $\pm$ 2.6 & 0.4 $\pm$ 0.3 & 109.5 $\pm$ 5.7 & 116.5 $\pm$ 5.2 & 136.0 $\pm$ 11.7 & 1.2 $\pm$ 0.9 \\
 & [500, $\infty$) & 35.4 $\pm$ 2.9 & 13.8 $\pm$ 2.5 & 0.2 $\pm$ $<$0.1 & 49.4 $\pm$ 3.9 & 52.8 $\pm$ 3.5 & 53.0 $\pm$ 7.3 & 1.0 $\pm$ 0.8 \\
\ttH 1W1b & [200, 300) & 1704.6 $\pm$ 49.9 & 190.7 $\pm$ 21.2 & 18.8 $\pm$ 16.8 & 1914.1 $\pm$ 56.8 & 1855.7 $\pm$ 41.2 & 1819.0 $\pm$ 42.6 & 5.7 $\pm$ 4.0 \\
 & [300, 400) & 395.6 $\pm$ 15.1 & 90.2 $\pm$ 12.7 & 4.3 $\pm$ 2.9 & 490.0 $\pm$ 19.9 & 485.0 $\pm$ 16.2 & 486.0 $\pm$ 22.0 & 2.9 $\pm$ 1.9 \\
 & [400, 500) & 56.2 $\pm$ 3.9 & 35.8 $\pm$ 6.5 & 0.8 $\pm$ 0.5 & 92.7 $\pm$ 7.7 & 103.7 $\pm$ 7.1 & 111.0 $\pm$ 10.5 & 0.9 $\pm$ 0.6 \\
 & [500, $\infty$) & 9.9 $\pm$ 1.3 & 13.9 $\pm$ 2.9 & 0.3 $\pm$ $<$0.1 & 24.1 $\pm$ 3.2 & 29.5 $\pm$ 3.0 & 37.0 $\pm$ 6.1 & 0.4 $\pm$ 0.3 \\
\ttH 1W2b & [200, 300) & 1295.8 $\pm$ 40.7 & 53.1 $\pm$ 5.7 & 5.6 $\pm$ 3.8 & 1354.5 $\pm$ 41.3 & 1311.6 $\pm$ 29.4 & 1276.0 $\pm$ 35.7 & 3.9 $\pm$ 3.2 \\
 & [300, 400) & 266.2 $\pm$ 11.8 & 27.2 $\pm$ 3.8 & 1.3 $\pm$ 0.9 & 294.7 $\pm$ 12.4 & 291.3 $\pm$ 9.9 & 298.0 $\pm$ 17.3 & 1.9 $\pm$ 1.6 \\
 & [400, 500) & 38.3 $\pm$ 3.3 & 8.1 $\pm$ 1.5 & 0.2 $\pm$ $<$0.1 & 46.6 $\pm$ 3.7 & 47.6 $\pm$ 3.1 & 47.0 $\pm$ 6.9 & 0.6 $\pm$ 0.4 \\
 & [500, $\infty$) & 6.0 $\pm$ 1.0 & 3.7 $\pm$ 0.7 & 0.1 $\pm$ $<$0.1 & 9.9 $\pm$ 1.2 & 11.3 $\pm$ 1.1 & 17.0 $\pm$ 4.1 & 0.2 $\pm$ $<$0.1 \\
\ttH 2Boosted1b & [200, 300) & 20.2 $\pm$ 3.6 & 3.8 $\pm$ 0.4 & 0.3 $\pm$ 0.3 & 24.3 $\pm$ 3.6 & 20.4 $\pm$ 2.6 & 14.0 $\pm$ 3.7 & 0.5 $\pm$ 0.3 \\
 & [300, $\infty$) & 6.3 $\pm$ 1.4 & 6.1 $\pm$ 0.9 & 0.1 $\pm$ $<$0.1 & 12.5 $\pm$ 1.7 & 12.9 $\pm$ 1.6 & 15.0 $\pm$ 3.9 & 0.5 $\pm$ 0.4 \\
\ttH 2Boosted2b & [200, 300) & 15.8 $\pm$ 2.9 & 3.9 $\pm$ 0.9 & 0.3 $\pm$ $<$0.1 & 20.0 $\pm$ 3.1 & 18.0 $\pm$ 2.4 & 15.0 $\pm$ 3.9 & 0.4 $\pm$ 0.3 \\
 & [300, $\infty$) & 5.4 $\pm$ 1.3 & 3.8 $\pm$ 0.5 & 0.1 $\pm$ $<$0.1 & 9.3 $\pm$ 1.4 & 8.6 $\pm$ 1.1 & 6.0 $\pm$ 2.4 & 0.5 $\pm$ 0.4 \\
\ttH 5j1b & [200, 300) & 5279.7 $\pm$ 114.4 & 1703.7 $\pm$ 82.8 & 99.1 $\pm$ 78.5 & 7082.4 $\pm$ 161.6 & 7122.6 $\pm$ 127.6 & 7207.0 $\pm$ 84.9 & 14.4 $\pm$ 7.7 \\
 & [300, 400) & 1135.0 $\pm$ 31.8 & 836.4 $\pm$ 50.0 & 22.5 $\pm$ 17.3 & 1994.0 $\pm$ 61.7 & 1960.9 $\pm$ 43.2 & 1907.0 $\pm$ 43.7 & 7.4 $\pm$ 3.8 \\
 & [400, 500) & 182.2 $\pm$ 9.0 & 267.5 $\pm$ 24.9 & 4.0 $\pm$ 2.8 & 453.6 $\pm$ 26.6 & 438.8 $\pm$ 16.2 & 427.0 $\pm$ 20.7 & 2.7 $\pm$ 1.4 \\
 & [500, $\infty$) & 54.2 $\pm$ 3.7 & 146.0 $\pm$ 20.3 & 1.5 $\pm$ 1.0 & 201.7 $\pm$ 20.6 & 226.2 $\pm$ 11.5 & 221.0 $\pm$ 14.9 & 1.5 $\pm$ 0.8 \\
\ttH 5j2b & [200, 300) & 1317.8 $\pm$ 47.3 & 350.0 $\pm$ 16.6 & 11.8 $\pm$ 8.5 & 1679.6 $\pm$ 50.9 & 1635.4 $\pm$ 33.9 & 1602.0 $\pm$ 40.0 & 6.3 $\pm$ 4.2 \\
 & [300, 400) & 188.7 $\pm$ 9.2 & 174.1 $\pm$ 10.4 & 2.7 $\pm$ 2.0 & 365.5 $\pm$ 14.1 & 363.3 $\pm$ 10.7 & 367.0 $\pm$ 19.2 & 2.9 $\pm$ 1.8 \\
 & [400, 500) & 33.6 $\pm$ 3.5 & 53.8 $\pm$ 5.1 & 0.5 $\pm$ 0.3 & 87.9 $\pm$ 6.2 & 86.3 $\pm$ 4.5 & 91.0 $\pm$ 9.5 & 0.9 $\pm$ 0.5 \\
 & [500, $\infty$) & 8.2 $\pm$ 1.4 & 24.6 $\pm$ 3.5 & 0.2 $\pm$ $<$0.1 & 33.0 $\pm$ 3.8 & 36.8 $\pm$ 2.5 & 36.0 $\pm$ 6.0 & 0.5 $\pm$ 0.3 \\
\ttH 6j1b & [200, 300) & 3851.5 $\pm$ 87.9 & 805.5 $\pm$ 38.8 & 85.9 $\pm$ 66.3 & 4742.9 $\pm$ 116.7 & 4672.6 $\pm$ 87.1 & 4632.0 $\pm$ 68.1 & 12.3 $\pm$ 8.1 \\
 & [300, 400) & 876.0 $\pm$ 27.5 & 438.8 $\pm$ 26.1 & 19.5 $\pm$ 13.4 & 1334.2 $\pm$ 40.2 & 1332.5 $\pm$ 30.4 & 1371.0 $\pm$ 37.0 & 6.7 $\pm$ 4.0 \\
 & [400, 500) & 179.6 $\pm$ 8.5 & 162.8 $\pm$ 15.4 & 3.4 $\pm$ 2.5 & 345.9 $\pm$ 17.8 & 330.9 $\pm$ 11.4 & 312.0 $\pm$ 17.7 & 2.4 $\pm$ 1.4 \\
 & [500, $\infty$) & 61.0 $\pm$ 4.0 & 98.2 $\pm$ 13.6 & 1.3 $\pm$ 1.0 & 160.5 $\pm$ 14.3 & 179.1 $\pm$ 8.4 & 197.0 $\pm$ 14.0 & 1.6 $\pm$ 0.8 \\
\ttH 6j2b & [200, 300) & 1214.0 $\pm$ 38.7 & 237.2 $\pm$ 11.4 & 15.6 $\pm$ 12.0 & 1466.8 $\pm$ 42.1 & 1433.1 $\pm$ 29.9 & 1404.0 $\pm$ 37.5 & 7.8 $\pm$ 6.1 \\
 & [300, 400) & 237.9 $\pm$ 12.0 & 118.8 $\pm$ 7.1 & 3.6 $\pm$ 2.9 & 360.3 $\pm$ 14.2 & 351.9 $\pm$ 10.8 & 341.0 $\pm$ 18.5 & 3.8 $\pm$ 2.9 \\
 & [400, 500) & 38.8 $\pm$ 3.8 & 40.9 $\pm$ 4.0 & 0.6 $\pm$ 0.4 & 80.3 $\pm$ 5.6 & 79.9 $\pm$ 4.3 & 91.0 $\pm$ 9.5 & 1.4 $\pm$ 1.0 \\
 & [500, $\infty$) & 12.9 $\pm$ 1.7 & 21.6 $\pm$ 3.0 & 0.2 $\pm$ $<$0.1 & 34.7 $\pm$ 3.5 & 38.1 $\pm$ 2.4 & 41.0 $\pm$ 6.4 & 0.7 $\pm$ 0.4 \\ [\cmsTabSkip]
 
\VH 2j0b & [200, 300) & 17753.9 $\pm$ 373.6 & 29102.3 $\pm$ 655.5 & 105.8 $\pm$ 68.3 & 46962.1 $\pm$ 757.6 & 47499.1 $\pm$ 460.7 & 47559.0 $\pm$ 218.1 & 185.6 $\pm$ 92.5 \\
 & [300, 400) & 2535.2 $\pm$ 69.4 & 5505.3 $\pm$ 155.0 & 16.8 $\pm$ 12.0 & 8057.3 $\pm$ 170.3 & 8075.7 $\pm$ 106.8 & 8106.0 $\pm$ 90.0 & 44.3 $\pm$ 23.0 \\
 & [400, 500) & 278.9 $\pm$ 16.1 & 684.1 $\pm$ 34.7 & 2.8 $\pm$ 1.8 & 965.8 $\pm$ 38.3 & 944.5 $\pm$ 26.7 & 938.0 $\pm$ 30.6 & 6.6 $\pm$ 3.4 \\
 & [500, $\infty$) & 19.2 $\pm$ 3.1 & 76.9 $\pm$ 8.1 & 0.9 $\pm$ 0.5 & 97.1 $\pm$ 8.7 & 95.7 $\pm$ 6.6 & 98.0 $\pm$ 9.9 & 0.6 $\pm$ 0.3 \\
\VH 2j1b & [200, 300) & 3020.1 $\pm$ 84.0 & 2490.4 $\pm$ 114.7 & 26.2 $\pm$ 24.5 & 5536.8 $\pm$ 144.3 & 5808.6 $\pm$ 111.1 & 5883.0 $\pm$ 76.7 & 20.3 $\pm$ 10.0 \\
 & [300, 400) & 360.1 $\pm$ 17.3 & 609.0 $\pm$ 44.1 & 3.6 $\pm$ 3.0 & 972.7 $\pm$ 47.5 & 962.3 $\pm$ 30.1 & 949.0 $\pm$ 30.8 & 5.2 $\pm$ 2.8 \\
 & [400, $\infty$) & 36.3 $\pm$ 4.5 & 66.7 $\pm$ 7.3 & 0.6 $\pm$ 0.5 & 103.7 $\pm$ 8.6 & 111.3 $\pm$ 7.7 & 120.0 $\pm$ 11.0 & 0.7 $\pm$ 0.4 \\
\VH 2j2b & [200, 300) & 209.4 $\pm$ 14.0 & 422.3 $\pm$ 46.6 & 2.0 $\pm$ 1.2 & 633.7 $\pm$ 48.6 & 620.1 $\pm$ 26.8 & 617.0 $\pm$ 24.8 & 10.8 $\pm$ 7.9 \\
 & [300, $\infty$) & 30.7 $\pm$ 3.5 & 102.6 $\pm$ 15.4 & 0.2 $\pm$ $<$0.1 & 133.6 $\pm$ 15.8 & 131.1 $\pm$ 9.8 & 128.0 $\pm$ 11.3 & 3.5 $\pm$ 2.5 \\
\end{tabular}}
\label{tab:SRYieldsTable}
\end{table*}

\begin{figure*}[htbp!]
    \centering
    \includegraphics[width=0.48\textwidth]{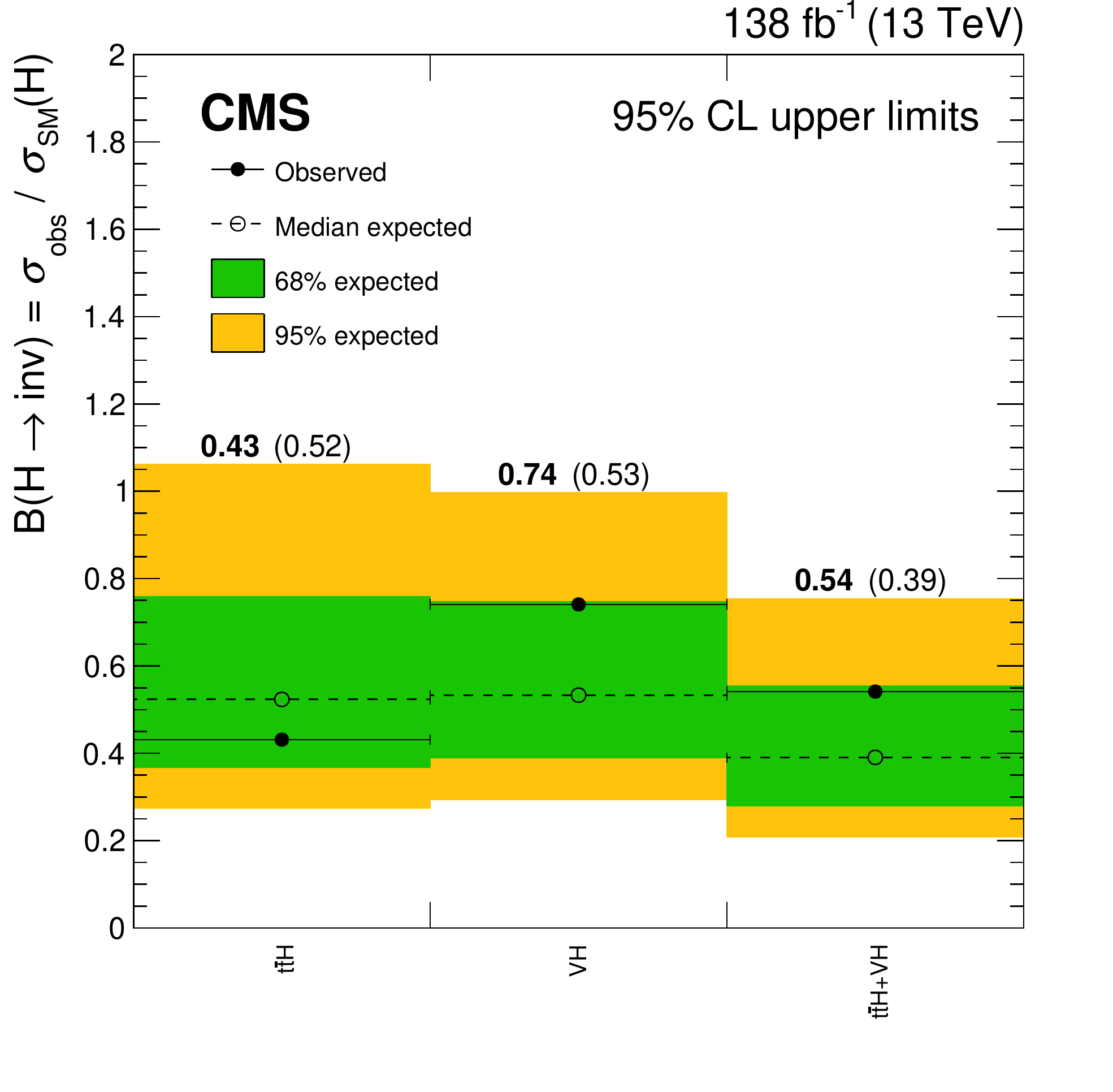}
    \includegraphics[width=0.48\textwidth]{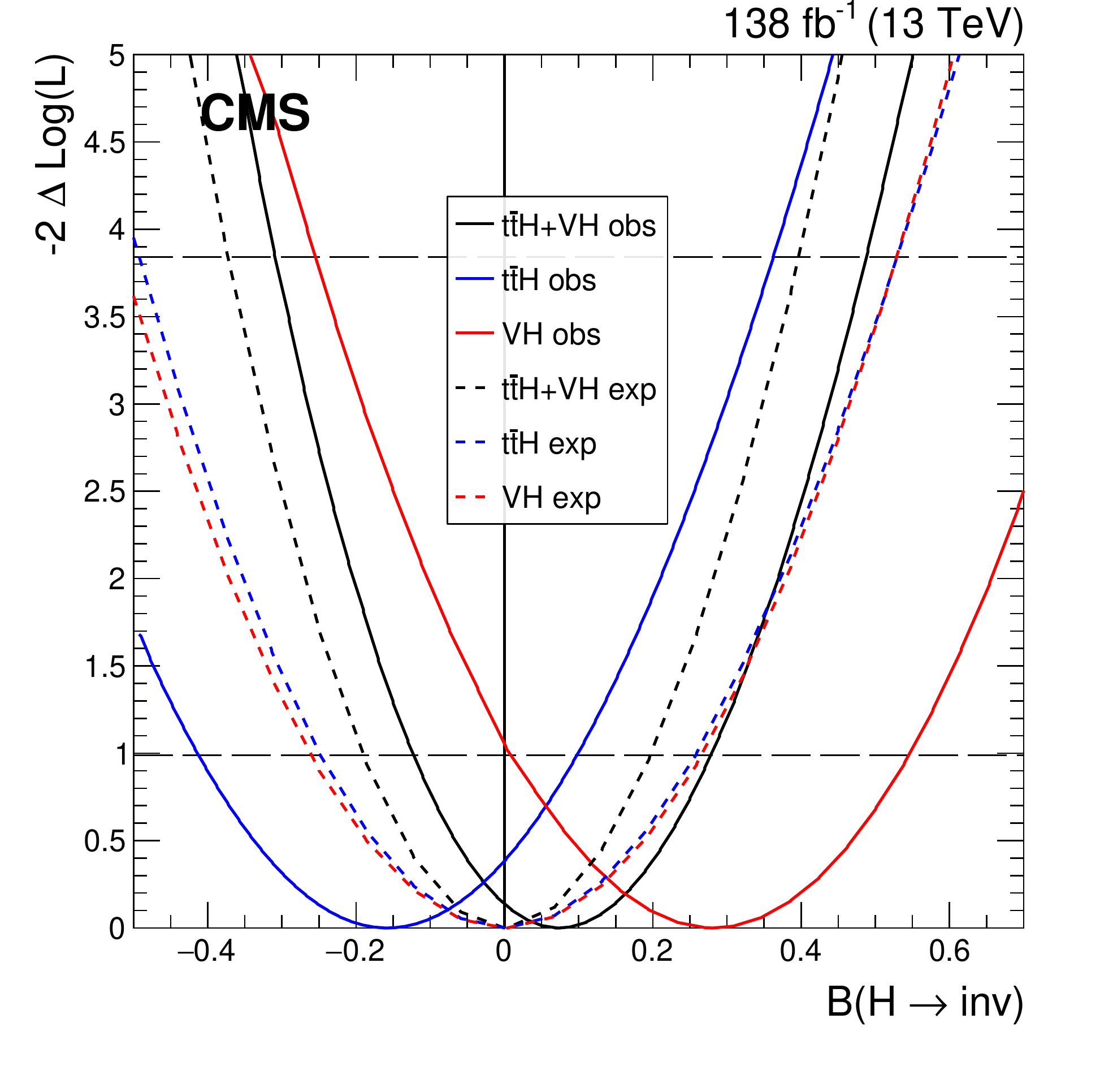}
    \caption{Left: Observed and expected limits at 95\% \CL for the \ttH and \VH categories using 2016--2018 data. Right: The profile likelihood scan corresponding to observed and expected (where $\bhinvlim=0$) limits in the fit to the \ttH and \VH categories.}
    \label{fig:Run2Plots}
\end{figure*}

The best-fit value for $\hat{\mu}$ and corresponding 68 and 95\% \CL confidence intervals are extracted following the procedure outlined in Ref.~\cite{Khachatryan:2014jba} and Ref.~\cite{CMS-NOTE-2011-005}. The computing of upper limits adheres to the CLs criterion~\cite{CLS1,CLS2} under the asymptotic approximation~\cite{Cowan:2010js}. The upper limits on \bhinvlim as extracted from the likelihood fit presented in Section~\ref{subsec:Likelihood_model} are found to be 0.43 (0.52 expected) and 0.74 (0.53 expected) at 95\% \CL for the \ttH and \VH categories, respectively, with a combined upper limit of 0.54 (0.39 expected). These results are shown in Fig.~\ref{fig:Run2Plots} together with the observed and expected profile likelihood distribution. The expected distribution assumes $\bhinvlim=0$. The results are compatible with the background expectation. The best-fit \bhinvlim for the \ttH and \VH categories is $\hat{\mu}=0.07_{-0.10}^{+0.10}(\text{stat.})$ $_{-0.17}^{+0.18}(\text{syst.})$ ($0.00_{-0.10}^{+0.10}(\text{stat.})$ $_{-0.16}^{+0.17}(\text{syst.})$ expected), where the pre-fit normalisation assumes that \bhinvlim$=1$. The systematic uncertainty with the largest impact on the \bhinvlim measurement for the \ttH and \VH categories using 2016--2018 data are those associated with the JES, while the statistical uncertainty contributes significantly to the overall uncertainty on \bhinvlim. The breakdown of the impacts into uncertainty groups are presented in Table~\ref{tab:impacts_table}, together with the expectation assuming \bhinvlim$=0$. The best-fit estimate for the \ttH (\VH) category is $\hat{\mu}=-0.16_{-0.26}^{+0.26}$ $(0.00_{-0.25}^{+0.26})$ ($\hat{\mu}=0.28_{-0.27}^{+0.27}$ $(0.00_{-0.26}^{+0.27})$).

\begin{table}[htb!]
\centering
\topcaption{The observed and expected impacts on \bhinvlim for different groups of uncertainties, where the expected results are produced with \bhinvlim$=0$.}
\begin{tabular}{lcc}
\multirow{2}{*}{Uncertainty group} & \multicolumn{2}{c}{Impact on \bhinvlim}  \\
   & Observed & Expected \\\hline
Jet energy calibration & $\pm0.11$ & $\pm0.11$ \\
Lepton veto & $\pm0.05$ & $^{+0.05}_{-0.04}$ \\
Lepton/photon identification & $\pm0.06$ & $\pm0.06$ \\
Theory & $^{+0.07}_{-0.06}$ & $^{+0.06}_{-0.05}$ \\
Integrated luminosity/pileup & $\pm0.02$ & $^{+0.02}_{-0.03}$ \\
QCD prediction & $\pm0.02$ & $\pm0.02$\\
Boosted object/\PQb jet tagging & $\pm0.02$ & $\pm0.02$ \\
Triggers & $\pm0.04$ & $\pm0.03$ \\
Stat. uncertainty of simulation & $\pm0.08$ & $\pm0.08$\\
Stat. uncertainty in data & $\pm0.10$ & $\pm0.10$
\end{tabular}
\label{tab:impacts_table}
\end{table}

\section{\texorpdfstring{Combined \hinv limits}%
                               {Combined PH to inv limits}}
\label{sec:Combination}

A variety of production modes of the Higgs boson can be used for searches for \hinv decays. A combination of the results of this analysis, analyses covering the years 2016--2018, and earlier published CMS combination results using Run~1 (years 2011--2012) and~2015 data~\cite{CMS:2016dhk} at $\sqrt{s}=7,~8,~$and 13\TeV, detailed in Table~\ref{tab:final_state_breakdown}, is performed by means of a combined likelihood fit in which systematic uncertainties are correlated across search regions where appropriate. Unless explicitly specified below, parameters of the individual likelihood functions are treated as independent.

\begin{table*}[htb]
\centering
\topcaption{Data sets and their respective integrated luminosities used for each production mode across Run~1 and Run~2. For some data-taking periods, no \hinv search have been performed for the given production mode, and are not included in the combination. }
\begin{tabular}{lllll}
Analysis tag & Production mode & \multicolumn{3}{c}{Integrated luminosity (fb$^{-1}$)} \\
 & & 7\TeV & 8\TeV & 13\TeV (Run~2) \\\hline
VBF-tagged & VBF & \NA & 19.2~\cite{CMS:2014gab} & 140~\cite{CMS:2016dhk}\cite{VBF-Run2-paper} \\ [\cmsTabSkip]

\multirow{4}{*}{\VH-tagged} & $\PZ(\Pell\Pell)\PH$ & 4.9~\cite{CMS:2014gab} & 19.7~\cite{CMS:2014gab} & 140~\cite{CMS:2016dhk}\cite{CMS:2020ulv} \\
 & ${\PZ(\bbbar)\PH}$ & \NA & 18.9~\cite{CMS:2014gab} & \NA \\
 & ${\PV(\text{jj})\PH}$ & \NA & 19.7~\cite{CMS:2016xus} & 140~\cite{CMS:2016dhk}[this paper] \\
 & Boosted \VH & \NA & \NA & 138~\cite{CMS_MonojetV} \\ [\cmsTabSkip]

\multirow{2}{*}{\ttH-tagged} & \ttH(hadronic) & \NA & \NA & 138 [this paper] \\
& \ttH(leptonic) & \NA & \NA & 138~\cite{SUS-19-009_PAPER,SUS-19-011_PAPER} \\ [\cmsTabSkip]

\ggH-tagged & \ggH & \NA & 19.7~\cite{CMS:2016xus} & 140~\cite{CMS:2016dhk}\cite{CMS_MonojetV} \\
\end{tabular}
\label{tab:final_state_breakdown}
\end{table*}

For the \ttH analysis with fully leptonic final states, a reinterpretation of the supersymmetry searches in the semileptonic and dileptonic \ttbar decay channels in Ref.~\cite{SUS-19-009_PAPER,SUS-19-011_PAPER} in the context of the \ttbar~$+$~DM model studied in Ref.~\cite{SUS-20-002_PAPER} has been performed. Another leptonic channel included in this combination is from the ${\PZ(\Pell\Pell)\PH}$ analysis~\cite{CMS:2020ulv} using 2016--2018 data.

Analyses with hadronic final states partially overlap in their phase space selection, and this must be accounted for in the statistical combination. Those affected by overlap are the VBF analysis~\cite{VBF-Run2-paper}, the analysis targetting hadronic \ggH and boosted \VH final states~\cite{CMS_MonojetV}, and the resolved \VH channel described in this paper.

To remove the overlap between the VBF analysis and \ggH/boosted \VH analysis, events are considered for rejection in the \ggH/boosted \VH analysis if they have at least two AK4 jets each with $\abs{\eta} < 4.7$. Specifically, an inversion of the VBF kinematic selection is applied similarly to the \ttH and resolved \VH analysis as described in Section~\ref{subsec:Baseline_selection}. These requirements mirror the selection used to enhance the characteristic VBF phase space in Ref.~\cite{VBF-Run2-paper}, with negligible effect on the sensitivity of the \ggH/boosted \VH analysis to \bhinvlim.

The overlap between the \ggH/boosted \VH analysis and the \VH 2j0b category of this analysis is driven by the low-purity \VH category of the boosted analysis. By removing events from the low-purity boosted \VH category that contain exactly two AK4 jets forming a dijet candidate with ${65<\mjj<120\GeV}$, there is negligible reduction in the exclusion sensitivity of that analysis. The overlap meanwhile is reduced from 30-40\% in the CR phase spaces to about 1\%.

The uncertainties in the overall cross section for the signal processes are treated as correlated amongst analysis channels, and amongst data sets with the same centre-of-mass energy. The uncertainties related to missing higher-order corrections, as well as PDF variations, are obtained from Ref.~\cite{deFlorian:2016spz}. In some of the channels, additional uncertainty contributions relating to signal acceptance modelling are considered. These are treated as uncorrelated amongst the different analysis channels.

The main sources of theoretical modelling uncertainties in the background estimate vary for the different analysis channels. The analyses preferentially select different phase space regions, and employ different assumptions for the modelling of theoretical uncertainties in transfer factors amongst different analysis regions. The resulting uncertainties are therefore treated as uncorrelated.

Significant correlations appear in the treatment of experimental uncertainties. The determination of the integrated luminosity estimate is affected by a number of sources of uncertainty, which are assumed to be correlated amongst all channels, and partially correlated amongst data sets. Some of the analysis channels share trigger requirements, and the uncertainties in the efficiencies of these common triggers are assumed to be correlated amongst channels and uncorrelated amongst data sets. Furthermore, analysis channels often share criteria used for identifying \PQb-tagged jets, as well as the hadronic decay products of tau leptons. The uncertainties in the efficiencies of these identification criteria are assumed to be correlated amongst channels using the same criteria in the same data set. Finally, uncertainties in the calibration of the JER and JES are treated as correlated amongst this analysis, the VBF, and the \ggH/boosted \VH channels. All other experimental uncertainties are assumed to be uncorrelated amongst channels. For earlier analyses using Run~1 and 2015 data, the correlation scheme established in Ref.~\cite{CMS:2016dhk} is used.

\begin{figure}[hbtp!]
    \centering
    \includegraphics[width=0.48\textwidth]{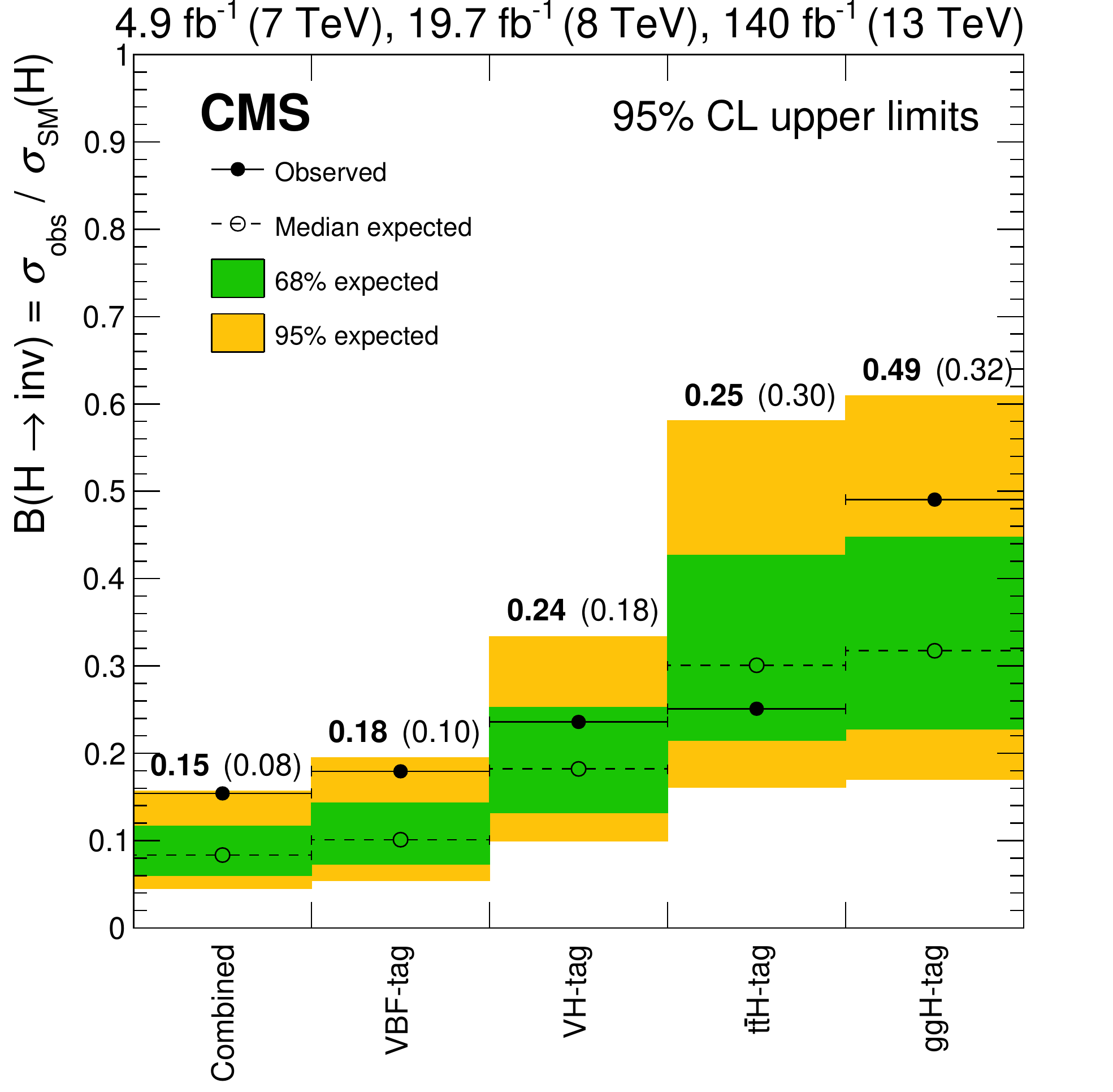}
    \includegraphics[width=0.48\textwidth]{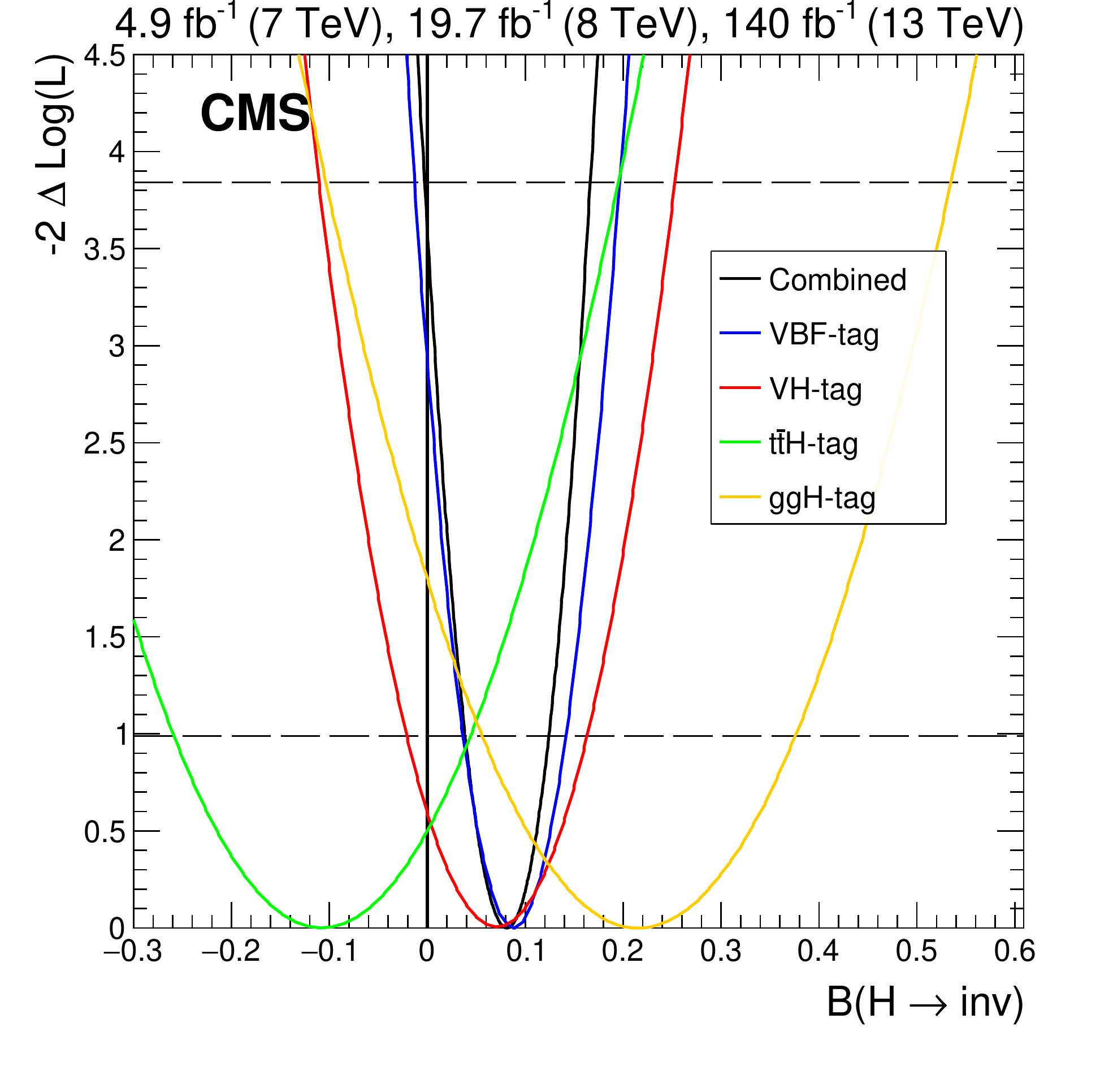}
    \caption{Left: Exclusion limits at 95\% \CL on \bhinvlim. The results are shown separately for each Higgs boson production mode as tagged by the input analyses for Run~1 and Run~2, as well as combined across modes. Right: Scan of the profile negative log-likelihood as a function of \bhinvlim broken down by the Higgs boson production mode as tagged by the input analyses for Run~1 and Run~2.
    }
    \label{fig:limsandllscan}
\end{figure}

Exclusion limits on \bhinvlim are calculated assuming SM production cross sections. The 2016--2018 data yields an overall limit of \binvRunTwoObs (\binvRunTwoExp expected). If the Run~1 and 2015 data-taking periods are included, values larger than \binvAlltimeObs (\binvAlltimeExp expected) are excluded at 95\% \CL. This value is dominated by the VBF channel, which yields a limit for \bhinvlim of 0.18 (0.10 expected). The limits for Run~1 and Run~2, separated by the Higgs boson production mode as tagged by the input analyses, are presented in Fig.~\ref{fig:limsandllscan}. The integrated luminosities of the Run~1 and Run~2 data sets~\cite{VBF-Run2-paper,CMS_MonojetV,CMS:2020ulv,SUS-20-002_PAPER,CMS:2016dhk} are described in Table~\ref{tab:final_state_breakdown}. The final combination represents an improvement in sensitivity of approximately 20\% relative to the most sensitive single channel (VBF).

Maximum likelihood fits to the individual production channels are performed, as well as to the combination of all channels. The dependence of the profile negative log-likelihood functions on the signal strength parameter $\hat{\mu}$ is shown in Fig.~\ref{fig:limsandllscan}~(right). The best-fit values of $\hat{\mu}$ for the individual production channels are compatible with one another and with the combined value of ${0.08^{+0.04}_{-0.04}}$, and the observed signal strength is compatible with the absence of a \hinv signal within two standard deviations. A breakdown of the best-fit values of $\hat{\mu}$ for each channel are presented in Table~\ref{llscan_vals_comb}. A saturated goodness-of-fit test is performed using the final combined likelihood function~\cite{Baker:1983tu}, yielding a probability of \combinedGofPValue that the S$+$B model is consistent with the observed results from the CMS experiment. Tabulated yields and fit results are provided in HEPData~\cite{HEPData}.

\begin{table}[htbp]
\topcaption{The observed best-fit estimates of \bhinvlim, for each analysis channel in the combination, and the 95\% \CL observed and expected (exp) upper limits on \bhinvlim.}
\centering
\begin{tabular}{ccc}
Channel & Best-fit \bhinvlim & \bhinvlim \\\hline
Combined & $0.08^{+0.04}_{-0.04}$ & 0.15 (0.08 exp) \\
VBF-tag & $0.09^{+0.05}_{-0.05}$ & 0.18 (0.10 exp) \\
\VH-tag & $0.07^{+0.09}_{-0.09}$ & 0.24 (0.18 exp) \\
\ttH-tag & $-0.11^{+0.15}_{-0.15}$ & 0.25 (0.30 exp) \\
\ggH-tag & $0.22^{+0.16}_{-0.16}$ & 0.49 (0.32 exp)
\end{tabular}
\label{llscan_vals_comb}
\end{table}

\begin{figure*}[hbt!]
    \centering
    \includegraphics[width=\cmsFigWidviii]{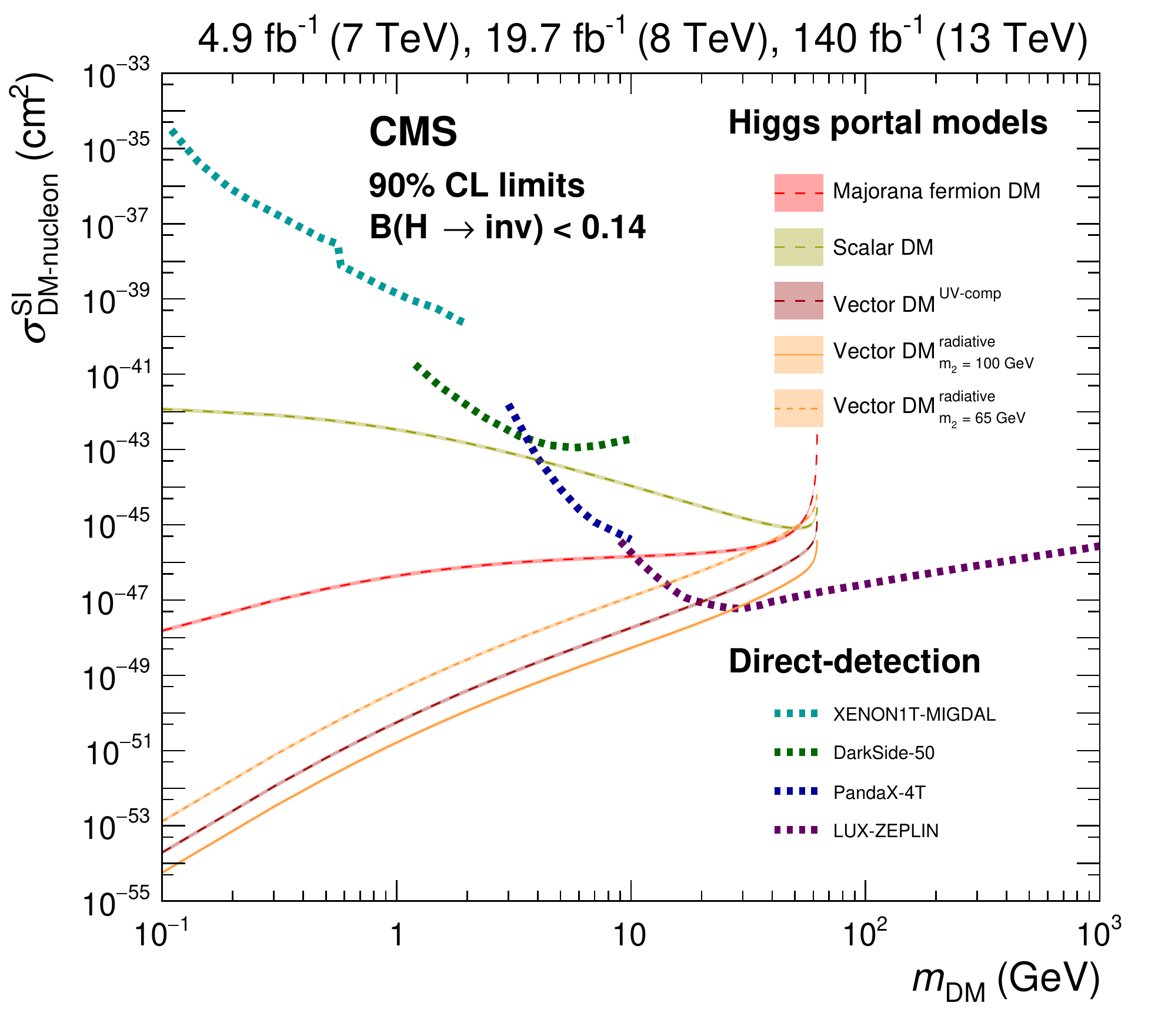}
    \caption{Upper limits on \sigmadmnuc as a function of DM candidate mass $m_{\text{DM}}$. Results are presented for a fermion (red) and scalar (yellow) DM candidate. In addition, a vector DM candidate is studied using two UV-complete approaches, the first denoted Vector DM$^{\text{UV-comp}}$~\cite{Baek:2012se} (burgundy), and the second a radiative portal version denoted Vector DM$^{\text{radiative}}_{m_{2}}$~\cite{DiFranzo:2015nli} (orange) with a dark Higgs boson mass of $m_2 = 65$ and $100\GeV$. Uncertainties are derived from Refs.~\cite{Young:2009zb,Toussaint:2009pz,Djouadi:2011aa}. Results are compared to direct-detection searches from XENON1T-Migdal~\cite{XENON:2019zpr}, DarkSide-50~\cite{DarkSide-50:2022qzh}, PandaX-4T~\cite{Wenbo:2022coa}, and LUX-ZEPLIN~\cite{LZ:2022ufs}.}
    \label{fig:HiggsPortal}
\end{figure*}

The upper limit on \bhinvlim is interpreted in the context of a set of Higgs portal models of DM interactions, where a stable weakly interacting massive particle (WIMP), such as a singlet scalar, fermion, or vector, has a substantial coupling to a Higgs boson of mass 125\GeV~\cite{Djouadi:2011aa,Baek:2012se}. The interaction of a WIMP with an atomic nucleus can occur via the exchange of a Higgs boson, and the resulting nuclear recoil is measured to obtain an upper bound on the spin-independent DM-nucleon scattering cross section, \sigmadmnuc. An effective field theory (EFT) approach is considered for scalar and fermionic WIMPs, while in the vectorial case two UV-complete DM models are considered, given the EFT appraoch violates unitarity~\cite{DiFranzo:2015nli,Zaazoua_2022}. The vector-spin WIMP model (Vector DM$^{\text{UV-comp}}$) described in Ref.~\cite{Baek:2012se}, and its radiative portal analogue (Vector DM$^{\text{radiative}}_{m_{2}}$) introduced in Ref.~\cite{DiFranzo:2015nli} for dark Higgs boson masses $m_2 = 65$ and $100\GeV$, and with a mixing angle between the SM and dark Higgs bosons $\theta = 0.2$, are presented. The results are compared to direct-detection searches, where in these experiments it is assumed DM particles interact with atomic nuclei. Direct-detection limits are reported by the XENON1T-Migdal \cite{XENON:2019zpr}, DarkSide-50 \cite{DarkSide-50:2022qzh}, Panda-X~4T \cite{Wenbo:2022coa}, and LUX-ZEPLIN \cite{LZ:2022ufs} experiments. Upper limits on \sigmadmnuc for DM masses ranging from $0.1\GeV$ to $m_{\PH}/2$ are presented in Fig.~\ref{fig:HiggsPortal} at the 90\% \CL using the full CMS data set. The uncertainties in \sigmadmnuc are obtained from the extrema of a coupling parametrisation factor as derived from lattice theory~\cite{Young:2009zb,Toussaint:2009pz,Djouadi:2011aa}. Results of the Higgs portal interpretation and direct-detection comparison are also provided in HEPData~\cite{HEPData}.

\begin{figure}[htb!]
    \centering
    \includegraphics[width=\cmsFigWidix]{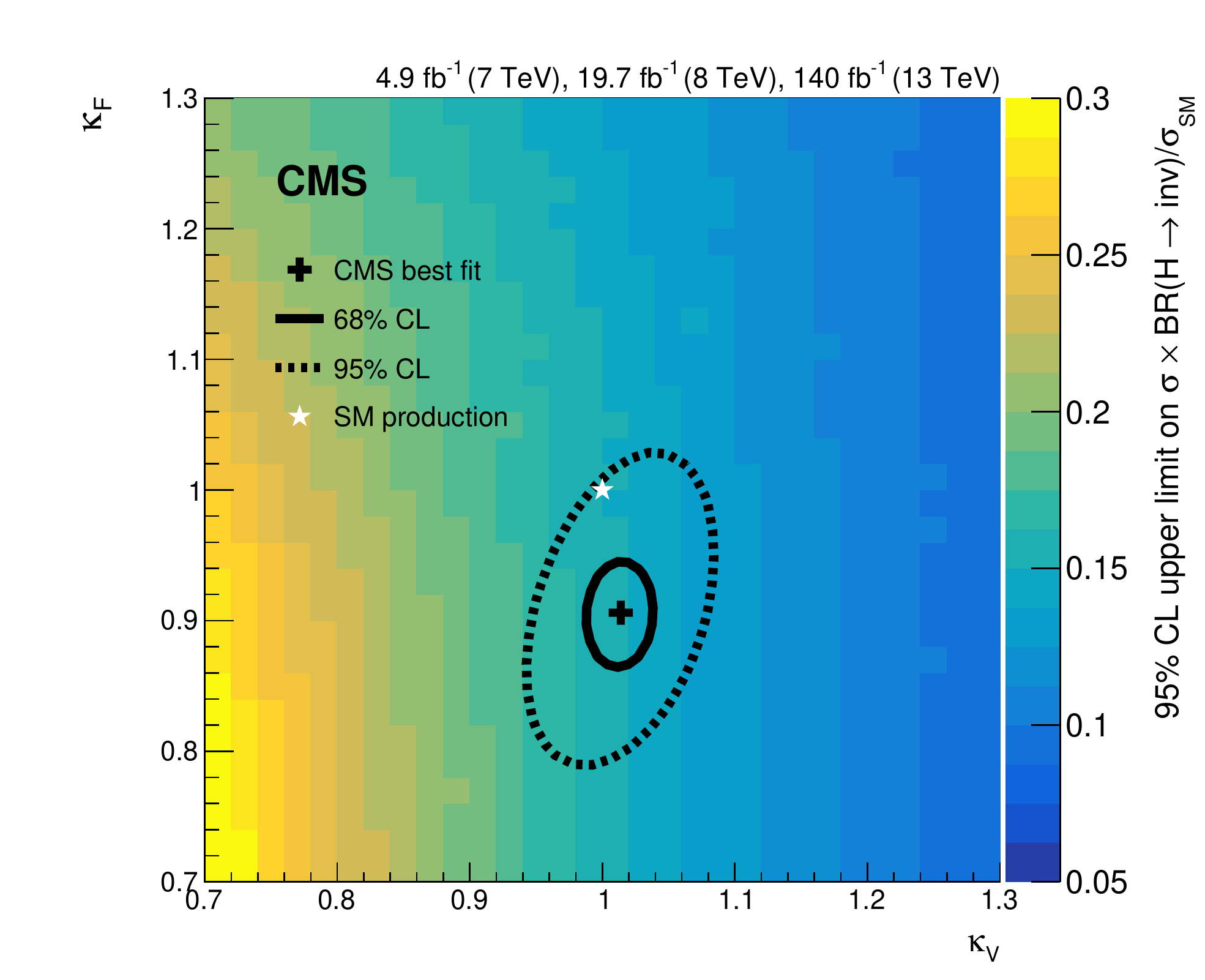}
    \caption{Observed 95\%~\CL upper limit on \bhinvlim as a function of coupling strength modifiers, $\kappa_{\text{V}}$ and $\kappa_{\text{F}}$, for a Higgs boson of mass 125\GeV. Best estimates for $\kappa_{\text{V}}$ and $\kappa_{\text{F}}$ from Ref.~\cite{CMS:2022dwd} are shown as a black cross, together with 68 and 95\%~\CL contours.}
    \label{fig:KappaContour}
\end{figure}

The sensitivity of the Run~1 and Run~2 combination depends on the cross sections assumed for the different Higgs boson production modes: VBF, \VH, \ggH, and \ttH. Cross sections can be parameterised by the coupling strength of the Higgs boson to \PV bosons and fermions. The cross sections can be directly scaled by coupling strength modifiers $\kappa_{\text{V}}$ and $\kappa_{\text{F}}$ to investigate BSM scenarios \cite{kappaV-kappaF}. In this context, the observed 95\% \CL upper limits on \bhinvlim are evaluated as a function of $\kappa_{\text{V}}$ and $\kappa_{\text{F}}$ and shown in Fig.~\ref{fig:KappaContour}. Best estimates of $\kappa_{\text{V}}$ and $\kappa_{\text{F}}$ from CMS~\cite{CMS:2022dwd} are presented with the 68 and 95\%~\CL contours. For the best estimate of $\kappa_{\text{V}}$ and $\kappa_{\text{F}}$ by CMS, the 95\%~\CL limit on \bhinvlim is found to be 0.15 and varies between 0.13 and 0.17 inside the 95\%~\CL contour.

\section{Summary}
\label{sec:Conclusion}

The results of a search for invisible decays of the Higgs boson produced in association with a top-antitop quark pair (\ttH) or a vector boson (\VH, where \PV stands for either a \PW or \PZ boson), which decays to a fully hadronic final state, are presented. The analysis is based on proton-proton collision data collected at $\sqrt{s}=13\TeV$ during the 2016--2018 data-taking period by the CMS experiment at the LHC, corresponding to an integrated luminosity of 138\fbinv. The \ttH production mechanism is investigated using final states containing \PQb jets, or boosted \PQt quarks or \PW bosons. The \VH production channel focuses on resolving a dijet pair with an invariant mass that is compatible with that of a \PW or \PZ boson. No significant excess of events is observed above the predicted SM background. A 95\% confidence level upper limit of 0.54 (0.39 expected) is set on the branching fraction of the decay of the Higgs boson to an invisible final state, \bhinvlim, assuming SM production cross sections.

The results are combined with previous \bhinvlim searches carried out at $\sqrt{s}=7$,~8, and 13\TeV in complementary production modes. The combined 95\% confidence level upper limit on \bhinvlim of 0.15 (0.08 expected) is obtained using Run~1 (2011--2012) and Run~2 (2015--2018) data. The combination represents an improvement in sensitivity of 20\% relative to the most sensitive single channel. The results are interpreted in the context of a set of Higgs portal models of dark matter interactions for dark matter masses in the range $0.1\GeV$ and “$m_{\PH}/2$”. Model-dependent exclusion limits are found to complement direct-detection experiments for light mass dark matter candidates.

\begin{acknowledgments}
    We congratulate our colleagues in the CERN accelerator departments for the excellent performance of the LHC and thank the technical and administrative staffs at CERN and at other CMS institutes for their contributions to the success of the CMS effort. In addition, we gratefully acknowledge the computing centres and personnel of the Worldwide LHC Computing Grid and other centres for delivering so effectively the computing infrastructure essential to our analyses. Finally, we acknowledge the enduring support for the construction and operation of the LHC, the CMS detector, and the supporting computing infrastructure provided by the following funding agencies: BMBWF and FWF (Austria); FNRS and FWO (Belgium); CNPq, CAPES, FAPERJ, FAPERGS, and FAPESP (Brazil); MES and BNSF (Bulgaria); CERN; CAS, MoST, and NSFC (China); MINCIENCIAS (Colombia); MSES and CSF (Croatia); RIF (Cyprus); SENESCYT (Ecuador); MoER, ERC PUT and ERDF (Estonia); Academy of Finland, MEC, and HIP (Finland); CEA and CNRS/IN2P3 (France); BMBF, DFG, and HGF (Germany); GSRI (Greece); NKFIH (Hungary); DAE and DST (India); IPM (Iran); SFI (Ireland); INFN (Italy); MSIP and NRF (Republic of Korea); MES (Latvia); LAS (Lithuania); MOE and UM (Malaysia); BUAP, CINVESTAV, CONACYT, LNS, SEP, and UASLP-FAI (Mexico); MOS (Montenegro); MBIE (New Zealand); PAEC (Pakistan); MES and NSC (Poland); FCT (Portugal); MESTD (Serbia); MCIN/AEI and PCTI (Spain); MOSTR (Sri Lanka); Swiss Funding Agencies (Switzerland); MST (Taipei); MHESI and NSTDA (Thailand); TUBITAK and TENMAK (Turkey); NASU (Ukraine); STFC (United Kingdom); DOE and NSF (USA).
    
    \hyphenation{Rachada-pisek} Individuals have received support from the Marie-Curie programme and the European Research Council and Horizon 2020 Grant, contract Nos.\ 675440, 724704, 752730, 758316, 765710, 824093, 884104, and COST Action CA16108 (European Union); the Leventis Foundation; the Alfred P.\ Sloan Foundation; the Alexander von Humboldt Foundation; the Belgian Federal Science Policy Office; the Fonds pour la Formation \`a la Recherche dans l'Industrie et dans l'Agriculture (FRIA-Belgium); the Agentschap voor Innovatie door Wetenschap en Technologie (IWT-Belgium); the F.R.S.-FNRS and FWO (Belgium) under the ``Excellence of Science -- EOS" -- be.h project n.\ 30820817; the Beijing Municipal Science \& Technology Commission, No. Z191100007219010; the Ministry of Education, Youth and Sports (MEYS) of the Czech Republic; the Hellenic Foundation for Research and Innovation (HFRI), Project Number 2288 (Greece); the Deutsche Forschungsgemeinschaft (DFG), under Germany's Excellence Strategy -- EXC 2121 ``Quantum Universe" -- 390833306, and under project number 400140256 - GRK2497; the Hungarian Academy of Sciences, the New National Excellence Program - \'UNKP, the NKFIH research grants K 124845, K 124850, K 128713, K 128786, K 129058, K 131991, K 133046, K 138136, K 143460, K 143477, 2020-2.2.1-ED-2021-00181, and TKP2021-NKTA-64 (Hungary); the Council of Science and Industrial Research, India; the Latvian Council of Science; the Ministry of Education and Science, project no. 2022/WK/14, and the National Science Center, contracts Opus 2021/41/B/ST2/01369 and 2021/43/B/ST2/01552 (Poland); the Funda\c{c}\~ao para a Ci\^encia e a Tecnologia, grant CEECIND/01334/2018 (Portugal); the National Priorities Research Program by Qatar National Research Fund; MCIN/AEI/10.13039/501100011033, ERDF ``a way of making Europe", and the Programa Estatal de Fomento de la Investigaci{\'o}n Cient{\'i}fica y T{\'e}cnica de Excelencia Mar\'{\i}a de Maeztu, grant MDM-2017-0765 and Programa Severo Ochoa del Principado de Asturias (Spain); the Chulalongkorn Academic into Its 2nd Century Project Advancement Project, and the National Science, Research and Innovation Fund via the Program Management Unit for Human Resources \& Institutional Development, Research and Innovation, grant B05F650021 (Thailand); the Kavli Foundation; the Nvidia Corporation; the SuperMicro Corporation; the Welch Foundation, contract C-1845; and the Weston Havens Foundation (USA).   
\end{acknowledgments}
\bibliography{auto_generated}

\providecommand{\href}[2]{#2}\begingroup\raggedright\begin{thebibliography}{100}%
\makeatletter
\providecommand{\hrefCMSnoop }[0]{\@secondoftwo}%
\makeatother
\providecommand{\doi}{\texttt{doi:}\begingroup \urlstyle{tt}\Url}

\bibitem{PhysRevLett.13.321}
\hrefCMSnoop {}{F.~Englert and R.~Brout, ``{B}roken symmetry and the mass of
  gauge vector mesons'',} \textit{ Phys. Rev. Lett.} \textbf{ 13} (1964) 321,
  \href{http://dx.doi.org/10.1103/PhysRevLett.13.321}{\doi{10.1103/PhysRevLett.13.321}}.

\bibitem{Higgs:1964ia}
\hrefCMSnoop {}{P.~W. Higgs, ``{B}roken symmetries, massless particles and
  gauge fields'',} \textit{ Phys. Lett.} \textbf{ 12} (1964) 132,
  \href{http://dx.doi.org/10.1016/0031-9163(64)91136-9}{\doi{10.1016/0031-9163(64)91136-9}}.

\bibitem{PhysRevLett.13.508}
\hrefCMSnoop {}{P.~W. Higgs, ``{B}roken symmetries and the masses of gauge
  bosons'',} \textit{ Phys. Rev. Lett.} \textbf{ 13} (1964) 508,
  \href{http://dx.doi.org/10.1103/PhysRevLett.13.508}{\doi{10.1103/PhysRevLett.13.508}}.

\bibitem{PhysRevLett.13.585}
\hrefCMSnoop {}{G.~S. Guralnik, C.~R. Hagen, and T.~W.~B. Kibble, ``{G}lobal
  conservation laws and massless particles'',} \textit{ Phys. Rev. Lett.}
  \textbf{ 13} (1964) 585,
  \href{http://dx.doi.org/10.1103/PhysRevLett.13.585}{\doi{10.1103/PhysRevLett.13.585}}.

\bibitem{Higgs:1966ev}
\hrefCMSnoop {}{P.~W. Higgs, ``{S}pontaneous symmetry breakdown without
  massless bosons'',} \textit{ Phys. Rev.} \textbf{ 145} (1966) 1156,
  \href{http://dx.doi.org/10.1103/PhysRev.145.1156}{\doi{10.1103/PhysRev.145.1156}}.

\bibitem{Kibble:1967sv}
\hrefCMSnoop {}{T.~W.~B. Kibble, ``{S}ymmetry breaking in non-abelian gauge
  theories'',} \textit{ Phys. Rev.} \textbf{ 155} (1967) 1554,
  \href{http://dx.doi.org/10.1103/PhysRev.155.1554}{\doi{10.1103/PhysRev.155.1554}}.

\bibitem{ATLAS:2012yve}
\hrefCMSnoop {}{{ATLAS Collaboration}, ``{O}bservation of a new particle in the
  search for the standard model {H}iggs boson with the {ATLAS} detector at the
  {LHC}'',} \textit{ Phys. Lett. B} \textbf{ 716} (2012) 1,
  \href{http://dx.doi.org/10.1016/j.physletb.2012.08.020}{\doi{10.1016/j.physletb.2012.08.020}},
  \href{http://www.arXiv.org/abs/1207.7214}{\texttt{arXiv:1207.7214}}.

\bibitem{CMS:2012qbp}
\hrefCMSnoop {}{{CMS Collaboration}, ``{O}bservation of a new boson at a mass
  of 125 {GeV} with the {CMS} experiment at the {LHC}'',} \textit{ Phys. Lett.
  B} \textbf{ 716} (2012) 30,
  \href{http://dx.doi.org/10.1016/j.physletb.2012.08.021}{\doi{10.1016/j.physletb.2012.08.021}},
  \href{http://www.arXiv.org/abs/1207.7235}{\texttt{arXiv:1207.7235}}.

\bibitem{CMS:2013btf}
\hrefCMSnoop {}{{CMS Collaboration}, ``{O}bservation of a new boson with mass
  near 125 {GeV} in pp collisions at $\sqrt{s}$ = 7 and 8 {TeV}'',} \textit{
  JHEP} \textbf{ 06} (2013) 081,
  \href{http://dx.doi.org/10.1007/JHEP06(2013)081}{\doi{10.1007/JHEP06(2013)081}},
  \href{http://www.arXiv.org/abs/1303.4571}{\texttt{arXiv:1303.4571}}.

\bibitem{ATLAS:2022vkf}
\hrefCMSnoop {}{{ATLAS Collaboration}, ``{A} detailed map of {H}iggs boson
  interactions by the {ATLAS} experiment ten years after the discovery'',}
  \textit{ Nature} \textbf{ 607} (2022) 52,
  \href{http://dx.doi.org/10.1038/s41586-022-04893-w}{\doi{10.1038/s41586-022-04893-w}},
  \href{http://www.arXiv.org/abs/2207.00092}{\texttt{arXiv:2207.00092}}.

\bibitem{CMS:2022dwd}
\hrefCMSnoop {}{{CMS Collaboration}, ``{A} portrait of the {H}iggs boson by the
  {CMS} experiment ten years after the discovery'',} \textit{ Nature} \textbf{
  607} (2022) 60,
  \href{http://dx.doi.org/10.1038/s41586-022-04892-x}{\doi{10.1038/s41586-022-04892-x}},
  \href{http://www.arXiv.org/abs/2207.00043}{\texttt{arXiv:2207.00043}}.

\bibitem{ParticleDataGroup:2022pth}
\hrefCMSnoop {}{{Particle Data Group}, ``{R}eview of particle physics'',}
  \textit{ PTEP} \textbf{ 2022} (2022) 083C01,
  \href{http://dx.doi.org/10.1093/ptep/ptac097}{\doi{10.1093/ptep/ptac097}}.

\bibitem{SHROCK1982250}
\hrefCMSnoop {}{R.~E. Shrock and M.~Suzuki, ``{I}nvisible decays of {H}iggs
  bosons'',} \textit{ Phys. Lett. B} \textbf{ 110} (1982) 250,
\href{http://dx.doi.org/10.1016/0370-2693(82)91247-3}{\doi{10.1016/0370-2693(82)91247-3}}.

\bibitem{Belanger:2001am}
G.~B{\'e}langer\hrefCMSnoop {}{ {et~al.}, ``{T}he {MSSM} invisible {H}iggs in
  the light of dark matter and g-2'',} \textit{ Phys. Lett. B} \textbf{ 519}
  (2001) 93,
  \href{http://dx.doi.org/10.1016/S0370-2693(01)00976-5}{\doi{10.1016/S0370-2693(01)00976-5}},
\href{http://www.arXiv.org/abs/hep-ph/0106275}{\texttt{arXiv:hep-ph/0106275}}.

\bibitem{Datta:2004jg}
\hrefCMSnoop {}{A.~Datta, K.~Huitu, J.~Laamanen, and B.~Mukhopadhyaya,
  ``{L}inear collider signals of an invisible {H}iggs boson in theories of
  large extra dimensions'',} \textit{ Phys. Rev. D} \textbf{ 70} (2004) 075003,
  \href{http://dx.doi.org/10.1103/PhysRevD.70.075003}{\doi{10.1103/PhysRevD.70.075003}},
\href{http://www.arXiv.org/abs/hep-ph/0404056}{\texttt{arXiv:hep-ph/0404056}}.

\bibitem{Dominici:2009pq}
\hrefCMSnoop {}{D.~Dominici and J.~F. Gunion, ``{I}nvisible {H}iggs decays from
  {H}iggs-graviscalar mixing'',} \textit{ Phys. Rev. D} \textbf{ 80} (2009)
  115006,
  \href{http://dx.doi.org/10.1103/PhysRevD.80.115006}{\doi{10.1103/PhysRevD.80.115006}},
\href{http://www.arXiv.org/abs/0902.1512}{\texttt{arXiv:0902.1512}}.

\bibitem{Argyropoulos:2021sav}
\hrefCMSnoop {}{S.~Argyropoulos, O.~Brandt, and U.~Haisch, ``{C}ollider
  searches for dark matter through the {H}iggs lens'',} \textit{ Symmetry}
  \textbf{ 13} (2021), no.~12, 2406,
  \href{http://dx.doi.org/10.3390/sym13122406}{\doi{10.3390/sym13122406}},
  \href{http://www.arXiv.org/abs/2109.13597}{\texttt{arXiv:2109.13597}}.

\bibitem{Kanemura:2010sh}
\hrefCMSnoop {}{S.~Kanemura, S.~Matsumoto, T.~Nabeshima, and N.~Okada, ``{C}an
  {WIMP} dark matter overcome the nightmare scenario?'',} \textit{ Phys. Rev.
  D} \textbf{ 82} (2010) 055026,
  \href{http://dx.doi.org/10.1103/PhysRevD.82.055026}{\doi{10.1103/PhysRevD.82.055026}},
  \href{http://www.arXiv.org/abs/1005.5651}{\texttt{arXiv:1005.5651}}.

\bibitem{Djouadi:2011aa}
\hrefCMSnoop {}{A.~Djouadi, O.~Lebedev, Y.~Mambrini, and J.~Quevillon,
  ``{I}mplications of {LHC} searches for {H}iggs--portal dark matter'',}
  \textit{ Phys. Lett. B} \textbf{ 709} (2012) 65,
  \href{http://dx.doi.org/10.1016/j.physletb.2012.01.062}{\doi{10.1016/j.physletb.2012.01.062}},
\href{http://www.arXiv.org/abs/1112.3299}{\texttt{arXiv:1112.3299}}.

\bibitem{Baek:2012se}
\hrefCMSnoop {}{S.~Baek, P.~Ko, W.-I. Park, and E.~Senaha, ``{H}iggs portal
  vector dark matter: revisited'',} \textit{ JHEP} \textbf{ 05} (2013) 036,
  \href{http://dx.doi.org/10.1007/JHEP05(2013)036}{\doi{10.1007/JHEP05(2013)036}},
\href{http://www.arXiv.org/abs/1212.2131}{\texttt{arXiv:1212.2131}}.

\bibitem{Djouadi:2012zc}
\hrefCMSnoop {}{A.~Djouadi, A.~Falkowski, Y.~Mambrini, and J.~Quevillon,
  ``{D}irect detection of {H}iggs--portal dark matter at the {LHC}'',} \textit{
  Eur. Phys. J. C} \textbf{ 73} (2013) 2455,
  \href{http://dx.doi.org/10.1140/epjc/s10052-013-2455-1}{\doi{10.1140/epjc/s10052-013-2455-1}},
\href{http://www.arXiv.org/abs/1205.3169}{\texttt{arXiv:1205.3169}}.

\bibitem{Beniwal:2015sdl}
A.~Beniwal\hrefCMSnoop {}{ {et~al.}, ``{C}ombined analysis of effective
  {H}iggs--portal dark matter models'',} \textit{ Phys. Rev. D} \textbf{ 93}
  (2016) 115016,
  \href{http://dx.doi.org/10.1103/PhysRevD.93.115016}{\doi{10.1103/PhysRevD.93.115016}},
\href{http://www.arXiv.org/abs/1512.06458}{\texttt{arXiv:1512.06458}}.

\bibitem{DiFranzo:2015nli}
\hrefCMSnoop {}{A.~DiFranzo, P.~J. Fox, and T.~M.~P. Tait, ``{V}ector dark
  matter through a radiative {H}iggs portal'',} \textit{ JHEP} \textbf{ 04}
  (2016) 135,
  \href{http://dx.doi.org/10.1007/JHEP04(2016)135}{\doi{10.1007/JHEP04(2016)135}},
  \href{http://www.arXiv.org/abs/1512.06853}{\texttt{arXiv:1512.06853}}.

\bibitem{ATLAS:2017nyv}
\hrefCMSnoop {}{{ATLAS Collaboration}, ``{S}earch for an invisibly decaying
  {H}iggs boson or dark matter candidates produced in association with a {$Z$}
  boson in $pp$ collisions at $\sqrt{s}$ = 13 {TeV} with the {ATLAS}
  detector'',} \textit{ Phys. Lett. B} \textbf{ 776} (2018) 318,
  \href{http://dx.doi.org/10.1016/j.physletb.2017.11.049}{\doi{10.1016/j.physletb.2017.11.049}},
  \href{http://www.arXiv.org/abs/1708.09624}{\texttt{arXiv:1708.09624}}.

\bibitem{ATLAS:2019cid}
\hrefCMSnoop {}{{ATLAS Collaboration}, ``{C}ombination of searches for
  invisible {H}iggs boson decays with the {ATLAS} experiment'',} \textit{ Phys.
  Rev. Lett.} \textbf{ 122} (2019) 231801,
  \href{http://dx.doi.org/10.1103/PhysRevLett.122.231801}{\doi{10.1103/PhysRevLett.122.231801}},
  \href{http://www.arXiv.org/abs/1904.05105}{\texttt{arXiv:1904.05105}}.

\bibitem{ATLAS:2021kxv}
\hrefCMSnoop {}{{ATLAS Collaboration}, ``{S}earch for new phenomena in events
  with an energetic jet and missing transverse momentum in pp collisions at
  $\sqrt {s}$ = 13 {TeV} with the {ATLAS} detector'',} \textit{ Phys. Rev. D}
  \textbf{ 103} (2021) 112006,
  \href{http://dx.doi.org/10.1103/PhysRevD.103.112006}{\doi{10.1103/PhysRevD.103.112006}},
  \href{http://www.arXiv.org/abs/2102.10874}{\texttt{arXiv:2102.10874}}.

\bibitem{ATLAS:2021gcn}
\hrefCMSnoop {}{{ATLAS Collaboration}, ``{S}earch for associated production of
  a z boson with an invisibly decaying {H}iggs boson or dark matter candidates
  at $\sqrt {s}$ = 13 {TeV} with the {ATLAS} detector'',} \textit{ Phys. Lett.
  B} \textbf{ 829} (2022) 137066,
  \href{http://dx.doi.org/10.1016/j.physletb.2022.137066}{\doi{10.1016/j.physletb.2022.137066}},
  \href{http://www.arXiv.org/abs/2111.08372}{\texttt{arXiv:2111.08372}}.

\bibitem{ATLAS-VBF-Run2}
\hrefCMSnoop {}{{ATLAS Collaboration}, ``{S}earch for invisible {H}iggs-boson
  decays in events with vector-boson fusion signatures using 139 fb$^{-1}$ of
  proton-proton data recorded by the {ATLAS} experiment'',} \textit{ JHEP}
  \textbf{ 08} (2022) 104,
  \href{http://dx.doi.org/10.1007/JHEP08(2022)104}{\doi{10.1007/JHEP08(2022)104}},
  \href{http://www.arXiv.org/abs/2202.07953}{\texttt{arXiv:2202.07953}}.

\bibitem{ATLAS:2022ygn}
\hrefCMSnoop {}{{ATLAS Collaboration}, ``{C}onstraints on spin-0 dark matter
  mediators and invisible {H}iggs decays using {ATLAS} 13 {TeV} pp collision
  data with two top quarks and missing transverse momentum in the final
  state'',}
  \href{http://www.arXiv.org/abs/2211.05426}{\texttt{arXiv:2211.05426}}.

\bibitem{CMS:2016dhk}
\hrefCMSnoop {}{{CMS Collaboration}, ``{S}earches for invisible decays of the
  {H}iggs boson in pp collisions at $\sqrt{s}$ = 7, 8, and 13 {TeV}'',}
  \textit{ JHEP} \textbf{ 02} (2017) 135,
  \href{http://dx.doi.org/10.1007/JHEP02(2017)135}{\doi{10.1007/JHEP02(2017)135}},
  \href{http://www.arXiv.org/abs/1610.09218}{\texttt{arXiv:1610.09218}}.

\bibitem{SUS-19-009_PAPER}
\hrefCMSnoop {}{{CMS Collaboration}, ``{S}earch for direct top squark pair
  production in events with one lepton, jets, and missing transverse momentum
  at 13 {TeV} with the {CMS} experiment'',} \textit{ JHEP} \textbf{ 05} (2020)
  032,
  \href{http://dx.doi.org/10.1007/JHEP05(2020)032}{\doi{10.1007/JHEP05(2020)032}},
  \href{http://www.arXiv.org/abs/1912.08887}{\texttt{arXiv:1912.08887}}.

\bibitem{SUS-19-011_PAPER}
\hrefCMSnoop {}{{CMS Collaboration}, ``{S}earch for top squark pair production
  using dilepton final states in pp collision data collected at $\sqrt{s}$ = 13
  {TeV}'',} \textit{ Eur. Phys. J. C} \textbf{ 81} (2021) 3,
  \href{http://dx.doi.org/10.1140/epjc/s10052-020-08701-5}{\doi{10.1140/epjc/s10052-020-08701-5}},
  \href{http://www.arXiv.org/abs/2008.05936}{\texttt{arXiv:2008.05936}}.

\bibitem{SUS-20-002_PAPER}
\hrefCMSnoop {}{{CMS Collaboration}, ``{C}ombined searches for the production
  of supersymmetric top quark partners in proton-proton collisions at
  $\sqrt{s}$ = 13 {TeV}'',} \textit{ Eur. Phys. J. C} \textbf{ 81} (2021) 970,
  \href{http://dx.doi.org/10.1140/epjc/s10052-021-09721-5}{\doi{10.1140/epjc/s10052-021-09721-5}},
  \href{http://www.arXiv.org/abs/2107.10892}{\texttt{arXiv:2107.10892}}.

\bibitem{CMS:2020ulv}
\hrefCMSnoop {}{{CMS Collaboration}, ``{S}earch for dark matter produced in
  association with a leptonically decaying z boson in proton-proton collisions
  at $\sqrt{s} =$ 13 {TeV}'',} \textit{ Eur. Phys. J. C} \textbf{ 81} (2021)
  13,
  \href{http://dx.doi.org/10.1140/epjc/s10052-020-08739-5}{\doi{10.1140/epjc/s10052-020-08739-5}},
  \href{http://www.arXiv.org/abs/2008.04735}{\texttt{arXiv:2008.04735}}.
  [Erratum: \DOI{10.1140/epjc/s10052-021-08959-3}].

\bibitem{CMS_MonojetV}
\hrefCMSnoop {}{{CMS Collaboration}, ``{S}earch for new particles in events
  with energetic jets and large missing transverse momentum in proton-proton
  collisions at $ \sqrt{s} $ = 13 {TeV}'',} \textit{ JHEP} \textbf{ 11} (2021)
  153,
  \href{http://dx.doi.org/10.1007/JHEP11(2021)153}{\doi{10.1007/JHEP11(2021)153}},
  \href{http://www.arXiv.org/abs/2107.13021}{\texttt{arXiv:2107.13021}}.

\bibitem{VBF-Run2-paper}
\hrefCMSnoop {}{{CMS Collaboration}, ``{S}earch for invisible decays of the
  {H}iggs boson produced via vector boson fusion in proton-proton collisions at
  $\sqrt{s} =$ 13 {TeV}'',} \textit{ Phys. Rev. D} \textbf{ 105} (2022) 092007,
  \href{http://dx.doi.org/10.1103/PhysRevD.105.092007}{\doi{10.1103/PhysRevD.105.092007}},
  \href{http://www.arXiv.org/abs/2201.11585}{\texttt{arXiv:2201.11585}}.

\bibitem{CMS:2020cmk}
\hrefCMSnoop {}{{CMS Collaboration}, ``{P}erformance of the {CMS} level-1
  trigger in proton-proton collisions at $\sqrt{s}$ = 13 {TeV}'',} \textit{
  JINST} \textbf{ 15} (2020) P10017,
  \href{http://dx.doi.org/10.1088/1748-0221/15/10/P10017}{\doi{10.1088/1748-0221/15/10/P10017}},
  \href{http://www.arXiv.org/abs/2006.10165}{\texttt{arXiv:2006.10165}}.

\bibitem{CMS:2016ngn}
\hrefCMSnoop {}{{CMS Collaboration}, ``{T}he {CMS} trigger system'',} \textit{
  JINST} \textbf{ 12} (2017) P01020,
  \href{http://dx.doi.org/10.1088/1748-0221/12/01/P01020}{\doi{10.1088/1748-0221/12/01/P01020}},
\href{http://www.arXiv.org/abs/1609.02366}{\texttt{arXiv:1609.02366}}.

\bibitem{CMS:2020uim}
\hrefCMSnoop {}{{CMS Collaboration}, ``{E}lectron and photon reconstruction and
  identification with the {CMS} experiment at the {CERN} {LHC}'',} \textit{
  JINST} \textbf{ 16} (2021) P05014,
  \href{http://dx.doi.org/10.1088/1748-0221/16/05/P05014}{\doi{10.1088/1748-0221/16/05/P05014}},
  \href{http://www.arXiv.org/abs/2012.06888}{\texttt{arXiv:2012.06888}}.

\bibitem{CMS:2018rym}
\hrefCMSnoop {}{{CMS Collaboration}, ``{P}erformance of the {CMS} muon detector
  and muon reconstruction with proton-proton collisions at $\sqrt{s}$ = 13
  {TeV}'',} \textit{ JINST} \textbf{ 13} (2018) P06015,
  \href{http://dx.doi.org/10.1088/1748-0221/13/06/P06015}{\doi{10.1088/1748-0221/13/06/P06015}},
  \href{http://www.arXiv.org/abs/1804.04528}{\texttt{arXiv:1804.04528}}.

\bibitem{CMS:2014pgm}
\hrefCMSnoop {}{{CMS Collaboration}, ``{D}escription and performance of track
  and primary-vertex reconstruction with the {CMS} tracker'',} \textit{ JINST}
  \textbf{ 9} (2014) P10009,
  \href{http://dx.doi.org/10.1088/1748-0221/9/10/P10009}{\doi{10.1088/1748-0221/9/10/P10009}},
  \href{http://www.arXiv.org/abs/1405.6569}{\texttt{arXiv:1405.6569}}.

\bibitem{CMS:2017yfk}
\hrefCMSnoop {}{{CMS Collaboration}, ``{P}article-flow reconstruction and
  global event description with the {CMS} detector'',} \textit{ JINST} \textbf{
  12} (2017) P10003,
  \href{http://dx.doi.org/10.1088/1748-0221/12/10/P10003}{\doi{10.1088/1748-0221/12/10/P10003}},
\href{http://www.arXiv.org/abs/1706.04965}{\texttt{arXiv:1706.04965}}.

\bibitem{CMS:2018jrd}
\hrefCMSnoop {}{{CMS Collaboration}, ``{P}erformance of reconstruction and
  identification of $\tau$ leptons decaying to hadrons and $\nu_\tau$ in pp
  collisions at $\sqrt{s}$ = 13 {TeV}'',} \textit{ JINST} \textbf{ 13} (2018)
  P10005,
  \href{http://dx.doi.org/10.1088/1748-0221/13/10/P10005}{\doi{10.1088/1748-0221/13/10/P10005}},
  \href{http://www.arXiv.org/abs/1809.02816}{\texttt{arXiv:1809.02816}}.

\bibitem{CMS:2016lmd}
\hrefCMSnoop {}{{CMS Collaboration}, ``{J}et energy scale and resolution in the
  {CMS} experiment in pp collisions at 8 {TeV}'',} \textit{ JINST} \textbf{ 12}
  (2017) P02014,
  \href{http://dx.doi.org/10.1088/1748-0221/12/02/P02014}{\doi{10.1088/1748-0221/12/02/P02014}},
  \href{http://www.arXiv.org/abs/1607.03663}{\texttt{arXiv:1607.03663}}.

\bibitem{CMS:2019ctu}
\hrefCMSnoop {}{{CMS Collaboration}, ``{P}erformance of missing transverse
  momentum reconstruction in proton-proton collisions at $\sqrt{s}$ = 13 {TeV}
  using the {CMS} detector'',} \textit{ JINST} \textbf{ 14} (2019) P07004,
  \href{http://dx.doi.org/10.1088/1748-0221/14/07/P07004}{\doi{10.1088/1748-0221/14/07/P07004}},
\href{http://www.arXiv.org/abs/1903.06078}{\texttt{arXiv:1903.06078}}.

\bibitem{CMS:2021xjt}
\hrefCMSnoop {}{{CMS Collaboration}, ``{P}recision luminosity measurement in
  proton-proton collisions at $\sqrt{s}$ = 13 {TeV} in 2015 and 2016 at
  {CMS}'',} \textit{ Eur. Phys. J. C} \textbf{ 81} (2021) 800,
  \href{http://dx.doi.org/10.1140/epjc/s10052-021-09538-2}{\doi{10.1140/epjc/s10052-021-09538-2}},
  \href{http://www.arXiv.org/abs/2104.01927}{\texttt{arXiv:2104.01927}}.

\bibitem{CMS-PAS-LUM-17-004}
\href {https://cds.cern.ch/record/2621960}{{CMS Collaboration}, ``{CMS}
  luminosity measurement for the 2017 data-taking period at $\sqrt{s}$ = 13
  {TeV}'',} CMS Physics Analysis Summary CMS-PAS-LUM-17-004, 2018.

\bibitem{CMS-PAS-LUM-18-002}
\href {https://cds.cern.ch/record/2676164}{{CMS Collaboration}, ``{CMS}
  luminosity measurement for the 2018 data-taking period at $\sqrt{s}$ = 13
  {TeV}'',} CMS Physics Analysis Summary CMS-PAS-LUM-18-002, 2019.

\bibitem{CMS:2008xjf}
\hrefCMSnoop {}{{CMS Collaboration}, ``{T}he {CMS} experiment at the {CERN}
  {LHC}'',} \textit{ JINST} \textbf{ 3} (2008) S08004,
  \href{http://dx.doi.org/10.1088/1748-0221/3/08/S08004}{\doi{10.1088/1748-0221/3/08/S08004}}.

\bibitem{Oleari:2010nx}
\hrefCMSnoop {}{C.~Oleari, ``{T}he {POWHEG-BOX}'',} \textit{ Nucl. Phys. B
  Proc. Suppl.} \textbf{ 205-206} (2010) 36,
  \href{http://dx.doi.org/10.1016/j.nuclphysbps.2010.08.016}{\doi{10.1016/j.nuclphysbps.2010.08.016}},
  \href{http://www.arXiv.org/abs/1007.3893}{\texttt{arXiv:1007.3893}}.

\bibitem{Alwall:2014hca}
J.~Alwall\hrefCMSnoop {}{ {et~al.}, ``{T}he automated computation of tree-level
  and next-to-leading order differential cross sections, and their matching to
  parton shower simulations'',} \textit{ JHEP} \textbf{ 07} (2014) 079,
  \href{http://dx.doi.org/10.1007/JHEP07(2014)079}{\doi{10.1007/JHEP07(2014)079}},
  \href{http://www.arXiv.org/abs/1405.0301}{\texttt{arXiv:1405.0301}}.

\bibitem{Sjostrand:2014zea}
T.~Sj{\"o}strand\hrefCMSnoop {}{ {et~al.}, ``{A}n introduction to {PYTHIA}
  8.2'',} \textit{ Comput. Phys. Commun.} \textbf{ 191} (2015) 159,
  \href{http://dx.doi.org/10.1016/j.cpc.2015.01.024}{\doi{10.1016/j.cpc.2015.01.024}},
  \href{http://www.arXiv.org/abs/1410.3012}{\texttt{arXiv:1410.3012}}.

\bibitem{PythiaTune}
\hrefCMSnoop {}{{CMS Collaboration}, ``{E}xtraction and validation of a new set
  of {CMS} {PYTHIA}8 tunes from underlying-event measurements'',} \textit{ Eur.
  Phys. J. C} \textbf{ 80} (2020) 4,
  \href{http://dx.doi.org/10.1140/epjc/s10052-019-7499-4}{\doi{10.1140/epjc/s10052-019-7499-4}},
  \href{http://www.arXiv.org/abs/1903.12179}{\texttt{arXiv:1903.12179}}.

\bibitem{Geant4}
\hrefCMSnoop {}{{GEANT4} Collaboration, ``{GEANT}4 --- a simulation toolkit'',}
  \textit{ Nucl. Instrum. Meth. A} \textbf{ 506} (2003) 250,
  \href{http://dx.doi.org/10.1016/S0168-9002(03)01368-8}{\doi{10.1016/S0168-9002(03)01368-8}}.

\bibitem{PDFs}
\hrefCMSnoop {}{{NNPDF} Collaboration, ``{P}arton distributions from
  high-precision collider data'',} \textit{ Eur. Phys. J. C} \textbf{ 77}
  (2017) 663,
  \href{http://dx.doi.org/10.1140/epjc/s10052-017-5199-5}{\doi{10.1140/epjc/s10052-017-5199-5}},
  \href{http://www.arXiv.org/abs/1706.00428}{\texttt{arXiv:1706.00428}}.

\bibitem{POWHEGttH}
\hrefCMSnoop {}{H.~B. Hartanto, B.~Jager, L.~Reina, and D.~Wackeroth, ``{H}iggs
  boson production in association with top quarks in the {POWHEG} {BOX}'',}
  \textit{ Phys. Rev. D} \textbf{ 91} (2015) 094003,
  \href{http://dx.doi.org/10.1103/PhysRevD.91.094003}{\doi{10.1103/PhysRevD.91.094003}},
  \href{http://www.arXiv.org/abs/1501.04498}{\texttt{arXiv:1501.04498}}.

\bibitem{POWHEGVBF}
\hrefCMSnoop {}{P.~Nason and C.~Oleari, ``{NLO} {H}iggs boson production via
  vector-boson fusion matched with shower in {POWHEG}'',} \textit{ JHEP}
  \textbf{ 02} (2010) 037,
  \href{http://dx.doi.org/10.1007/JHEP02(2010)037}{\doi{10.1007/JHEP02(2010)037}},
  \href{http://www.arXiv.org/abs/0911.5299}{\texttt{arXiv:0911.5299}}.

\bibitem{POWHEGVH}
\hrefCMSnoop {}{G.~Luisoni, P.~Nason, C.~Oleari, and F.~Tramontano,
  ``{HW$^{\pm}$}/{HZ} $+$ 0 and 1 jet at {NLO} with the {POWHEG} {BOX}
  interfaced to {G}o{S}am and their merging within {M}i{NLO}'',} \textit{ JHEP}
  \textbf{ 10} (2013) 083,
  \href{http://dx.doi.org/10.1007/JHEP10(2013)083}{\doi{10.1007/JHEP10(2013)083}},
  \href{http://www.arXiv.org/abs/1306.2542}{\texttt{arXiv:1306.2542}}.

\bibitem{POWHEGggF}
\hrefCMSnoop {}{E.~Bagnaschi, G.~Degrassi, P.~Slavich, and A.~Vicini, ``{H}iggs
  production via gluon fusion in the {POWHEG} approach in the {SM} and in the
  {MSSM}'',} \textit{ JHEP} \textbf{ 02} (2012) 088,
  \href{http://dx.doi.org/10.1007/JHEP02(2012)088}{\doi{10.1007/JHEP02(2012)088}},
  \href{http://www.arXiv.org/abs/1111.2854}{\texttt{arXiv:1111.2854}}.

\bibitem{deFlorian:2016spz}
\hrefCMSnoop {}{{LHC Higgs Cross Section Working Group}, ``{H}andbook of {LHC}
  {H}iggs cross sections: 4. deciphering the nature of the {H}iggs sector'',}
  {CERN} {R}eport CERN-2017-002-M, 2016.
\newblock
  \href{http://dx.doi.org/10.23731/CYRM-2017-002}{\doi{10.23731/CYRM-2017-002}},
  \href{http://www.arXiv.org/abs/1610.07922}{\texttt{arXiv:1610.07922}}.

\bibitem{Frederix:2012ps}
\hrefCMSnoop {}{R.~Frederix and S.~Frixione, ``{M}erging meets matching in
  {MC@NLO}'',} \textit{ JHEP} \textbf{ 12} (2012) 061,
  \href{http://dx.doi.org/10.1007/JHEP12(2012)061}{\doi{10.1007/JHEP12(2012)061}},
  \href{http://www.arXiv.org/abs/1209.6215}{\texttt{arXiv:1209.6215}}.

\bibitem{Zanoli:2021iyp}
S.~Zanoli\hrefCMSnoop {}{ {et~al.}, ``{N}ext-to-next-to-leading order event
  generation for {VH} production with {H} ${\rightarrow b\overline{b}}$
  decay'',} \textit{ JHEP} \textbf{ 07} (2022) 008,
  \href{http://dx.doi.org/10.1007/JHEP07(2022)008}{\doi{10.1007/JHEP07(2022)008}},
  \href{http://www.arXiv.org/abs/2112.04168}{\texttt{arXiv:2112.04168}}.

\bibitem{Mangano:2006rw}
\hrefCMSnoop {}{M.~L. Mangano, M.~Moretti, F.~Piccinini, and M.~Treccani,
  ``{M}atching matrix elements and shower evolution for top-quark production in
  hadronic collisions'',} \textit{ JHEP} \textbf{ 01} (2007) 013,
  \href{http://dx.doi.org/10.1088/1126-6708/2007/01/013}{\doi{10.1088/1126-6708/2007/01/013}},
  \href{http://www.arXiv.org/abs/hep-ph/0611129}{\texttt{arXiv:hep-ph/0611129}}.

\bibitem{POWHEG:doubletop}
\hrefCMSnoop {}{J.~M. Campbell, R.~K. Ellis, P.~Nason, and E.~Re, ``{T}op-pair
  production and decay at {NLO} matched with parton showers'',} \textit{ JHEP}
  \textbf{ 04} (2015) 114,
  \href{http://dx.doi.org/10.1007/JHEP04(2015)114}{\doi{10.1007/JHEP04(2015)114}},
  \href{http://www.arXiv.org/abs/1412.1828}{\texttt{arXiv:1412.1828}}.

\bibitem{Alioli:2009je}
\hrefCMSnoop {}{S.~Alioli, P.~Nason, C.~Oleari, and E.~Re, ``{NLO} single-top
  production matched with shower in {POWHEG}: $s$- and $t$-channel
  contributions'',} \textit{ JHEP} \textbf{ 09} (2009) 111,
  \href{http://dx.doi.org/10.1088/1126-6708/2009/09/111}{\doi{10.1088/1126-6708/2009/09/111}},
  \href{http://www.arXiv.org/abs/0907.4076}{\texttt{arXiv:0907.4076}}.
  [Erratum: \DOI{10.1007/JHEP02(2010)011}].

\bibitem{POWHEG:singletop}
\hrefCMSnoop {}{E.~Re, ``{S}ingle-top {W}t-channel production matched with
  parton showers using the {POWHEG} method'',} \textit{ Eur. Phys. J. C}
  \textbf{ 71} (2011) 1547,
  \href{http://dx.doi.org/10.1140/epjc/s10052-011-1547-z}{\doi{10.1140/epjc/s10052-011-1547-z}},
  \href{http://www.arXiv.org/abs/1009.2450}{\texttt{arXiv:1009.2450}}.

\bibitem{Czakon:2017wor}
M.~Czakon\hrefCMSnoop {}{ {et~al.}, ``{T}op-pair production at the {LHC}
  through {NNLO} {QCD} and {NLO} {EW}'',} \textit{ JHEP} \textbf{ 10} (2017)
  186,
  \href{http://dx.doi.org/10.1007/JHEP10(2017)186}{\doi{10.1007/JHEP10(2017)186}},
  \href{http://www.arXiv.org/abs/1705.04105}{\texttt{arXiv:1705.04105}}.

\bibitem{Artoisenet:2012st}
\hrefCMSnoop {}{P.~Artoisenet, R.~Frederix, O.~Mattelaer, and R.~Rietkerk,
  ``{A}utomatic spin-entangled decays of heavy resonances in monte carlo
  simulations'',} \textit{ JHEP} \textbf{ 03} (2013) 015,
  \href{http://dx.doi.org/10.1007/JHEP03(2013)015}{\doi{10.1007/JHEP03(2013)015}},
  \href{http://www.arXiv.org/abs/1212.3460}{\texttt{arXiv:1212.3460}}.

\bibitem{POWHEG:WW}
\hrefCMSnoop {}{T.~Melia, P.~Nason, R.~Rontsch, and G.~Zanderighi,
  ``{W$^+$W$^-$}, {WZ} and {ZZ} production in the {POWHEG} {BOX}'',} \textit{
  JHEP} \textbf{ 11} (2011) 078,
  \href{http://dx.doi.org/10.1007/JHEP11(2011)078}{\doi{10.1007/JHEP11(2011)078}},
  \href{http://www.arXiv.org/abs/1107.5051}{\texttt{arXiv:1107.5051}}.

\bibitem{CACCIARI2008119}
\hrefCMSnoop {}{M.~Cacciari and G.~P. Salam, ``{P}ileup subtraction using jet
  areas'',} \textit{ Phys. Lett. B} \textbf{ 659} (2008) 119,
  \href{http://dx.doi.org/10.1016/j.physletb.2007.09.077}{\doi{10.1016/j.physletb.2007.09.077}},
  \href{http://www.arXiv.org/abs/0707.1378}{\texttt{arXiv:0707.1378}}.

\bibitem{CMS-TDR-15-02}
\href {http://cds.cern.ch/record/2020886}{{CMS Collaboration}, ``{T}echnical
  proposal for the phase-{II} upgrade of the compact muon solenoid'',} CMS
  Technical Proposal CERN-LHCC-2015-010, CMS-TDR-15-02, 2015.

\bibitem{AntiKt}
\hrefCMSnoop {}{M.~Cacciari, G.~P. Salam, and G.~Soyez, ``{T}he
  anti-$\mathrm{k_{T}}$ jet clustering algorithm'',} \textit{ JHEP} \textbf{
  04} (2008) 063,
  \href{http://dx.doi.org/10.1088/1126-6708/2008/04/063}{\doi{10.1088/1126-6708/2008/04/063}},
  \href{http://www.arXiv.org/abs/0802.1189}{\texttt{arXiv:0802.1189}}.

\bibitem{Cacciari:2011ma}
\hrefCMSnoop {}{M.~Cacciari, G.~P. Salam, and G.~Soyez, ``{F}astjet user
  manual'',} \textit{ Eur. Phys. J. C} \textbf{ 72} (2012) 1896,
  \href{http://dx.doi.org/10.1140/epjc/s10052-012-1896-2}{\doi{10.1140/epjc/s10052-012-1896-2}},
  \href{http://www.arXiv.org/abs/1111.6097}{\texttt{arXiv:1111.6097}}.

\bibitem{CMS_PAS_JME-14-001}
\href {https://cds.cern.ch/record/1751454}{{CMS Collaboration}, ``{S}tudy of
  pileup removal algorithms for jets'',} CMS Physics Analysis Summary
  CMS-PAS-JME-14-001, 2014.

\bibitem{Sirunyan:2017ezt}
\hrefCMSnoop {}{{CMS Collaboration}, ``{I}dentification of heavy-flavour jets
  with the {CMS} detector in pp collisions at 13 {TeV}'',} \textit{ JINST}
  \textbf{ 13} (2018) P05011,
  \href{http://dx.doi.org/10.1088/1748-0221/13/05/P05011}{\doi{10.1088/1748-0221/13/05/P05011}},
\href{http://www.arXiv.org/abs/1712.07158}{\texttt{arXiv:1712.07158}}.

\bibitem{Sirunyan:2020foa}
\hrefCMSnoop {}{{CMS Collaboration}, ``{P}ileup mitigation at {CMS} in 13 {TeV}
  data'',} \textit{ JINST} \textbf{ 15} (2020) P09018,
  \href{http://dx.doi.org/10.1088/1748-0221/15/09/P09018}{\doi{10.1088/1748-0221/15/09/P09018}},
  \href{http://www.arXiv.org/abs/2003.00503}{\texttt{arXiv:2003.00503}}.

\bibitem{Bertolini:2014bba}
\hrefCMSnoop {}{D.~Bertolini, P.~Harris, M.~Low, and N.~Tran, ``{P}ileup per
  particle identification'',} \textit{ JHEP} \textbf{ 10} (2014) 059,
  \href{http://dx.doi.org/10.1007/JHEP10(2014)059}{\doi{10.1007/JHEP10(2014)059}},
  \href{http://www.arXiv.org/abs/1407.6013}{\texttt{arXiv:1407.6013}}.

\bibitem{Larkoski:2014wba}
\hrefCMSnoop {}{A.~J. Larkoski, S.~Marzani, G.~Soyez, and J.~Thaler, ``{S}oft
  drop'',} \textit{ JHEP} \textbf{ 05} (2014) 146,
  \href{http://dx.doi.org/10.1007/JHEP05(2014)146}{\doi{10.1007/JHEP05(2014)146}},
  \href{http://www.arXiv.org/abs/1402.2657}{\texttt{arXiv:1402.2657}}.

\bibitem{Dasgupta:2013ihk}
\hrefCMSnoop {}{M.~Dasgupta, A.~Fregoso, S.~Marzani, and G.~P. Salam,
  ``{T}owards an understanding of jet substructure'',} \textit{ JHEP} \textbf{
  09} (2013) 029,
  \href{http://dx.doi.org/10.1007/JHEP09(2013)029}{\doi{10.1007/JHEP09(2013)029}},
\href{http://www.arXiv.org/abs/1307.0007}{\texttt{arXiv:1307.0007}}.

\bibitem{Butterworth:2008iy}
\hrefCMSnoop {}{J.~M. Butterworth, A.~R. Davison, M.~Rubin, and G.~P. Salam,
  ``{J}et substructure as a new {H}iggs search channel at the {LHC}'',}
  \textit{ Phys. Rev. Lett.} \textbf{ 100} (2008) 242001,
  \href{http://dx.doi.org/10.1103/PhysRevLett.100.242001}{\doi{10.1103/PhysRevLett.100.242001}},
\href{http://www.arXiv.org/abs/0802.2470}{\texttt{arXiv:0802.2470}}.

\bibitem{Thaler:2010tr}
\hrefCMSnoop {}{J.~Thaler and K.~Van~Tilburg, ``{I}dentifying boosted objects
  with {$N$}-subjettiness'',} \textit{ JHEP} \textbf{ 03} (2011) 015,
  \href{http://dx.doi.org/10.1007/JHEP03(2011)015}{\doi{10.1007/JHEP03(2011)015}},
\href{http://www.arXiv.org/abs/1011.2268}{\texttt{arXiv:1011.2268}}.

\bibitem{CMS-PAS-JME-18-002}
\hrefCMSnoop {}{{CMS Collaboration}, ``{I}dentification of heavy, energetic,
  hadronically decaying particles using machine-learning techniques'',}
  \textit{ JINST} \textbf{ 15} (2020) P06005,
  \href{http://dx.doi.org/10.1088/1748-0221/15/06/P06005}{\doi{10.1088/1748-0221/15/06/P06005}},
  \href{http://www.arXiv.org/abs/2004.08262}{\texttt{arXiv:2004.08262}}.

\bibitem{Cowan:2010js}
\hrefCMSnoop {}{G.~Cowan, K.~Cranmer, E.~Gross, and O.~Vitells, ``{A}symptotic
  formulae for likelihood-based tests of new physics'',} \textit{ Eur. Phys. J.
  C} \textbf{ 71} (2011) 1554,
  \href{http://dx.doi.org/10.1140/epjc/s10052-011-1554-0}{\doi{10.1140/epjc/s10052-011-1554-0}},
  \href{http://www.arXiv.org/abs/1007.1727}{\texttt{arXiv:1007.1727}}.
  [Erratum: \DOI{10.1140/epjc/s10052-013-2501-z}].

\bibitem{sakuma2019alternative}
\hrefCMSnoop {}{T.~Sakuma, H.~Fl{\"a}cher, and D.~Smith, ``{A}lternative
  angular variables for suppression of {QCD} multijet events in new physics
  searches with missing transverse momentum at the {LHC}'',} 2018.
  \href{http://www.arXiv.org/abs/1803.07942}{\texttt{arXiv:1803.07942}}.

\bibitem{CMS:2015xaf}
\hrefCMSnoop {}{{CMS Collaboration}, ``{P}erformance of electron reconstruction
  and selection with the {CMS} detector in proton-proton collisions at
  $\sqrt{s}$ = 8 {TeV}'',} \textit{ JINST} \textbf{ 10} (2015) P06005,
  \href{http://dx.doi.org/10.1088/1748-0221/10/06/P06005}{\doi{10.1088/1748-0221/10/06/P06005}},
  \href{http://www.arXiv.org/abs/1502.02701}{\texttt{arXiv:1502.02701}}.

\bibitem{Lindert:2017olm}
\hrefCMSnoop {}{J.~M. Lindert, S.~Pozzorini, R.~Boughezal {et~al.}, ``{P}recise
  predictions for {$V+$} jets dark matter backgrounds'',} \textit{ Eur. Phys.
  J. C} \textbf{ 77} (2017) 829,
  \href{http://dx.doi.org/10.1140/epjc/s10052-017-5389-1}{\doi{10.1140/epjc/s10052-017-5389-1}},
  \href{http://www.arXiv.org/abs/1705.04664}{\texttt{arXiv:1705.04664}}.

\bibitem{Khachatryan:2014jba}
\hrefCMSnoop {}{{CMS Collaboration}, ``{P}recise determination of the mass of
  the {H}iggs boson and tests of compatibility of its couplings with the
  standard model predictions using proton collisions at 7 and 8 {TeV}'',}
  \textit{ Eur. Phys. J. C} \textbf{ 75} (2015) 212,
  \href{http://dx.doi.org/10.1140/epjc/s10052-015-3351-7}{\doi{10.1140/epjc/s10052-015-3351-7}},
\href{http://www.arXiv.org/abs/1412.8662}{\texttt{arXiv:1412.8662}}.

\bibitem{CMS-NOTE-2011-005}
\href {https://cds.cern.ch/record/1379837}{{ATLAS and CMS Collaborations, and
  LHC Higgs Combination Group}, ``{P}rocedure for the {LHC} {H}iggs boson
  search combination in summer 2011'',} {T}echnical report CMS-NOTE-2011-005,
  ATL-PHYS-PUB-2011-11, 2011.

\bibitem{CLS1}
\hrefCMSnoop {}{A.~L. Read, ``{P}resentation of search results: The
  {CL$_{\text{s}}$} technique'',} \textit{ J. Phys. G} \textbf{ 28} (2002)
  2693,
\href{http://dx.doi.org/10.1088/0954-3899/28/10/313}{\doi{10.1088/0954-3899/28/10/313}}.

\bibitem{CLS2}
\hrefCMSnoop {}{T.~Junk, ``{C}onfidence level computation for combining
  searches with small statistics'',} \textit{ Nucl. Instrum. Meth. A} \textbf{
  434} (1999) 435,
  \href{http://dx.doi.org/10.1016/S0168-9002(99)00498-2}{\doi{10.1016/S0168-9002(99)00498-2}},
\href{http://www.arXiv.org/abs/hep-ex/9902006}{\texttt{arXiv:hep-ex/9902006}}.

\bibitem{CMS:2014gab}
\hrefCMSnoop {}{{CMS Collaboration}, ``{S}earch for invisible decays of {H}iggs
  bosons in the vector boson fusion and associated {ZH} production modes'',}
  \textit{ Eur. Phys. J. C} \textbf{ 74} (2014) 2980,
  \href{http://dx.doi.org/10.1140/epjc/s10052-014-2980-6}{\doi{10.1140/epjc/s10052-014-2980-6}},
  \href{http://www.arXiv.org/abs/1404.1344}{\texttt{arXiv:1404.1344}}.

\bibitem{CMS:2016xus}
\hrefCMSnoop {}{{CMS Collaboration}, ``{S}earch for dark matter in
  proton-proton collisions at 8 {TeV} with missing transverse momentum and
  vector boson tagged jets'',} \textit{ JHEP} \textbf{ 12} (2016) 083,
  \href{http://dx.doi.org/10.1007/JHEP12(2016)083}{\doi{10.1007/JHEP12(2016)083}},
  \href{http://www.arXiv.org/abs/1607.05764}{\texttt{arXiv:1607.05764}}.
  [Erratum: \DOI{10.1007/JHEP08(2017)035}].

\bibitem{Baker:1983tu}
\hrefCMSnoop {}{S.~Baker and R.~D. Cousins, ``{C}larification of the use of chi
  square and likelihood functions in fits to histograms'',} \textit{ Nucl.
  Instrum. Meth.} \textbf{ 221} (1984) 437,
  \href{http://dx.doi.org/10.1016/0167-5087(84)90016-4}{\doi{10.1016/0167-5087(84)90016-4}}.

\bibitem{HEPData}
\hrefCMSnoop {}{}{HEPData} record for this analysis, 2023.
\newblock
  \href{http://dx.doi.org/10.17182/hepdata.137761}{\doi{10.17182/hepdata.137761}}.

\bibitem{Zaazoua_2022}
\hrefCMSnoop {}{M.~Zaazoua, L.~Truong, K.~A. Assamagan, and F.~Fassi, ``{H}iggs
  portal vector dark matter interpretation: Review of effective field theory
  approach and ultraviolet complete models'',} \textit{ LHEP} \textbf{ 2022}
  (2022) 270,
  \href{http://dx.doi.org/10.31526/lhep.2022.270}{\doi{10.31526/lhep.2022.270}},
  \href{http://www.arXiv.org/abs/2107.01252}{\texttt{arXiv:2107.01252}}.

\bibitem{XENON:2019zpr}
\hrefCMSnoop {}{{XENON} Collaboration, ``{S}earch for light dark matter
  interactions enhanced by the {M}igdal effect or {B}remsstrahlung in
  {XENON1T}'',} \textit{ Phys. Rev. Lett.} \textbf{ 123} (2019) 241803,
  \href{http://dx.doi.org/10.1103/PhysRevLett.123.241803}{\doi{10.1103/PhysRevLett.123.241803}},
  \href{http://www.arXiv.org/abs/1907.12771}{\texttt{arXiv:1907.12771}}.

\bibitem{DarkSide-50:2022qzh}
\hrefCMSnoop {}{{DarkSide-50} Collaboration, ``{S}earch for low-mass dark
  matter {WIMP}s with 12 ton-day exposure of {D}ark{S}ide-50'',} 2022.
  \href{http://www.arXiv.org/abs/2207.11966}{\texttt{arXiv:2207.11966}}.

\bibitem{Wenbo:2022coa}
\hrefCMSnoop {}{{PandaX} Collaboration, ``{A} first search for solar $^8${B}
  neutrino in the {PandaX-4T} experiment using neutrino-nucleus coherent
  scattering'',} 2022.
  \href{http://www.arXiv.org/abs/2207.04883}{\texttt{arXiv:2207.04883}}.

\bibitem{LZ:2022ufs}
\hrefCMSnoop {}{{LZ} Collaboration, ``{F}irst dark matter search results from
  the {LUX-ZEPLIN} ({LZ}) experiment'',} 2022.
  \href{http://www.arXiv.org/abs/2207.03764}{\texttt{arXiv:2207.03764}}.

\bibitem{Young:2009zb}
\hrefCMSnoop {}{R.~D. Young and A.~W. Thomas, ``{O}ctet baryon masses and sigma
  terms from an {SU}(3) chiral extrapolation'',} \textit{ Phys. Rev. D}
  \textbf{ 81} (2010) 014503,
  \href{http://dx.doi.org/10.1103/PhysRevD.81.014503}{\doi{10.1103/PhysRevD.81.014503}},
  \href{http://www.arXiv.org/abs/0901.3310}{\texttt{arXiv:0901.3310}}.

\bibitem{Toussaint:2009pz}
\hrefCMSnoop {}{{MILC} Collaboration, ``{T}he strange quark condensate in the
  nucleon in 2+1 flavor {QCD}'',} \textit{ Phys. Rev. Lett.} \textbf{ 103}
  (2009) 122002,
  \href{http://dx.doi.org/10.1103/PhysRevLett.103.122002}{\doi{10.1103/PhysRevLett.103.122002}},
  \href{http://www.arXiv.org/abs/0905.2432}{\texttt{arXiv:0905.2432}}.

\bibitem{kappaV-kappaF}
\hrefCMSnoop {}{{LHC Higgs Cross Section Working Group}, ``{H}andbook of {LHC}
  {H}iggs cross sections: 3. {H}iggs properties'',} {T}echnical report, 2013.
\newblock
  \href{http://dx.doi.org/10.5170/CERN-2013-004}{\doi{10.5170/CERN-2013-004}},
  \href{http://www.arXiv.org/abs/1307.1347}{\texttt{arXiv:1307.1347}}.

\end{thebibliography}\endgroup
\cleardoublepage \appendix\section{The CMS Collaboration \label{app:collab}}\begin{sloppypar}\hyphenpenalty=5000\widowpenalty=500\clubpenalty=5000
\cmsinstitute{Yerevan Physics Institute, Yerevan, Armenia}
{\tolerance=6000
A.~Tumasyan\cmsAuthorMark{1}\cmsorcid{0009-0000-0684-6742}
\par}
\cmsinstitute{Institut f\"{u}r Hochenergiephysik, Vienna, Austria}
{\tolerance=6000
W.~Adam\cmsorcid{0000-0001-9099-4341}, J.W.~Andrejkovic, T.~Bergauer\cmsorcid{0000-0002-5786-0293}, S.~Chatterjee\cmsorcid{0000-0003-2660-0349}, K.~Damanakis\cmsorcid{0000-0001-5389-2872}, M.~Dragicevic\cmsorcid{0000-0003-1967-6783}, A.~Escalante~Del~Valle\cmsorcid{0000-0002-9702-6359}, P.S.~Hussain\cmsorcid{0000-0002-4825-5278}, M.~Jeitler\cmsAuthorMark{2}\cmsorcid{0000-0002-5141-9560}, N.~Krammer\cmsorcid{0000-0002-0548-0985}, L.~Lechner\cmsorcid{0000-0002-3065-1141}, D.~Liko\cmsorcid{0000-0002-3380-473X}, I.~Mikulec\cmsorcid{0000-0003-0385-2746}, P.~Paulitsch, J.~Schieck\cmsAuthorMark{2}\cmsorcid{0000-0002-1058-8093}, R.~Sch\"{o}fbeck\cmsorcid{0000-0002-2332-8784}, D.~Schwarz\cmsorcid{0000-0002-3821-7331}, M.~Sonawane\cmsorcid{0000-0003-0510-7010}, S.~Templ\cmsorcid{0000-0003-3137-5692}, W.~Waltenberger\cmsorcid{0000-0002-6215-7228}, C.-E.~Wulz\cmsAuthorMark{2}\cmsorcid{0000-0001-9226-5812}
% ...
\par}
\vskip\cmsinstskip
\dag:~Deceased\\
$^{1}$Also at Yerevan State University, Yerevan, Armenia\\
$^{2}$Also at TU Wien, Vienna, Austria\\
% ...
\end{sloppypar}
\end{document}